\documentclass[11pt,a4paper]{article}
\pdfoutput=1

\usepackage{jheppub}
\usepackage{latexsym}
\usepackage{revsymb}
\usepackage{multirow}
\usepackage{color}
\usepackage[usenames,dvipsnames,svgnames,table]{xcolor}

\usepackage{graphicx}% Include  files
\usepackage{epsfig}  % Include figure files
\usepackage{epsf}    % Include figure files
\usepackage{dcolumn}% Align table columns on decimal point
\usepackage{bm}% bold math
\usepackage{dcolumn}% Align table columns on decimal point
\usepackage{textcomp}% Align table columns on decimal point
\usepackage{float}
\usepackage{hypcap}
\usepackage[]{hyperref}
\usepackage{adjustbox}
\usepackage{subfigure}
\usepackage{alphalph}
\usepackage{morefloats}
\usepackage{textcomp}
\newcommand{\be}{\begin{equation}}
\newcommand{\ee}{\end{equation}}
\newcommand{\ba}{\begin{eqnarray}}
\newcommand{\ea}{\end{eqnarray}}
\def\nue{{\nu_e}}
\def\anue{{\bar\nu_e}}
\def\numu{{\nu_{\mu}}}
\def\anumu{{\bar\nu_{\mu}}}
\def\nutau{{\nu_{\tau}}}

\def\anu{{\bar\nu}}
\newcommand{\ms}{\Delta m^2_{21}}
\newcommand{\ma}{\Delta m^2_{31}}

\newcommand{\sa}{\sin^2 \theta_{23}}

\newcommand{\dcp}{\delta_{\mathrm{CP}}}
\newcommand{\tmt}{\theta_{23}}
\newcommand{\tet}{\theta_{13}}
\newcommand{\tem}{\theta_{12}}

\def\nue{{\nu_e}}
\def\anue{{\bar\nu_e}}
\def\numu{{\nu_{\mu}}}
\def\anumu{{\bar\nu_{\mu}}}
\def\nutau{{\nu_{\tau}}}

\newcommand{\ie}{{\it i.e.}}

\preprint{IP/BBSR/2015-3}

\title{Exploring Flavor-Dependent Long-Range Forces in Long-Baseline
Neutrino Oscillation Experiments} 

\author[a]{Sabya Sachi Chatterjee,}
\author[a,b]{Arnab Dasgupta,}
\author[a]{Sanjib Kumar Agarwalla}

\affiliation[a]{Institute of Physics, Sachivalaya Marg, Sainik School Post, Bhubaneswar 751005, India}
\affiliation[b]{Centre for Theoretical Physics, Jamia Millia Islamia - Central University, \\ Jamia Nagar, New Delhi - 110025, India} 

\emailAdd{sabya@iopb.res.in}
\emailAdd{arnab.d@iopb.res.in}
\emailAdd{sanjib@iopb.res.in}

\abstract{
The Standard Model gauge group can be extended with minimal matter
content by introducing anomaly free U(1) symmetry, such as $L_e-L_{\mu}$ 
or $L_e-L_{\tau}$. If the neutral gauge boson corresponding to this abelian
symmetry is ultra-light, then it will give rise to flavor-dependent long-range
leptonic force, which can have significant impact on neutrino oscillations.
For an instance, the electrons inside the Sun can generate a flavor-dependent
long-range potential at the Earth surface, which can suppress the 
$\numu \to \nue$ appearance probability in terrestrial experiments. 
The sign of this potential is opposite for anti-neutrinos, and affects the 
oscillations of (anti-)neutrinos in different fashion. This feature 
invokes fake CP-asymmetry like the SM matter effect and can severely 
affect the leptonic CP-violation searches in long-baseline experiments. 
In this paper, we study in detail the possible impacts of these long-range 
flavor-diagonal neutral current interactions due to $L_e-L_{\mu}$ symmetry,
when (anti-)neutrinos travel from Fermilab to Homestake (1300 km) 
and CERN to Pyh\"asalmi (2290 km) in the context of future high-precision 
superbeam facilities, DUNE and LBNO respectively. If there is no signal of
long-range force, DUNE (LBNO) can place stringent 
constraint on the effective gauge coupling 
$\alpha_{e\mu} < 1.9 \times 10^{-53}~(7.8 \times 10^{-54})$ 
at 90\% C.L., which is almost 30 (70) times better than the existing bound 
from the Super-Kamiokande experiment. We also observe that if
$\alpha_{e\mu} \geq 2 \times 10^{-52}$, the CP-violation 
discovery reach of these future facilities vanishes completely.
The mass hierarchy measurement remains robust in DUNE (LBNO)
if $\alpha_{e\mu} < 5 \times 10^{-52}~(10^{-52})$.
}
\arxivnumber{1509.03517}

\begin{document}
\maketitle
\flushbottom
%============

%==========================
\section{Introduction and Motivation}
\label{introduction}
%==========================

Over the past two decades or so, active attempts have been made 
both in experimental and theoretical fronts to improve our knowledge 
about neutrinos \cite{Mohapatra:2005wg,Strumia:2006db,GonzalezGarcia:2007ib}. 
The three most important fundamental issues that have taken center stage 
in neutrino physics as a part of these activities are the following.
First issue: how tiny is the neutrino mass? Second issue: can a neutrino
turn into its own anti-particle? Third issue: do different neutrino flavors
oscillate into one another? To shed light on the first issue, recently, 
the Planck Collaboration has reported an upper limit on the sum of 
all the neutrino mass eigenvalues of $\sum m_i < 0.23$ eV 
at $95\%$ C.L. in combination with the WMAP polarization and 
baryon acoustic oscillation (BAO) measurements \cite{Ade:2013zuv}.
Here, the sum runs over all the neutrino mass eigenstates which 
are in thermal equilibrium in the early Universe.
As far as the second issue is concerned, the hunt for neutrinoless
double beta decay process is still on which violates the total lepton number 
and requires Majorana neutrinos \cite{Avignone:2007fu,Vergados:2012xy,Rodejohann:2012xd,Pas:2015eia}.
The third question has been answered positively only in recent years
\cite{Bilenky:2014eza,Gil-Botella:2015qaa,Wang:2015rma}
and now, we have compelling evidence in favor of neutrino flavor oscillation
\cite{Pontecorvo:1967fh,Gribov:1968kq}, suggesting that leptonic flavors 
are not symmetries of Nature. This entails that neutrinos are massive and 
they mix with each other, providing an evidence for physics beyond 
the Standard Model (SM) \cite{Agashe:2014kda,Hewett:2012ns}.
This milestone has been achieved due to several fantastic world-class 
oscillation experiments involving neutrinos from the Sun
\cite{Cleveland:1998nv,Altmann:2005ix,Hosaka:2005um,Ahmad:2002jz,Aharmim:2008kc,Aharmim:2009gd,Arpesella:2008mt}, 
the Earth's atmosphere \cite{Fukuda:1998mi,Ashie:2005ik,Wendell:2010md},
nuclear reactors 
\cite{Araki:2004mb,Abe:2008aa,An:2012eh,An:2013uza,An:2013zwz,An:2015rpe,Ahn:2012nd,Abe:2011fz,Abe:2012tg,Abe:2013sxa,Abe:2014bwa}, 
and accelerators 
\cite{Ahn:2006zza,Adamson:2008zt,Adamson:2011qu,Adamson:2013ue,Adamson:2013whj,Abe:2011sj,Abe:2013xua,Abe:2013hdq,Abe:2014ugx,Abe:2015awa} 
which have enabled us to obtain a clear understanding of the lepton mixing pattern in three-flavor scenario.
Using the standard Particle Data Group convention \cite{Agashe:2014kda}, 
we parametrize the vacuum Pontecorvo-Maki-Nakagawa-Sakata (PMNS) matrix 
\cite{Pontecorvo:1957vz,Maki:1962mu,Pontecorvo:1967fh} in terms of the 
three mixing angles: $\tem$, $\tmt$, $\tet$, and 
one Dirac-type CP phase $\dcp$ (ignoring Majorana phases).
In a three-flavor framework, the transition probability also depends on two independent 
mass-squared differences: 
$\Delta m^2_{21} \equiv m_2^2-m_1^2$ in the solar sector and 
$\Delta m^2_{32} \equiv m_3^2-m_2^2$ in the atmospheric sector 
where $m_{3}$ corresponds to the neutrino mass eigenstate 
with the smallest electron component.
The smallest lepton mixing angle $\tet$ connects these two sectors
and determines the impact of the sub-leading three-flavor effects 
\cite{Pascoli:2013wca, Agarwalla:2013hma,Agarwalla:2014fva}.
All the neutrino oscillation data available till date have been explained
quite successfully in terms of these mass-mixing parameters
\cite{Gonzalez-Garcia:2014bfa,Capozzi:2013csa,Forero:2014bxa},
excluding few anomalous results obtained at very-short-baseline 
experiments \cite{Abazajian:2012ys}. 

Neutrinos acquire additional phases while travelling in matter, the
so-called `MSW effect' \cite{Wolfenstein:1977ue,Mikheev:1986gs,Mikheev:1986wj,Blennow:2013rca}
which determined the ordering of the 1-2 mass splitting using solar neutrinos.
In light of large $\theta_{13}$, matter effects are also going to play
an important role in presently running and future long-baseline 
\cite{Pascoli:2013wca,Feldman:2013vca,Agarwalla:2014fva} 
neutrino oscillation experiments to settle the remaining unsolved issues,
namely, the neutrino mass hierarchy\footnote{Two patterns of neutrino masses 
are possible: $m_3 > m_2 > m_1$, called normal hierarchy (NH) where $\ma > 0$ 
and $m_2 > m_1 > m_3$, called inverted hierarchy (IH) where $\ma < 0$.},
possibility of leptonic CP-violation if $\dcp$ differs from both $0^\circ$ and $180^\circ$,
and the octant degeneracy of $\theta_{23}$ \cite{Fogli:1996pv} 
if $\theta_{23}$ turns out to be non-maximal.
The combined data from the current off-axis $\nu_e$ appearance experiments,
T2K \cite{Itow:2001ee,Abe:2011ks} and NO$\nu$A \cite{Ayres:2002ws,Ayres:2004js,Ayres:2007tu},
hold promise of providing a first hint for these missing links for only favorable ranges of oscillation parameters
\cite{Huber:2009cw,Agarwalla:2012bv,Agarwalla:2013ju,Chatterjee:2013qus,Machado:2013kya,Ghosh:2014dba,Abe:2014tzr}. 
Hence, future facilities consisting of intense, high power, on-axis wide-band beams 
and large smart detectors have been proposed to cover the entire parameter space 
at unprecedented confidence level \cite{Feldman:2013vca}. 
Future superbeam long-baseline facilities with liquid argon detectors,
Deep Underground Neutrino Experiment (DUNE) 
\cite{Diwan:2003bp,Barger:2007yw,Huber:2010dx,Akiri:2011dv,Adams:2013qkq} 
in the United States with a baseline of 1300 km from Fermilab to Homestake mine
in South Dakota and Long-Baseline Neutrino Oscillation Experiment (LBNO) in Europe 
involving a path length of 2290 km between CERN and Pyh\"asalmi mine in Finland
\cite{Agarwalla:2011hh,Stahl:2012exa,::2013kaa,Agarwalla:2014tca}
are the two major candidates in this roadmap which are capable enough to claim the discovery 
for the above mentioned issues \cite{Agarwalla:2013hma,Agarwalla:2014fva}.
For both the DUNE and LBNO baselines, the matter effects are quite significant which 
break the eight-fold degeneracies \cite{Barger:2001yr,Minakata:2002qi} among the 
various oscillation parameters and improve the physics reach by considerable amount.

Apart from the SM $W$-exchange interaction in matter, there may well be 
flavor-dependent, vector-like, leptonic long-range force (LRF), like those 
mediated by the $L_e-L_{\mu,\tau}$ gauge boson which is very light and 
neutral, leading to new non-universal flavor-diagonal neutral current (FDNC) 
interactions of the neutrinos which can give rise to non-trivial three-neutrino 
mixing effects in terrestrial experiments, and could affect 
the neutrino propagation through matter 
\cite{Joshipura:2003jh,Grifols:2003gy,GonzalezGarcia:2006vp,PhysRevD.75.093005,Samanta:2010zh}.
Can we constrain/discover these long-range FDNC interactions in
upcoming long-baseline neutrino experiments? If this LRF exists in Nature, 
can it become fatal in our attempts to resolve the remaining unknowns in 
neutrino oscillation? In this paper, we attempt to address these pressing issues
in the context of future high-precision superbeam facilities, DUNE and LBNO.

This paper is organized as follows. We start section~\ref{lrf} with a discussion 
on flavor-dependent LRF and how it arises from abelian gauged 
$L_e-L_{\mu,\tau}$ symmetry. Then, we discuss the strength of the 
long-range potential $V_{e\mu/e\tau}$ at the Earth surface generated 
by the electrons inside the Sun. This is followed by a brief discussion 
on the current constraints that we have on the effective gauge couplings
$\alpha_{e\mu, \, e\tau}$ of the  $L_e-L_{\mu,\tau}$ symmetry from
various neutrino oscillation experiments. In section~\ref{lrf-oscillation}, 
we study in detail the three-flavor oscillation picture in presence of 
long-range potential. We derive the compact analytical expressions 
for the effective oscillation parameters in case of the $L_e-L_{\mu}$ 
symmetry, and also present a simple expression for the resonance 
energy, where 1-3 mixing angle in matter becomes $45^{\circ}$ in 
the presence of long-range potential. Next, we demonstrate the 
accuracy of our approximate analytical probability expressions 
by comparing it with the exact numerical results. We also discuss
some salient features of the appearance and disappearance 
probabilities (calculated numerically) for the Fermilab-Homestake 
and CERN-Pyh\"asalmi baselines in the presence of long-range 
potential towards the end of this section. We give the similar plots 
for the anti-neutrino case in appendix~\ref{probability-discusions-anti-neutrino-case}.
At the beginning of section~\ref{event-study-with-lrf}, we give a brief
description of the main experimental features of the DUNE and 
LBNO set-ups. Then, we study the impact of the long-range potential
due to $L_e-L_{\mu}$ symmetry on the expected event spectra and total
event rates for the DUNE and LBNO experiments. 
In section~\ref{simulation-method}, we describe our simulation strategy.
Next, we derive the expected constraints on $\alpha_{e\mu}$ 
from DUNE and LBNO in section~\ref{expected-constraints}, and estimate
the discovery reach for $\alpha_{e\mu}$ in section~\ref{expected-discovery}.
In section~\ref{CPV-Impact}, we study how the long-range potential affects
the CP-violation search of these future facilities. Section~\ref{MH-Impact}
is devoted to assess the impact of LRF on mass hierarchy measurements.
Finally, we summarize and draw our conclusions in 
section~\ref{summary-conclusions}.

%=============================================================
\section{Flavor-Dependent Long-Range Forces from Gauged $U(1)$ Symmetries}
\label{lrf}
%=============================================================

The SM gauge group $SU(3)_C \times SU(2)_L \times U(1)_Y$ can be 
extended with minimal matter content by introducing anomaly free 
$U(1)$ symmetries under which the SM remains invariant and 
renormalizable \cite{Langacker:2008yv}. There are three lepton flavor 
combinations: $L_e-L_{\mu}$, $L_e-L_{\tau}$, and $L_{\mu}-L_{\tau}$ 
which can be gauged in an anomaly free way with the particle content of 
the SM \cite{Foot:1990mn,He:1991qd,Foot:1994vd}. These $U(1)$ 
gauge symmetries have to be broken in Nature in order to allow 
the different neutrino species to mix among each other giving rise 
to neutrino oscillation as demanded by the recent data 
\cite{Agashe:2014kda,Gonzalez-Garcia:2014bfa,Capozzi:2013csa,Forero:2014bxa}.
This can be understood from the following example. If we assume that 
neutrino masses are generated by effective five-dimensional operator
following say, seesaw mechanism, then this operator would remain invariant 
under these $U(1)$ symmetries if they are exact. This will ultimately 
give us an effective neutrino mass matrix containing a Dirac and a 
Majorana neutrino which remain unmixed and there will be no neutrino
oscillation. For more discussions on these issues, see references 
\cite{Joshipura:2003jh,PhysRevD.75.093005}. 
Now, there are two possibilities for the extra gauge boson associated 
with this abelian symmetry\footnote{Models with these symmetries 
necessarily possess a Higgs sector which also discriminates among 
different lepton families \cite{Foot:2005uc}, but we will only focus here 
on the effect of the extra gauge boson.}: either it can be very heavy or 
it is very light but in both the cases, it couples to matter very feebly 
to escape direct detection. If the electrically neutral gauge boson 
($Z^{\prime}$) corresponding to a flavor-diagonal generator of 
this new gauge group is relatively heavy\footnote{For an example, 
the $Z^{\prime}$ in gauged $L_e-L_{\mu,\tau}$ can be produced 
in $e^{+}e^{-}$ collisions and subsequently decay into $e^{+}e^{-}$
or $\mu^{+}\mu^{-}$ or $\tau^{+}\tau^{-}$ pairs and can be constrained
using the data from LEP/LEP2 
\cite{Acciarri:2000uh,Abbiendi:2003dh,Abdallah:2005ph,Schael:2006wu,Honda:2007wv}.}, 
then the phenomenological consequences can be quite interesting 
\cite{Foot:1990mn,He:1991qd,Foot:1994vd,Dutta:1994dx,Langacker:2008yv}.
On the other hand, if the mass of the gauge boson is very light, then it can
introduce LRF with terrestrial range (greater than or equal to the Sun-Earth
distance) without call upon extremely low 
mass scales \cite{Joshipura:2003jh,Grifols:2003gy}.
This LRF is composition dependent 
(depends on the leptonic content and the mass of an object) 
and violate the universality of free fall which could be tested 
in the classic lunar ranging \cite{Williams:2004qba,Williams:1995nq} 
and E\"ot\"vos type gravity experiments 
\cite{Adelberger:2003zx,Dolgov:1999gk}.
In fact, this idea was given long back by Lee and Yang 
\cite{Lee:1955vk} and later, using this idea, Okun \cite{Okun:1969ey,Okun:1995dn}
gave a 2$\sigma$ bound of $\alpha < 3.4 \times 10^{-49}$
($\alpha$ denotes the strength of the long-range potential)
for a range of the Sun-Earth distance or more. Now, the light
gauge boson $Z^{\prime}$ should have a mass 
$m_{Z^{\prime}} \sim g \, \langle v \rangle$ where $g$ is the gauge 
coupling of the new $U(1)$ symmetry and it should be 
$\lesssim 0.6 \times 10^{-24}$ since $g \approx \sqrt{\alpha}$
and $\langle v \rangle$ is the vacuum expectation value of the 
symmetry breaking scale. If the range of the LRF is comparable
to the Sun-Earth distance ($\approx$ $1.5\times10^{13}\;\textrm{cm}$)
then $m_{Z^{\prime}} \sim$ 1/($1.5\times10^{13}\;\textrm{cm}$)
= $1.3 \times 10^{-18} \, \textrm{eV}$ which means 
$\langle v \rangle$ $\gtrsim$ 2 MeV.

%=============================================
\subsection{Abelian Gauged $L_e-L_{\mu,\tau}$ Symmetries}
\label{le-lmu-le-ltau-symmetries}
%=============================================

If the extra $U(1)$ corresponds to $L_e-L_{\mu}$ or $L_e-L_{\tau}$ 
flavor combination\footnote{Due to the absence of muons or tau 
leptons inside the Sun or Earth, we do not consider gauged 
$L_{\mu}-L_{\tau}$ symmetry here.}, 
the coupling of the solar electron to the 
$L_e-L_{\mu,\tau}$ gauge boson generates a flavor-dependent 
long-range potential for neutrinos \cite{Grifols:1993rs,Grifols:1996fk,Horvat:1996mt} 
and can have significant impact on neutrino oscillations 
\cite{Joshipura:2003jh,Grifols:2003gy,GonzalezGarcia:2006vp,PhysRevD.75.093005,Samanta:2010zh} 
inspite of such stringent constraints on $\alpha$. 
This is caused due to the fact that the ($L_e-L_{\mu,\tau}$)-charge 
of the electron neutrino is opposite to that of muon or tau flavor,
leading to new non-universal FDNC interactions of the neutrinos 
in matter on top of the SM $W$-exchange interactions between 
the matter electrons and the propagating electron neutrinos. 
These new flavor-dependent FDNC interactions alter the `running' 
of the oscillation parameters in matter by considerable amount 
\cite{Agarwalla:2015cta}. Another important point is that the large number 
of electrons inside the Sun and the long-range nature of interaction 
balance the smallness of coupling and generate noticeable potential. 
As an example, the electrons inside the Sun give rise to a potential 
$V_{e\mu/e\tau}$ at the Earth surface which is given by 
\cite{Joshipura:2003jh,PhysRevD.75.093005}
\begin{equation} 
V_{e\mu/e\tau}(R_{SE}) =
\alpha_{e\mu/e\tau} \frac{N_e}{R_{SE}}
\approx 1.3 \times 10^{-11} \mathrm{eV}
\left(\frac{\alpha_{e\mu/e\tau}}{10^{-50}}\right) \;,
\label{eq:v-lrf}
\end{equation}
where $\alpha_{e\mu/e\tau} = \frac{g^2_{e\mu/e\tau}}{4\pi}$,
and $g_{e\mu/e\tau}$ is the gauge coupling of the $L_e-L_{\mu,\tau}$
symmetry. In Eq.~(\ref{eq:v-lrf}), $N_e$ ($\approx 10^{57}$) is the
total number of electrons inside the Sun \cite{bahcall:1989} and
$R_{SE}$ is the Sun-Earth distance $\approx$ 
$1.5\times10^{13}\;\textrm{cm} = 7.6\times10^{26}\;\textrm{GeV}^{-1}$. 
Here $\alpha_{e\mu/e\tau}$ can be identified as the `fine-structure constant'
of the $U(1)_{L_e-L_{\mu,\tau}}$ symmetry and its value is 
positive\footnote{In our work, we consider the case of a light
vector boson exchange which makes sure that $\alpha_{e\mu/e\tau}$
is positive. It means that for an example, the force between an isolated 
electron and $\nue$ is repulsive.}.
The corresponding potential due to the electrons inside the Earth 
with a long-range force having the Earth-radius range 
($R_E \sim$ 6400 km) is roughly one order of magnitude smaller
compared to the solar long-range potential and can be safely 
neglected\footnote{The possibility of the local screening of the 
leptonic force (generated due to the solar electrons) by the 
cosmic anti-neutrinos is also negligible over the Sun-Earth
distance \cite{Joshipura:2003jh}.} \cite{Joshipura:2003jh,PhysRevD.75.093005}.
The long-range potential $V_{e\mu/e\tau}$ in Eq.~(\ref{eq:v-lrf}) 
appears with a negative sign for anti-neutrinos and can affect the
neutrino and anti-neutrino oscillation probabilities in different fashion.
This feature can invoke fake CP-asymmetry like the SM matter effect
and can influence the CP-violation search in long-baseline experiments.
This is one of the important findings of our paper and we will discuss this
issue in detail in the later section. Now, it would be quite 
interesting to compare the strength of the potential given in 
Eq.~(\ref{eq:v-lrf}) with the quantity $\Delta m^2/2E$ which governs 
the neutrino oscillation probability. For long-baseline neutrinos, 
$\Delta m^2/2E \sim 10^{-12} \, \mathrm{eV}$ 
(assuming $\Delta m^2 \sim 2 \times 10^{-3}$ eV$^2$ and $E \sim$ 1 GeV)
which is comparable to $V_{e\mu/e\tau}$ even for 
$\alpha_{e\mu/e\tau} \sim 10^{-51}$ and can affect the
long-baseline experiments significantly which we are going to
explore in this paper in the context of upcoming facilities 
DUNE and LBNO.

%=================================================================
\subsection{Existing Phenomenological Constraints on $L_e-L_{\mu,\tau}$ Parameters}
\label{current-bounds}
%=================================================================

There are phenomenological bounds on the effective gauge coupling 
$\alpha_{e\mu/e\tau}$ of the $L_e-L_{\mu,\tau}$ abelian 
symmetry\footnote{Flavor-dependent long-range leptonic forces
can also be generated via the unavoidable mixing of light $Z^{\prime}$ 
boson of the $L_{\mu}-L_{\tau}$ symmetry with the $Z$ boson of the SM.
This issue was discussed in the context of the MINOS long-baseline 
experiment in \cite{Heeck:2010pg,Davoudiasl:2011sz}.}
using data from various neutrino oscillation experiments.
It was shown in \cite{Joshipura:2003jh} that $L_e-L_{\mu,\tau}$
potential at the Earth due to the huge number of electrons inside
the Sun suppresses the atmospheric neutrino $\numu \to \nutau$
oscillations which enabled them to place tight constraints on
$\alpha_{e\mu/e\tau}$ using the oscillation of multi-GeV neutrinos 
observed at the Super-Kamiokande (SK) experiment. They obtained an
upper bound of $\alpha_{e\mu} < 5.5 \times 10^{-52}$ and
$\alpha_{e\tau} < 6.4 \times 10^{-52}$ at 
90\% C.L. \cite{Joshipura:2003jh}. In \cite{PhysRevD.75.093005},
the authors performed a global fit to the solar neutrino and KamLAND 
data including the flavor-dependent LRF. They quoted an upper bound 
of  $\alpha_{e\mu} < 3.4 \times 10^{-53}$ and 
$\alpha_{e\tau} < 2.5 \times 10^{-53}$ at 
3$\sigma$ C.L. assuming $\theta_{13}$ = 0$^{\circ}$ 
\cite{PhysRevD.75.093005}. A similar analysis was performed
in \cite{GonzalezGarcia:2006vp} to place the constraints 
on LRF mediated by vector and non-vector (scalar or tensor) 
neutral bosons where the authors assumed one mass scale 
dominance. The proposed 50 kt magnetized iron calorimeter 
(ICAL) detector at the India-based Neutrino Observatory (INO) 
can also probe the existence of LRF by observing the 
atmospheric neutrinos and anti-neutrinos separately over 
a wide range of energies and baselines \cite{Samanta:2010zh}.
With an exposure of one Mton.yr and using the muon momentum
as observable, ICAL would be able to constrain $\alpha_{e\mu/e\tau}$
$\lesssim$ $1.65 \times 10^{-53}$ at 3$\sigma$ C.L. \cite{Samanta:2010zh}.

%==========================================================
\section{Three-Flavor Oscillation Picture in Presence of Long-Range Potential}
\label{lrf-oscillation}
%==========================================================

In this section, we focus our attention to study the impact of flavor-dependent 
long-range leptonic potential (due to the electrons inside the Sun) at the Earth
surface when neutrinos travel through the Earth matter. In a three-flavor framework,
the long-range potential of Eq.~(\ref{eq:v-lrf}) due to $L_e-L_{\mu}$ symmetry
modifies the effective Hamiltonian for neutrino propagation in Earth matter 
in the flavor basis to
\begin{equation}
H_f \;=\; 
\left(
U
\begin{bmatrix}
0 & 0 & 0 \\
0 & \frac{\Delta m_{21}^2}{2E} & 0 \\
0 & 0 & \frac{\Delta m_{31}^2}{2E}
\end{bmatrix}
U^\dagger
+
\begin{bmatrix}
V_{CC} & 0 & 0 \\
0 & 0 & 0 \\
0 & 0 & 0
\end{bmatrix}
+
\begin{bmatrix}
V_{e\mu} & 0 & 0 \\
0 & - V_{e\mu} & 0 \\
0 & 0 & 0
\end{bmatrix}
\right) \;,
\label{eq:flavor-Hamiltonian}
\end{equation}
where $U$ is the vacuum PMNS matrix 
\cite{Pontecorvo:1957vz,Maki:1962mu,Pontecorvo:1967fh}
which can be parametrized as  
\begin{equation}
U =  R_{23}(\theta_{23},0)\;R_{13}(\theta_{13},\dcp)\;R_{12}(\theta_{12},0) \;.
\label{eq:U-parametrization}
\end{equation}
In the above equation, $E$ is the neutrino energy and $V_{CC}$ is the Earth matter
potential which appears in the form
\begin{equation}
V_{CC} = \sqrt{2} G_F N_e \simeq 7.6 \times Y_e \times \frac{\rho}{10^{14}
\mbox{g}/\mbox{cm}^3} \, \mbox{eV}\,,
\label{eq:matter_density}
\end{equation}
where $G_F$ is the Fermi coupling constant, $N_e$ is the electron number 
density inside the Earth, $\rho$ is the matter density, and 
$Y_e = \frac{N_e}{N_p + N_n}$ is the relative 
electron number density. $N_p$, $N_n$ are the proton and neutron densities 
in Earth matter respectively. In an electrically neutral, isoscalar medium, 
we have $N_e = N_p = N_n$ and $Y_e$ comes out to be 0.5.
In Eq.~(\ref{eq:flavor-Hamiltonian}), $V_{e\mu}$ is the long-range potential 
due to $L_e-L_{\mu}$ symmetry and its strength is given by Eq.~(\ref{eq:v-lrf}). 
In case of $L_e-L_{\tau}$ symmetry, the contribution due to long-range potential
in Eq.~(\ref{eq:flavor-Hamiltonian}) takes the form 
Diag\,$\left(V_{e\tau},0,-V_{e\tau}\right)$. Note that 
the strength of the long-range potential $V_{e\mu/e\tau}$ 
does not depend on the Earth matter density\footnote{This is 
in contrast to the usual non-standard interactions (NSI's) whose
strengths are proportional to the Earth matter density.} 
and takes the same value for any baseline on the Earth. 
In case of anti-neutrino propagation, we have to reverse the sign of 
$V_{CC}$, $V_{e\mu}$ (or $V_{e\tau}$), and the CP phase $\dcp$.

%=========================================
\begin{table}[t]
\begin{center}
{
\newcommand{\mc}[3]{\multicolumn{#1}{#2}{#3}}
\newcommand{\mr}[3]{\multirow{#1}{#2}{#3}}
\begin{adjustbox}{width=0.95\textwidth}
\begin{tabular}{|c|c|c|c|c|c|}
\hline
\mr{2}{*}{Set-up} & \mr{2}{*}{$1^{\rm st}$ osc. max. (GeV) } & 
\mr{2}{*}{$\frac{\Delta m_{31}^2}{2E}$ (eV)} & 
\mr{2}{*}{$V_{CC}$ (eV)} & 
\mc{2}{c|}{$V_{e\mu}$ (eV)} \\
\cline{5-6}
 & & & & $\alpha_{e\mu}=10^{-52}$ & $\alpha_{e\mu}=10^{-53}$ \\
\hline
\mr{2}{*}{DUNE} & \mr{2}{*}{2.56} & \mr{2}{*}{4.8$\times10^{-13}$} & 
\mr{2}{*}{1.1$\times10^{-13}$} &\mr{2}{*}{1.3$\times10^{-13}$} &
\mr{2}{*}{1.3$\times10^{-14}$} \\
 & & & &  &  \\
\hline
\mr{2}{*}{LBNO} & \mr{2}{*}{4.54} & \mr{2}{*}{2.7$\times10^{-13}$} & 
\mr{2}{*}{1.3$\times10^{-13}$} &\mr{2}{*}{1.3$\times10^{-13}$} &
\mr{2}{*}{1.3$\times10^{-14}$} \\
 & & & &  &  \\
\hline
\end{tabular}
\end{adjustbox}
}
\caption{The second column shows the first oscillation maxima for  
Fermilab--Homestake and CERN--Pyh\"asalmi baselines considering
$\Delta m_{31}^2 = 2.44 \times 10^{-3}$ eV$^{2}$. The fourth column 
depicts the strength of the Earth matter potentials for these baselines.
The estimate of the long-range potentials $V_{e\mu}$ for two different
values of $\alpha_{e\mu}$ is given in the last column.}
\label{tab:compare}
\end{center}
\end{table}
%===========================================

%===========================================
\begin{table}[t]
\begin{center}
{
\newcommand{\mc}[4]{\multicolumn{#1}{#2}{#3}{#4}}
\newcommand{\mr}[3]{\multirow{#1}{#2}{#3}}
\begin{tabular}{|c |c|c|c|}
\hline
\mr{2}{*}{Parameter} & \mr{2}{*}{Best-fit and 1$\sigma$ error} & \mr{2}{*}{True value} & \mr{2}{*}{Marginalization range} \\
  & & & \\
\hline
\mr{2}{*}{$\sin^2{\theta_{12}}$} & \mr{2}{*}{$0.304^{+ 0.013}_{-0.012}$} & \mr{2}{*}{0.304} & \mr{2}{*}{Not marginalized} \\
  & & & \\
\hline
\mr{2}{*}{$\sin^2{\theta_{13}}$} & \mr{2}{*}{$0.0218^{+ 0.0010}_{-0.0010}$} & \mr{2}{*}{$0.0218$} & \mr{2}{*}{Not marginalized} \\ 
  & & & \\
\hline
\mr{2}{*}{$\sin^2{\theta_{23}}$} & \mr{2}{*}{$0.452^{+ 0.052}_{-0.028}$} & \mr{2}{*}{0.50} & \mr{2}{*}{[0.38, 0.64]} \\
  & & & \\
\hline
\mr{2}{*}{$\dcp/^{\circ}$} & \mr{2}{*}{$306^{+ 39}_{-70}$} & \mr{2}{*}{[0, 360]} & \mr{2}{*}{[0, 360]} \\
  & & & \\
\hline
\mr{2}{*}{$\frac{\Delta{m^2_{21}}}{10^{-5} \, \rm{eV}^2}$} & \mr{2}{*}{$7.5^{+ 0.19}_{-0.17}$} & \mr{2}{*}{7.50} & \mr{2}{*}{Not marginalized} \\
  & & & \\
\hline
\mr{2}{*}{$\frac{\Delta{m^2_{31}}}{10^{-3} \, \rm{eV}^2}$} & \mr{2}{*}{$2.457^{+ 0.047}_{-0.047}$}  & \mr{2}{*}{2.44} &\mr{2}{*}{Not marginalized} \\
  & & &\\
\hline
\mr{2}{*}{$\frac{\Delta{m^2_{\mu\mu}}}{10^{-3} \, \rm{eV}^2}$} & \mr{2}{*}{$2.410^{+ 0.051}_{-0.056}$} & \mr{2}{*}{$2.40$} & \mr{2}{*}{Not marginalized} \\
  & & & \\
\hline
\end{tabular}
}
\caption{The second column shows the current best-fit values and 1$\sigma$
uncertainties on the three-flavor oscillation parameters assuming normal hierarchy 
in the fit \cite{Gonzalez-Garcia:2014bfa}. The third column shows the true values 
of the oscillation parameters used to simulate the `observed' data set.
The fourth column depicts the range over which $\sin^2\theta_{23}$ and $\dcp$
are varied while minimizing the $\chi^{2}$ to obtain the final results.
In our calculations, $\Delta{m^2_{\mu\mu}}$ serves as an input parameter and 
then we estimate the value of $\Delta{m^2_{31}}$ using the relation given in 
Eq.~(\ref{eq:parkedef}) (see text for details). In the third column, 
we take $\dcp = 0^{\circ}$ to calculate the value of $\Delta{m^2_{31}}$
from $\Delta{m^2_{\mu\mu}}$.}
\label{tab:benchmark-parameters}
\end{center}
\end{table}
%=======================================

To judge the impact of the long-range potential $V_{e\mu}$ in long-baseline
experiments with multi-GeV neutrinos, we need to compare its strength with 
the two other main terms in Eq.~(\ref{eq:flavor-Hamiltonian}) which are
$\frac{\Delta m_{31}^2}{2E}$ and the Earth matter potential $V_{CC}$.
In Table~\ref{tab:compare}, we compare the strengths of these three quantities 
which control the `running' of the oscillation parameters in matter. 
The first oscillation maximum for the DUNE (LBNO) experiment 
occurs at 2.56 GeV (4.54 GeV) assuming 
$\Delta m_{31}^2 = 2.44 \times 10^{-3}$ eV$^{2}$ 
(see Table~\ref{tab:benchmark-parameters}).
In the fourth column of Table~\ref{tab:compare}, the values of $V_{CC}$ have 
been estimated using Eq.~(\ref{eq:matter_density}) where we take the 
line-averaged constant Earth matter densities\footnote{These line-averaged 
constant Earth matter densities have been estimated using the Preliminary Reference 
Earth Model (PREM) \cite{PREM:1981}.} for both the baselines: 
$\rho=2.87\,\mathrm{g/cm^3}$ for the DUNE baseline and 
$\rho=3.54\,\mathrm{g/cm^3}$ for the LBNO baseline.
Table~\ref{tab:compare} shows that around first oscillation
maximum, the strengths of the terms $\frac{\Delta m_{31}^2}{2E}$, 
$V_{CC}$, and $V_{e\mu}$ are comparable for both the set-ups under 
consideration even if $\alpha_{e\mu}$ is as small as $10^{-52}$. 
It means that these three terms can interfere with each other 
having significant impact on the oscillation probability 
which we are going to study in this section with the help of 
analytical expressions. Before we start deriving our approximate
analytical expressions for the effective mass-squared differences 
and mixing angles in matter in the presence of long-range potential,
let us take a look at the current status of the oscillation parameters.
The second column of Table~\ref{tab:benchmark-parameters}
shows the present best-fit values and 1$\sigma$ errors on the 
three-flavor oscillation parameters assuming normal hierarchy 
in the fit \cite{Gonzalez-Garcia:2014bfa}. We use the benchmark 
values of the various oscillation parameters as given in the third 
column of Table~\ref{tab:benchmark-parameters} to draw the
oscillation probability plots in this section and to generate the
`observed' data set while estimating the physics reach of the 
experimental set-ups. The ranges over which $\sin^2\theta_{23}$ 
and $\dcp$ are marginalized while minimizing the $\chi^{2}$ are 
given in the fourth column which we will discuss in detail while 
describing the simulation method in the later section.
In Table~\ref{tab:benchmark-parameters}, $\Delta{m^2_{\mu\mu}}$ 
is the effective mass-squared difference measured by the accelerator 
experiments using $\numu \rightarrow \numu$ disappearance 
channel \cite{Nichol:2013caa,Adamson:2013whj} and it is a linear 
combination of $\ma$ and $\ms$. The value of $\ma$ is estimated from 
$\Delta{m^2_{\mu\mu}}$ using the relation 
\cite{Nunokawa:2005nx,deGouvea:2005hk}
\begin{equation}
\ma = \Delta{m^2_{\mu\mu}} + \ms (\cos^2\tem - \cos\dcp \, \sin\tet \, \sin2\tem \, \tan\tmt) \,.
\label{eq:parkedef}
\end{equation}
The value of $\ma$ is calculated separately for NH and IH using 
the above equation assuming $\Delta{m^2_{\mu\mu}} = 
\pm \,\, 2.4 \times 10^{-3}$ eV$^2$ where positive (negative) sign 
is for NH (IH). Note that through out this paper, whenever we vary
$\dcp$ or $\theta_{23}$ or both, we calculate a new value for 
$\ma$ using Eq.~(\ref{eq:parkedef}).

%=======================================================
\subsection{Analytical Expressions for the Effective Oscillation Parameters}
\label{analytical-expressions}
%========================================================

In a CP-conserving scenario ($\dcp = 0^{\circ}$), the effective Hamiltonian 
in the flavor basis given in Eq.~(\ref{eq:flavor-Hamiltonian}) takes the form
\begin{equation}
H_f = R_{23}\left(\theta_{23}\right)R_{13}\left(\theta_{13}\right)R_{12}\left(\theta_{12}\right)H_0R_{12}^T\left(\theta_{12}\right)R_{13}^T\left(\theta_{13}\right)R_{23}^T\left(\theta_{23}\right)+V \,,
\label{eq:new-flavor-Hamiltonian}
\end{equation}
where $H_0$ = Diag\,$\left(0,\Delta_{21},\Delta_{31}\right)$ with
$\Delta_{21} \equiv \Delta m_{21}^2/2E$ and $\Delta_{31} \equiv \Delta m_{31}^2/2E$.
In the above equation, $V$ = Diag\,$\left(V_{CC} + V_{e\mu},-V_{e\mu},0\right)$ 
for $L_e-L_{\mu}$ symmetry. We can rewrite $H_f$ in 
Eq.~(\ref{eq:new-flavor-Hamiltonian}) as
\begin{eqnarray}
H_f =
\Delta_{31}
\left(\begin{array}{lll}
a_{11} & a_{12} & a_{13} \\
a_{12} & a_{22} & a_{23} \\
a_{13} & a_{23} & a_{33}
\end{array}\right) \,,
\label{eq:rewrite-Hamiltonian}
\end{eqnarray}
where
\begin{equation}
a_{11} = A+W+\sin^2\theta_{13}+\alpha\cos^2\theta_{13}\sin^2\theta_{12} \,,
\label{eq:a11}
\end{equation}
\begin{equation}
a_{12} = \frac{1}{\sqrt{2}}\left[\cos\theta_{13}(\alpha \cos\theta_{12}\sin\theta_{12}+\sin\theta_{13}-\alpha \sin^2\theta_{12}\sin\theta_{13})\right] \,,
\label{eq:a12}
\end{equation}
\begin{equation}
a_{13} =  \frac{1}{\sqrt{2}}\left[\cos\theta_{13}(-\alpha\cos\theta_{12}\sin\theta_{12}+\sin\theta_{13}-\alpha\sin^2\theta_{12}\sin\theta_{13})\right] \,,
\label{eq:a13}
\end{equation}
\begin{equation}
a_{22} = \frac{1}{2}\left[\alpha\cos^2\theta_{12}+\cos^2\theta_{13}-2\alpha\cos\theta_{12}\sin\theta_{12}\sin\theta_{13}+\alpha \sin^2\theta_{12}\sin^2\theta_{13}-2W\right] \,,
\label{eq:a22}
\end{equation}
\begin{equation}
a_{23} = \frac{1}{2}\left[\cos^2\theta_{13}-\alpha \cos^2\theta_{12}+\alpha\sin^2\theta_{12}\sin^2\theta_{13}\right] \,,
\label{eq:a23}
\end{equation}
\begin{equation}
a_{33} = \frac{1}{2}\left[\cos^2\theta_{13}+\alpha \cos^2\theta_{12}+\alpha\sin\theta_{13}(\sin2\theta_{12}+\sin^2\theta_{12}\sin\theta_{13})\right] \,.
\label{eq:a33}
\end{equation}
In the above equations, we introduce the terms $A$, $W$, and $\alpha$ which are defined as
\begin{equation}
A \equiv \frac{V_{CC}}{\Delta_{31}} = \frac{2EV_{CC}}{\Delta m_{31}^2} \,\,, \,\,
W \equiv\frac{V_{e\mu}}{\Delta_{31}} = \frac{2EV_{e\mu}}{\Delta m_{31}^2} \,\,, \,\,
\alpha \equiv \frac{\Delta m_{21}^2}{\Delta m_{31}^2} \,,
\label{eq:definitions}
\end{equation}
and we assume that the vacuum value of $\theta_{23}$ is $45^{\circ}$.
Note that we have kept the terms of all orders in $\sin\theta_{13}$ and $\alpha$
which are quite important in light of the large value of 1-3 mixing angle 
as was measured recently by the modern reactor experiments.
Now, we need to diagonalize the effective Hamiltonian $H_f$ 
in Eq.~(\ref{eq:rewrite-Hamiltonian}) to find the effective mass-squared
differences and mixing angles in the presence of the 
Earth matter potential ($V_{CC}$) and the long-range potential ($V_{e\mu}$).
We can almost diagonalize $H_f$ with the help of a unitary matrix 
\begin{equation}
\tilde{U} \equiv R_{23}\left(\theta_{23}^m\right)R_{13}\left(\theta_{13}^m\right)R_{12}\left(\theta_{12}^m\right) \,,
\label{eq:new-U}
\end{equation}
such that
\begin{equation}
\tilde{U}^{T}H_{f}\tilde{U} \simeq \mathrm{Diag} \left(m_{1,m}^2/2E, \, m_{2,m}^2/2E, \, m_{3,m}^2/2E\right) \,,
\label{eq:diagonal-matrix}
\end{equation}
where off-diagonal terms are quite small and can be safely neglected.
The lower right $2\times2$ block in Eq.~(\ref{eq:rewrite-Hamiltonian}) 
gives us the angle $\theta_{23}^m$ which has the form
\begin{equation}
\tan2\theta_{23}^m = \frac{\cos^2\theta_{13}-\alpha \cos^2\theta_{12}+\alpha \sin^2\theta_{12}\sin^2\theta_{13}}{W+\alpha \sin 2\theta_{12}\sin\theta_{13}} \,.
\label{eq:theta23-m}
\end{equation}
The mixing angles $\theta_{13}^m$ and $\theta_{12}^m$ can be obtained by 
subsequent diagonalizations of the (1,3) and (1,2) blocks respectively and
we get the following expressions 
\begin{equation}
\tan2\theta_{13}^m = \frac{\sin2\theta_{13}(1-\alpha \sin^2\theta_{12})\left(\cos\theta_{23}^m+\sin\theta_{23}^m\right)-\alpha \sin2\theta_{12} \cos\theta_{13}\left(\cos\theta_{23}^m-\sin\theta_{23}^m\right)}{\sqrt{2}\left(\lambda_3 - A-W-\sin^2\theta_{13}-\alpha \sin^2\theta_{12}\cos^2\theta_{13} \right)} \,,
\label{eq:theta13-m}
\end{equation}
and
\begin{eqnarray}
\tan2\theta_{12}^m=  \hspace{12.5cm} \nonumber \\
\frac{\cos\theta_{13}^m\left[\sin2\theta_{13}(1-\alpha\sin^2\theta_{12})\left(\cos\theta_{23}^m-\sin\theta_{23}^m\right)+\alpha \sin 2\theta_{12}\cos\theta_{13}(\cos\theta_{23}^m+\sin\theta_{23}^m)\right]}{\sqrt{2}\left(\lambda_2-\lambda_1\right)}  \,, \nonumber \\
\label{eq:theta12-m}
\end{eqnarray}
where
\begin{equation}
\lambda_3 = \frac{1}{2}\biggl[\cos^2\theta_{13}+\alpha\cos^2\theta_{12}+\alpha\sin^2\theta_{12}\sin^2\theta_{13}-W+\frac{W+ \alpha\sin2\theta_{12}\sin\theta_{13}}{\cos2\theta_{23}^m}\biggr] \,,
\label{eq:lambda-3}
\end{equation}
\begin{equation}
\lambda_2  = \frac{1}{2}\biggl[\cos^2\theta_{13}+\alpha\cos^2\theta_{12}+\alpha\sin^2\theta_{12}\sin^2\theta_{13}-W-\frac{W+ \alpha\sin2\theta_{12}\sin\theta_{13}}{\cos2\theta_{23}^m}\biggr] \,,
\label{eq:lambda-2}
\end{equation}
and
\begin{align}
\lambda_1 = \hspace{15cm} \nonumber \\
\frac{1}{2}\biggl[\left(\lambda_3+A+W+\sin^2\theta_{13}+\alpha\cos^2\theta_{13}\sin^2\theta_{12}\right) \hspace{7cm} \nonumber \\
-\frac{\left(\lambda_3-A-W-\sin^2\theta_{13}-\alpha\cos^2\theta_{13}\sin^2\theta_{12}\right)}{\cos2\theta_{13}^m}\biggr] \,. \hspace{2cm} \nonumber \\
\label{eq:lambda-1}
\end{align}
The eigenvalues $m_{i,m}^2/2E \, (i = 1,2,3)$ are given by the expressions
\begin{align}
m_{3,m}^2/2E = \hspace{15.3cm} \nonumber \\
\frac{\Delta_{31}}{2}\biggl[\left(\lambda_3+A+W+\sin^2\theta_{13}+\alpha\cos^2\theta_{13}\sin^2\theta_{12}\right)\hspace{7cm} \nonumber \\
+\frac{\left(\lambda_3-A-W-\sin^2\theta_{13}-\alpha\cos^2\theta_{13}\sin^2\theta_{12}\right)}{\cos2\theta_{13}^m}\biggr] \,, \hspace{3.5cm} \nonumber \\
\label{eq:m3m-square-by-2E}
\end{align}
\begin{equation}
m_{2,m}^2/2E = \frac{\Delta_{31}}{2}\left[\lambda_1+\lambda_2-\frac{\left(\lambda_1-\lambda_2\right)}{\cos2\theta_{12}^m}\right] \,,
\label{eq:m2m-square-by-2E}
\end{equation}
and
\begin{equation}
m_{1,m}^2/2E = \frac{\Delta_{31}}{2}\left[\lambda_1+\lambda_2+\frac{\left(\lambda_1-\lambda_2\right)}{\cos2\theta_{12}^m}\right] \,.
\label{eq:m1m-square-by-2E}
\end{equation}
%

%---------------------------------------------------------------------------------------------------------------------------------
\begin{figure}[H]
\centerline{
\includegraphics[width=0.33\textwidth]{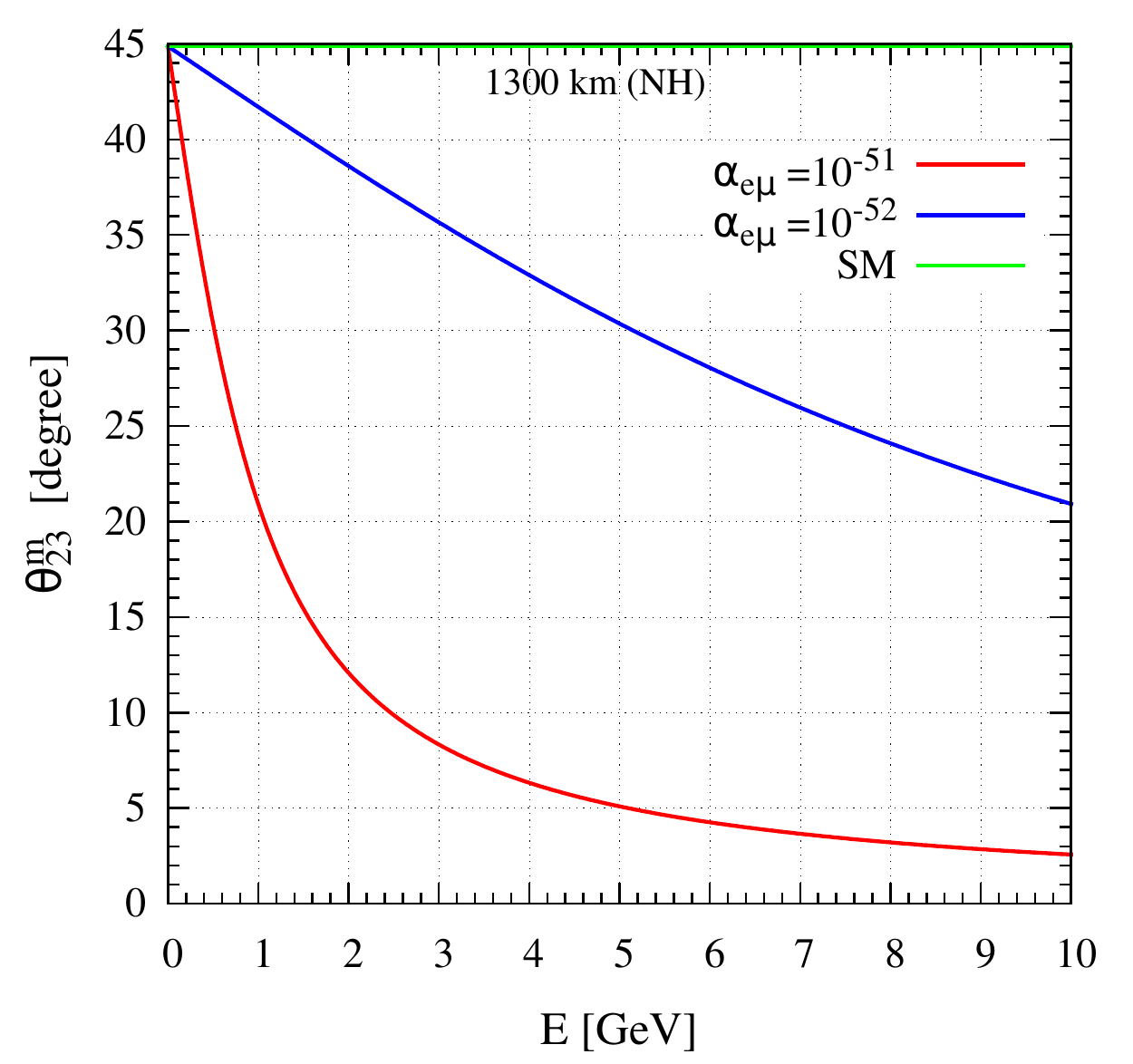}
\includegraphics[width=0.33\textwidth]{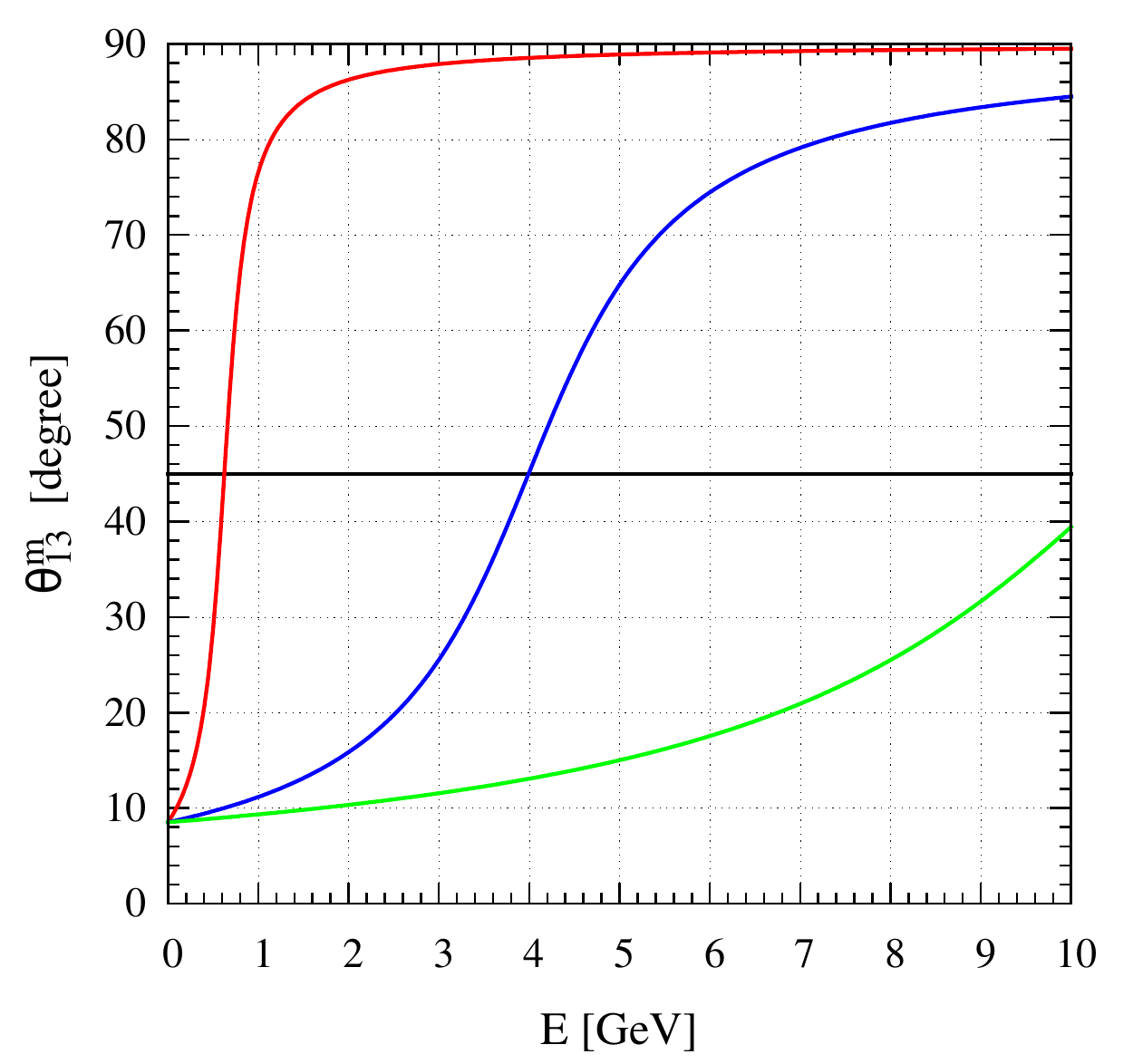}
\includegraphics[width=0.33\textwidth]{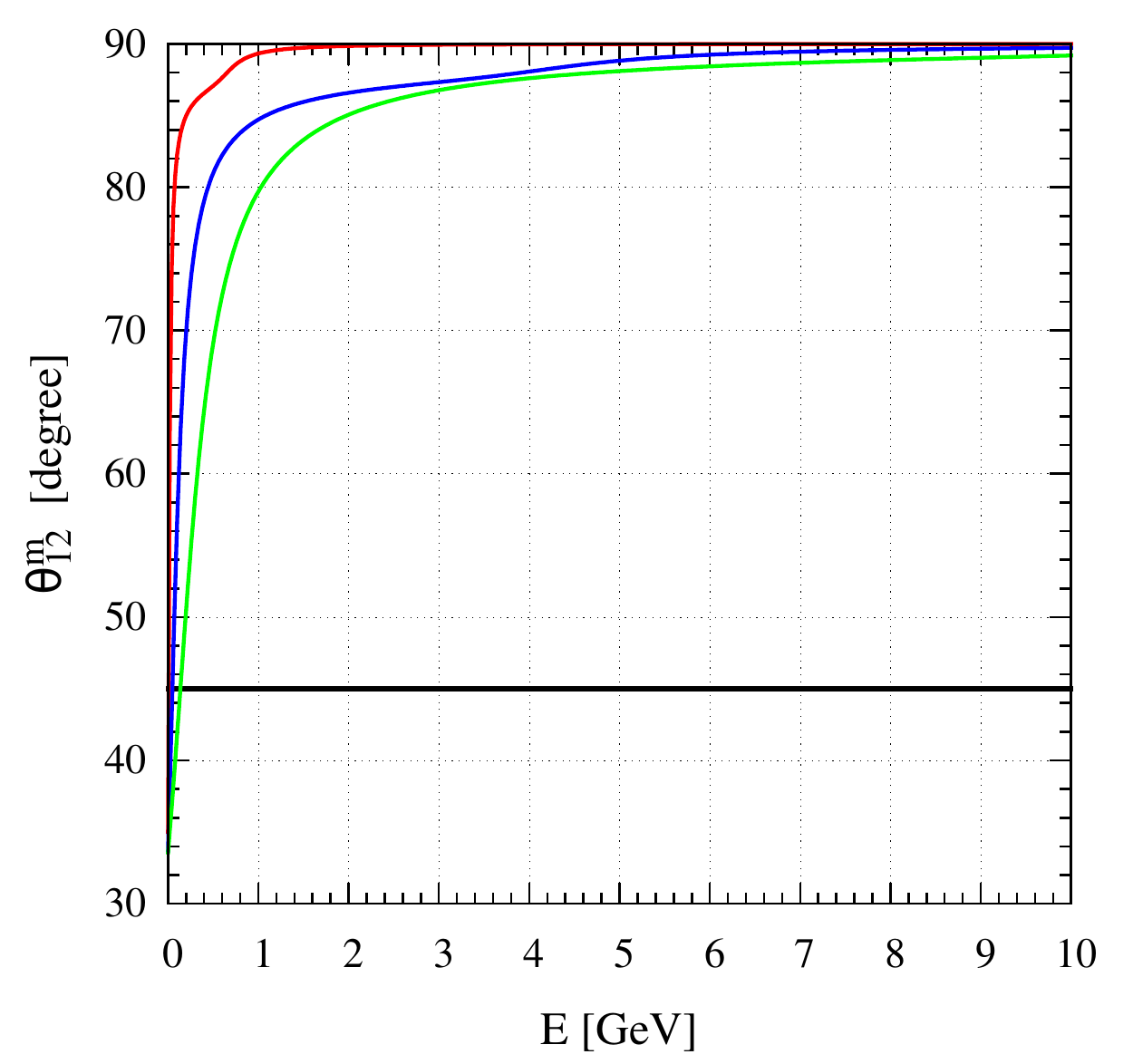}
}
\caption{The variations in the effective mixing angles with the neutrino energy $E$
in the presence of the Earth matter potential ($V_{CC}$) and long-range potential 
($V_{e\mu}$). The left, middle, and right panels show the `running' of 
$\theta_{23}^m$, $\theta_{13}^m$, and $\theta_{12}^m$ respectively. 
Here, we take $L$ = 1300 km which corresponds to the Fermilab--Homestake 
baseline and assume NH. Plots are given for three different choices 
of the effective gauge coupling $\alpha_{e\mu}$: 0 (the SM case), $10^{-52}$, 
and $10^{-51}$. The vacuum values of the oscillation parameters are taken from 
the third column of Table~\ref{tab:benchmark-parameters} and 
we consider $\dcp = 0^{\circ}$.}
\label{fig:running-mixing-angles}
\end{figure}
%----------------------------------------------------------------------------------------------------------------------------------

%---------------------------------------------------------------------------------------------------------------------
\begin{figure}[H]
\centerline{
\includegraphics[width=0.49\textwidth]{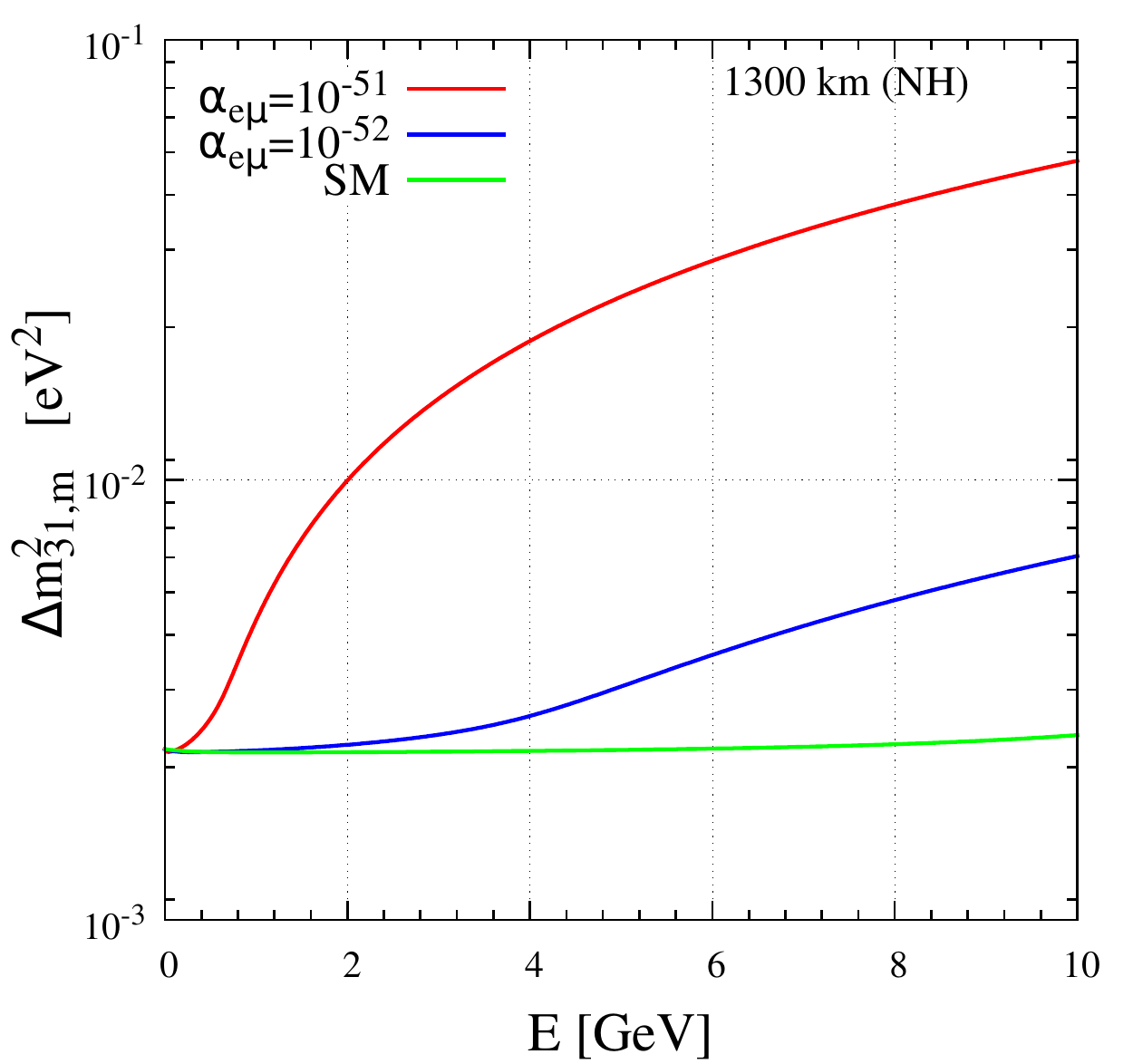}
\includegraphics[width=0.49\textwidth]{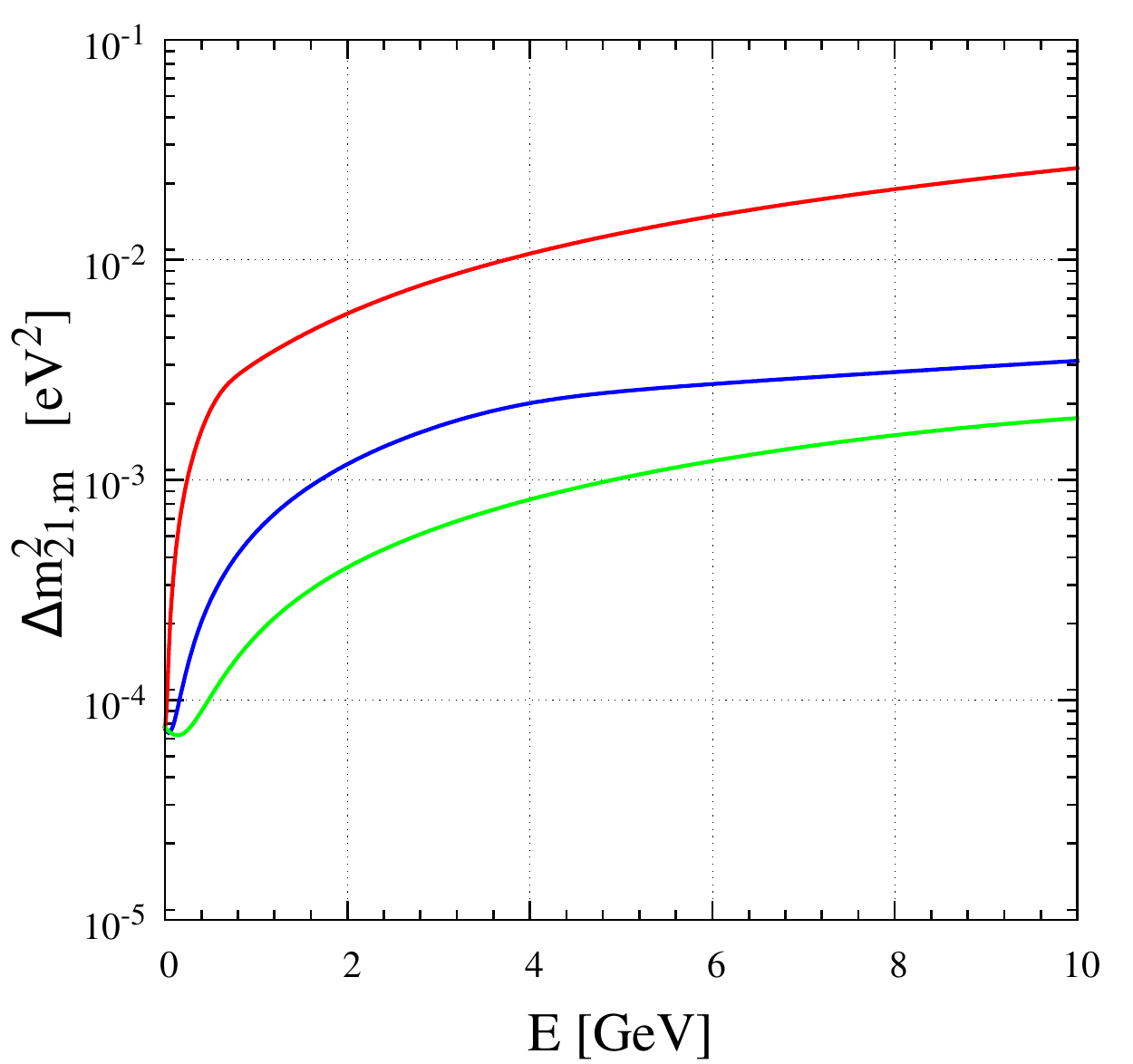}
}
\caption{The variations in the effective mass-squared differences with the neutrino energy $E$
in the presence of $V_{CC}$ and $V_{e\mu}$. Left panel shows the `running' of 
$\Delta m_{31,m}^2 (\equiv m_{3,m}^2 - m_{1,m}^2$) while right panel is for
$\Delta m_{21,m}^2 (\equiv m_{2,m}^2 - m_{1,m}^2$). Here, we take $L$ = 1300 km 
and assume NH. We give plots for three different choices of the effective gauge coupling 
$\alpha_{e\mu}$: 0 (the SM case), $10^{-52}$, and $10^{-51}$. 
The vacuum values of the oscillation parameters are taken from the third column of 
Table~\ref{tab:benchmark-parameters} and we consider $\dcp = 0^{\circ}$.}
\label{fig:running-mass-squared-differences}
\end{figure}
%--------------------------------------------------------------------------------------------------------------------

With the help of Eq.~(\ref{eq:theta23-m}), Eq.~(\ref{eq:theta13-m}), and Eq.~(\ref{eq:theta12-m}),
we plot the `running' of the effective mixing angles $\theta_{23}^m$, $\theta_{13}^m$, and 
$\theta_{12}^m$ respectively in Fig.~\ref{fig:running-mixing-angles} as functions of the neutrino
energy $E$ in the presence of $V_{CC}$ and $V_{e\mu}$.
Here, we consider $L$ = 1300 km and NH. We give the plots for three different choices 
of the effective gauge coupling $\alpha_{e\mu}$: 0 (the SM case), $10^{-52}$, 
and $10^{-51}$. The vacuum values of the oscillation parameters are taken from 
the third column of Table~\ref{tab:benchmark-parameters}.
We do the same for the effective mass-squared differences 
(see Eqs.~(\ref{eq:m3m-square-by-2E}), (\ref{eq:m2m-square-by-2E}), 
and (\ref{eq:m1m-square-by-2E})) in Fig.~\ref{fig:running-mass-squared-differences}.
The extreme right panel of Fig.~\ref{fig:running-mixing-angles} shows that 
$\theta_{12}^m$ approaches to $90^{\circ}$ very rapidly with increasing $E$
in the presence of $V_{CC}$ and this behavior does not change much when
we introduce $V_{e\mu}$. This is not the case for $\theta_{23}^m$ and
$\theta_{13}^m$. The long-range potential $V_{e\mu}$ affects the
`running' of $\theta_{23}^m$ (see extreme left panel of Fig.~\ref{fig:running-mixing-angles})
significantly. As we go to higher energies, $\theta_{23}^m$ deviates from $45^{\circ}$ and
its value decreases very sharply depending on the strength of $\alpha_{e\mu}$.
In case of $\theta_{13}^m$, the effect of $V_{e\mu}$ is quite opposite as compared to 
$\theta_{23}^m$ as can be seen from the middle panel of 
Fig.~\ref{fig:running-mixing-angles}.
Assuming NH, as we increase $E$, $\theta_{13}^m$ quickly reaches to $45^{\circ}$ 
(resonance point) in the presence of $V_{e\mu}$ and then finally approaches 
toward $90^{\circ}$ as we further increase $E$. These two opposite behaviors 
of $\theta_{23}^m$ and $\theta_{13}^m$ alter the amplitudes and the locations 
of oscillation maxima in the transition probability substantially for non-zero 
$\alpha_{e\mu}$ which we discuss in the later part of this section. 
For $\alpha_{e\mu}$ = $10^{-52}$
($10^{-51}$), the resonance occurs around 4 GeV (0.6 GeV) for 1300 km baseline.
We can obtain an analytical expression for the resonance energy demanding 
$\theta_{13}^m$ = $45^{\circ}$ in Eq.~(\ref{eq:theta13-m}). In one mass scale 
dominance approximation where $\Delta m_{21}^2$ can be neglected
$\ie$ assuming $\alpha = 0$, the condition for the resonance energy 
($E_{res}$) takes the form:
\begin{equation}
\lambda_3 = A + W + \sin^2\theta_{13} \,.
\label{eq:resonance-energy-condition}
\end{equation}
Now, putting $\alpha = 0$ in Eqs.~(\ref{eq:lambda-3}) and (\ref{eq:theta23-m}),
we get a simplified expression for $\lambda_3$ which has the following form
\begin{equation}
\lambda_3 = \frac{1}{2}\left[\cos^2\theta_{13} - W + \sqrt{W^2+ (\cos^2\theta_{13})^2} \right]
\simeq \frac{1}{2}\left[2\cos^2\theta_{13} - W \right] \,,
\label{eq:simplified-lambda_3}
\end{equation}
since at resonance energies, the term $W^2$ is small compared to $\cos^4\theta_{13}$
and can be safely neglected. Now, comparing Eq.~(\ref{eq:resonance-energy-condition}) 
and Eq.~(\ref{eq:simplified-lambda_3}), we get a simple and compact expression for the
resonance energy
\begin{equation}
E_{res} = \frac{\Delta m_{31}^2\cos2\theta_{13}}{2V_{cc}+3V_{e\mu}} \,.
\label{eq:resonance-energy}
\end{equation}
In the absence of long-range potential $V_{e\mu}$, Eq.~(\ref{eq:resonance-energy})
gives us the standard expression for the resonance energy in the SM framework.
Eq.~(\ref{eq:resonance-energy}) suggests that for a given baseline, in the presence 
of $V_{e\mu}$, the resonance occurs at lower energy as compared to the 
SM case and this is exactly what we observe in the middle panel of 
Fig.~\ref{fig:running-mixing-angles}. The right panel of 
Fig.~\ref{fig:running-mass-squared-differences}
demonstrates that the effective solar mass-squared difference $\Delta m_{21,m}^2$
increases with energy and can be comparable to the vacuum value of 
$\Delta m_{31}^2$ at higher energies in the SM case. In the presence of $V_{e\mu}$, 
$\Delta m_{21,m}^2$ increases with energy even faster and can become quite large
at higher energies depending on the strength of $\alpha_{e\mu}$.
In the SM framework, the effective atmospheric mass-squared difference 
$\Delta m_{31,m}^2$ does not run with energy for this choice of 
baseline (see left panel of Fig.~\ref{fig:running-mass-squared-differences}).
But, in the presence of $V_{e\mu}$, the value of $\Delta m_{31,m}^2$
enhances a lot depending on the strength of $\alpha_{e\mu}$ as energy 
is increased. Next, to check the accuracy of our approximate analytical results, 
we compare the oscillation probabilities calculated with our approximate 
effective running mixing angles and mass-squared differences with those 
calculated numerically for the same baseline and line-averaged constant 
matter density along it.

%================================================
\subsection{Demonstration of the Accuracy of the Approximation}
\label{analytical-vs-numerical}
%================================================

%-----------------------------------------------------------------------------------------------
\begin{figure}[t]
\centerline{
\includegraphics[width=0.49\textwidth]{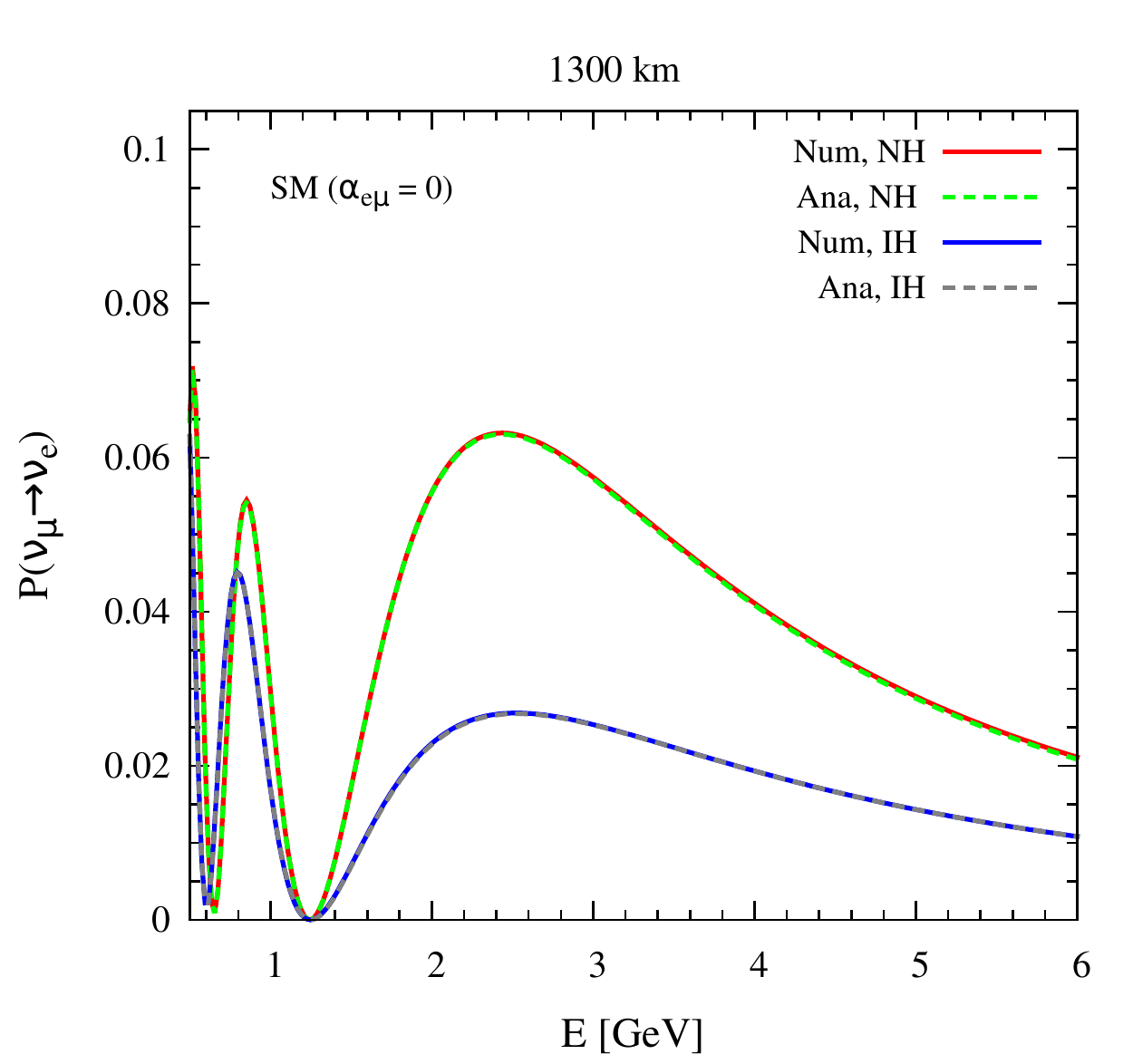}
\includegraphics[width=0.49\textwidth]{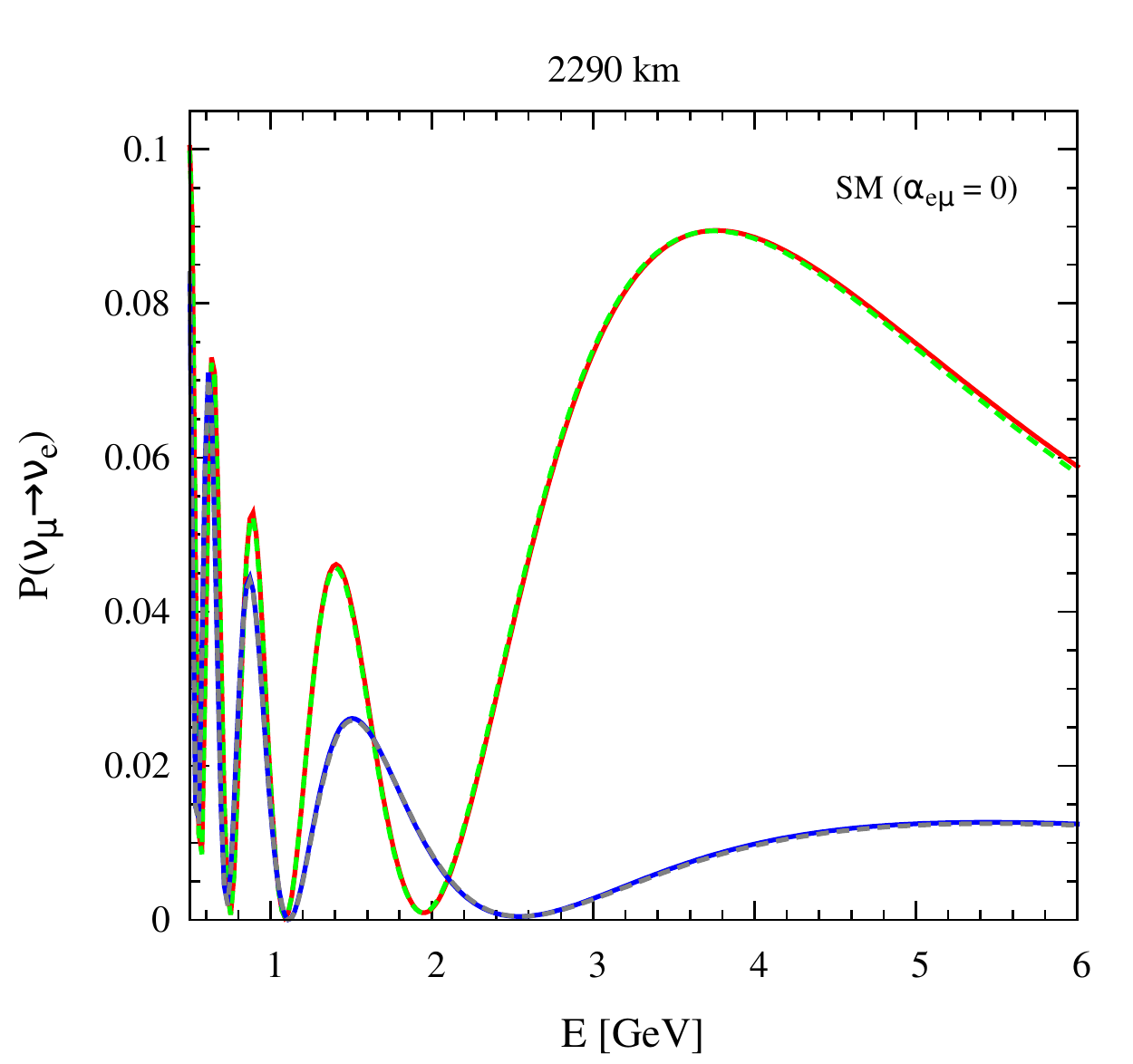}
}
\centerline{
\includegraphics[width=0.49\textwidth]{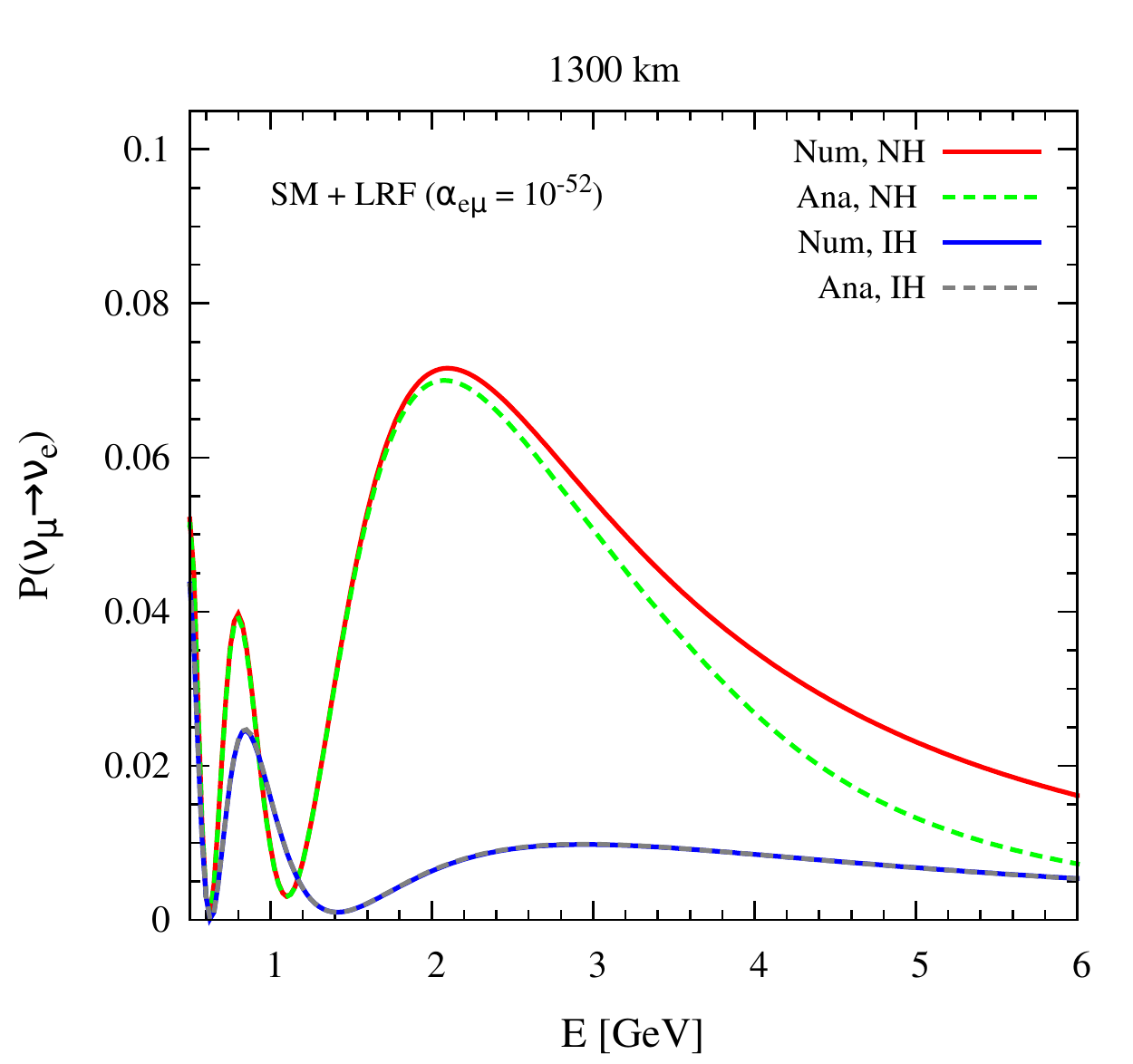}
\includegraphics[width=0.49\textwidth]{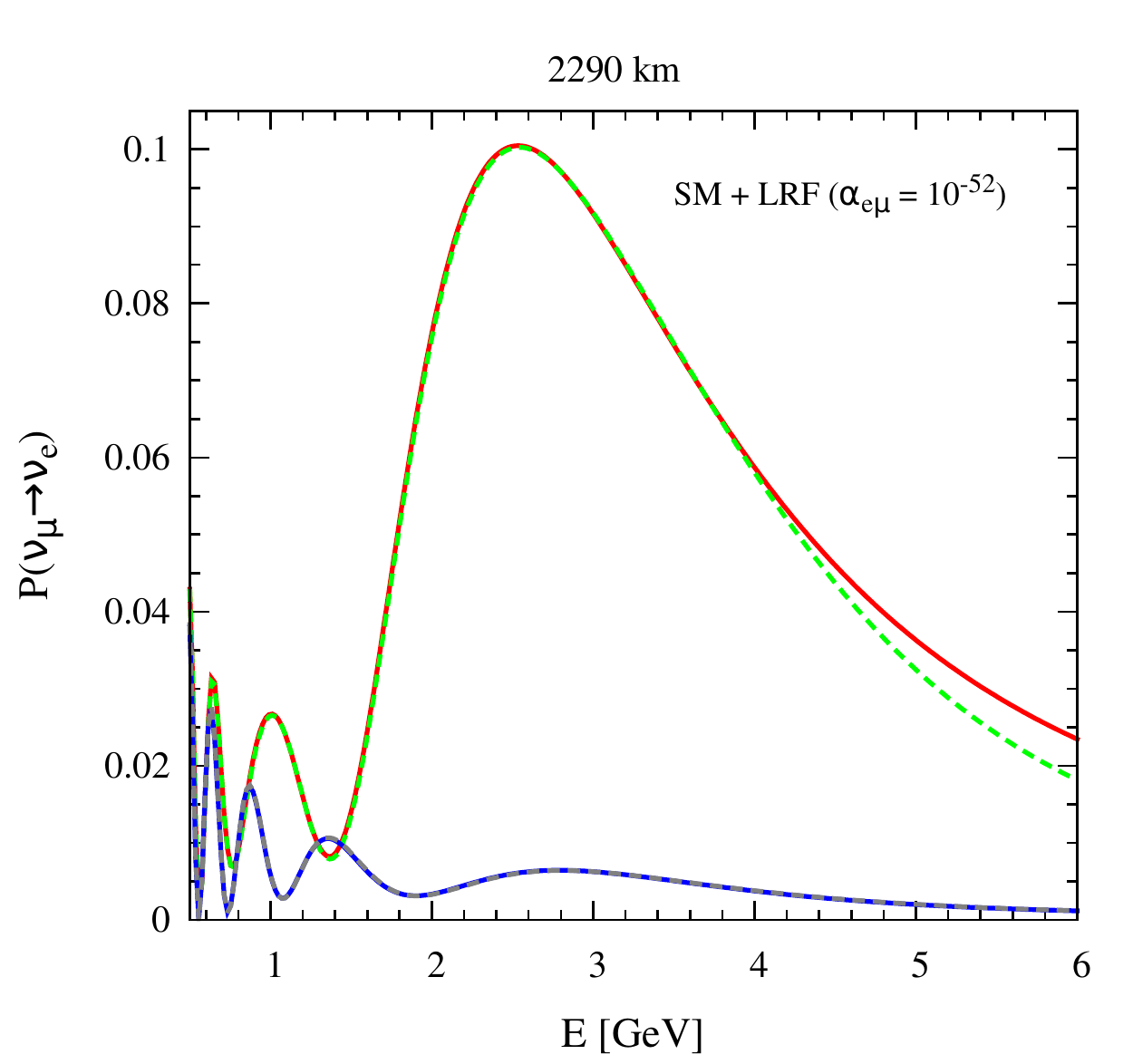}
}
\caption{$\numu \rightarrow \nue$ transition probability as a function 
of neutrino energy $E$ in GeV for 1300 km (2290 km) baseline in 
left (right) panels. The upper panels are for the SM case without 
long-range potential. The lower panels correspond to 
$\alpha_{e\mu}=10^{-52}$. In all the panels, we compare our analytical 
expressions (dashed curves) to the exact numerical results (solid curves) 
for NH and IH. The vacuum values of the oscillation parameters are taken 
from the third column of Table~\ref{tab:benchmark-parameters} and 
we take $\dcp = 0^{\circ}$.}
\label{fig:compare-appearance-neutrino}
\end{figure}
%----------------------------------------------------------------------------------------------

%--------------------------------------------------------------------------------------------
\begin{figure}[t]
\centerline{
\includegraphics[width=0.49\textwidth]{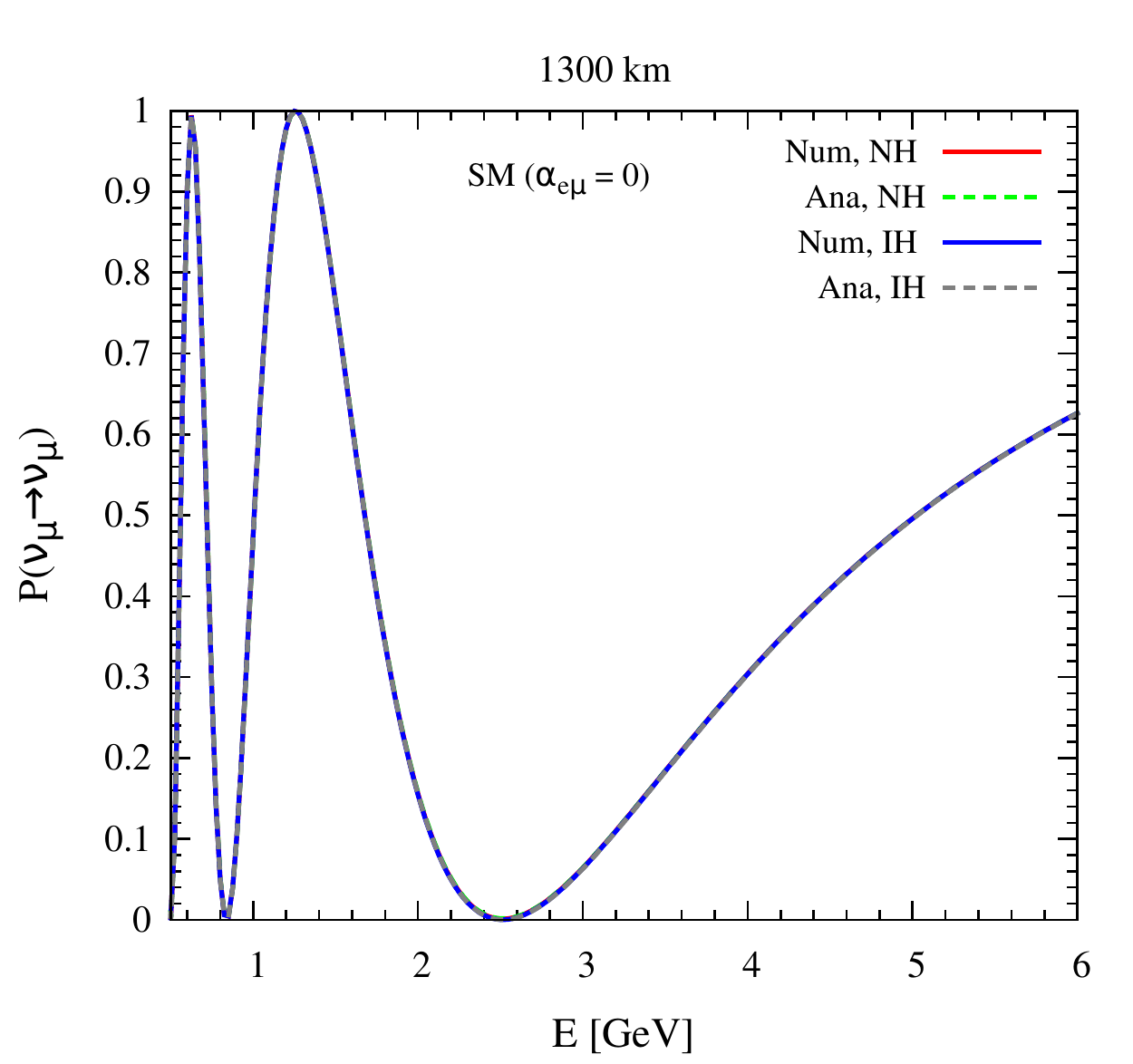}
\includegraphics[width=0.49\textwidth]{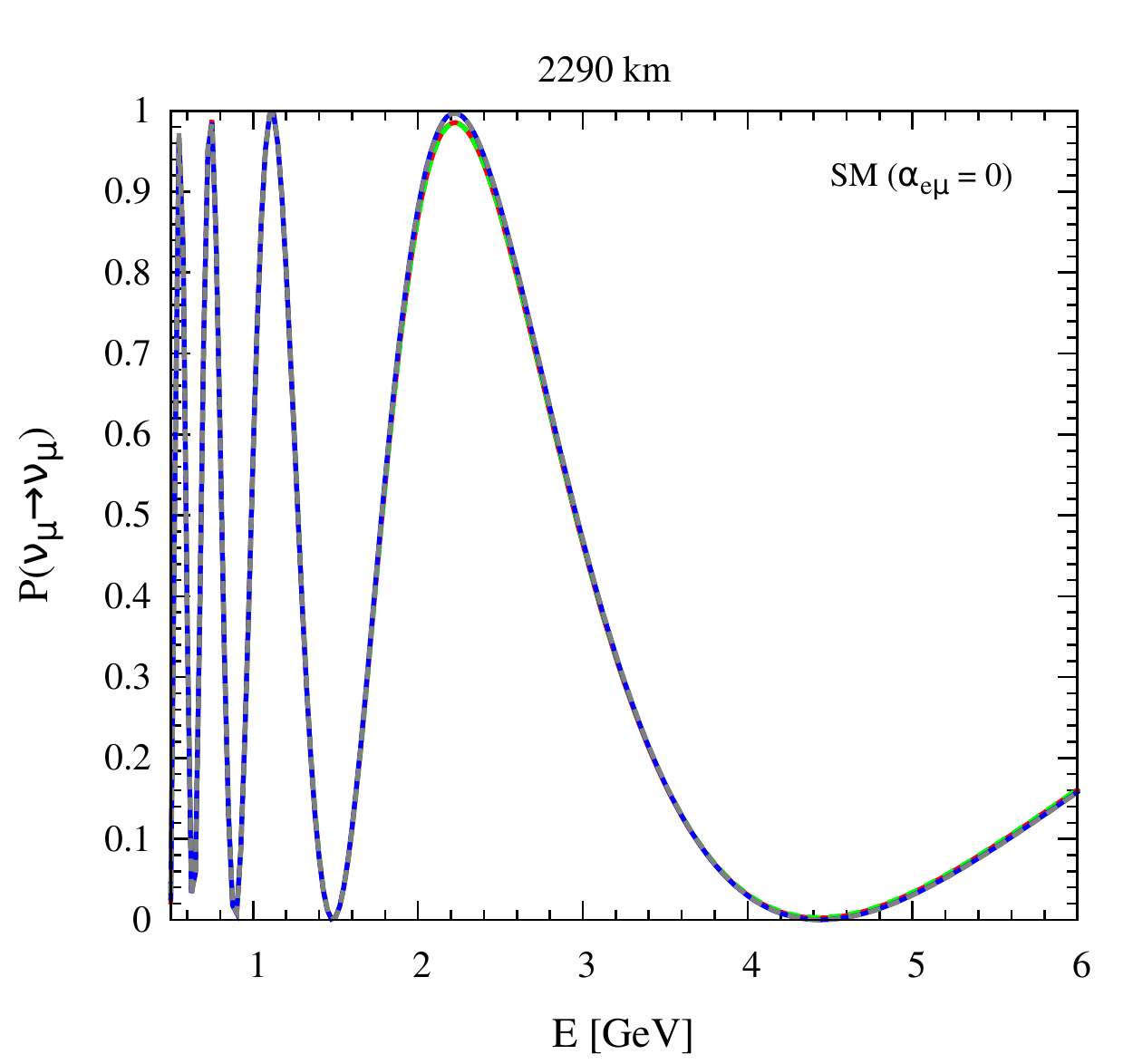}
}
\centerline{
\includegraphics[width=0.49\textwidth]{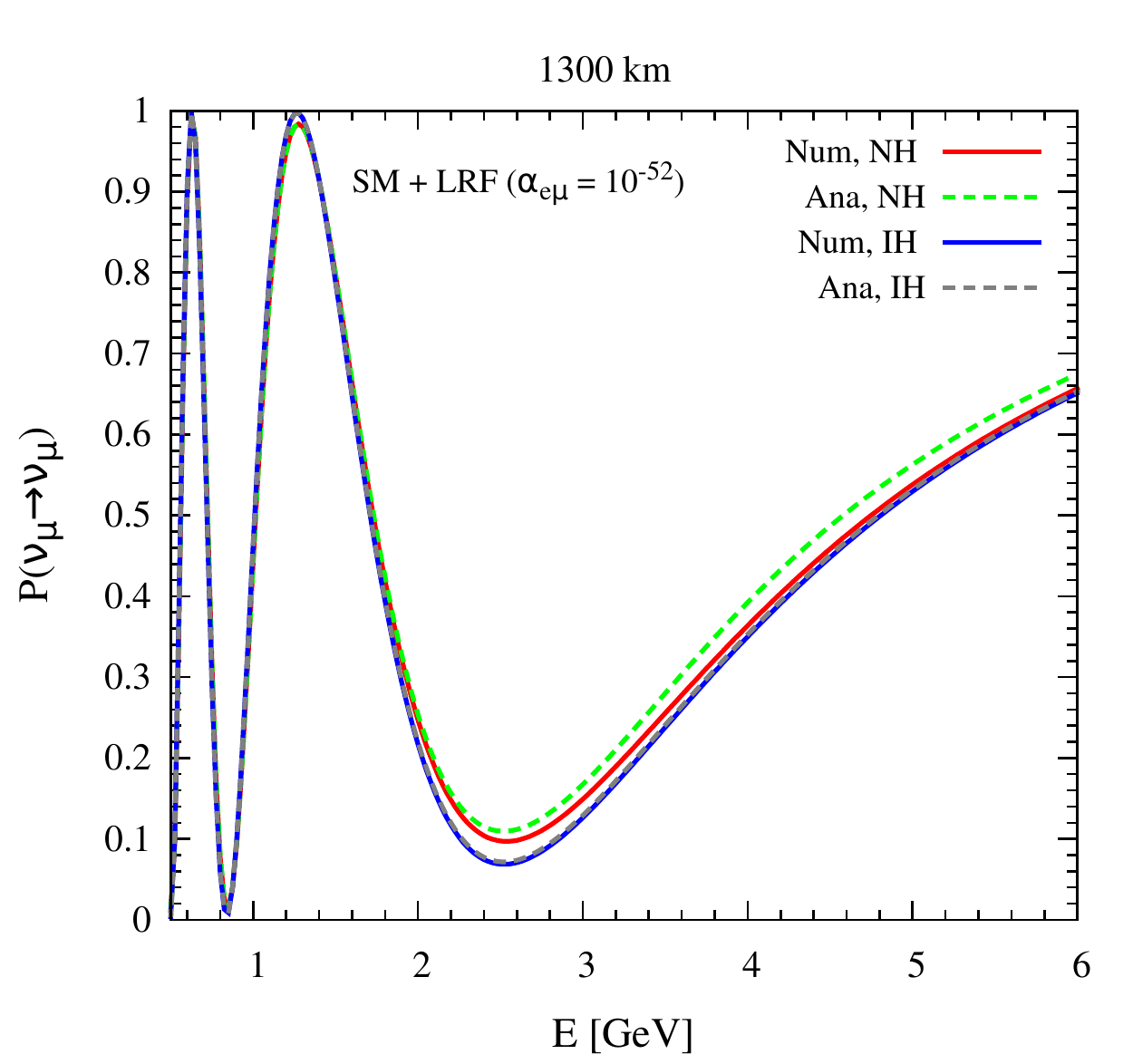}
\includegraphics[width=0.49\textwidth]{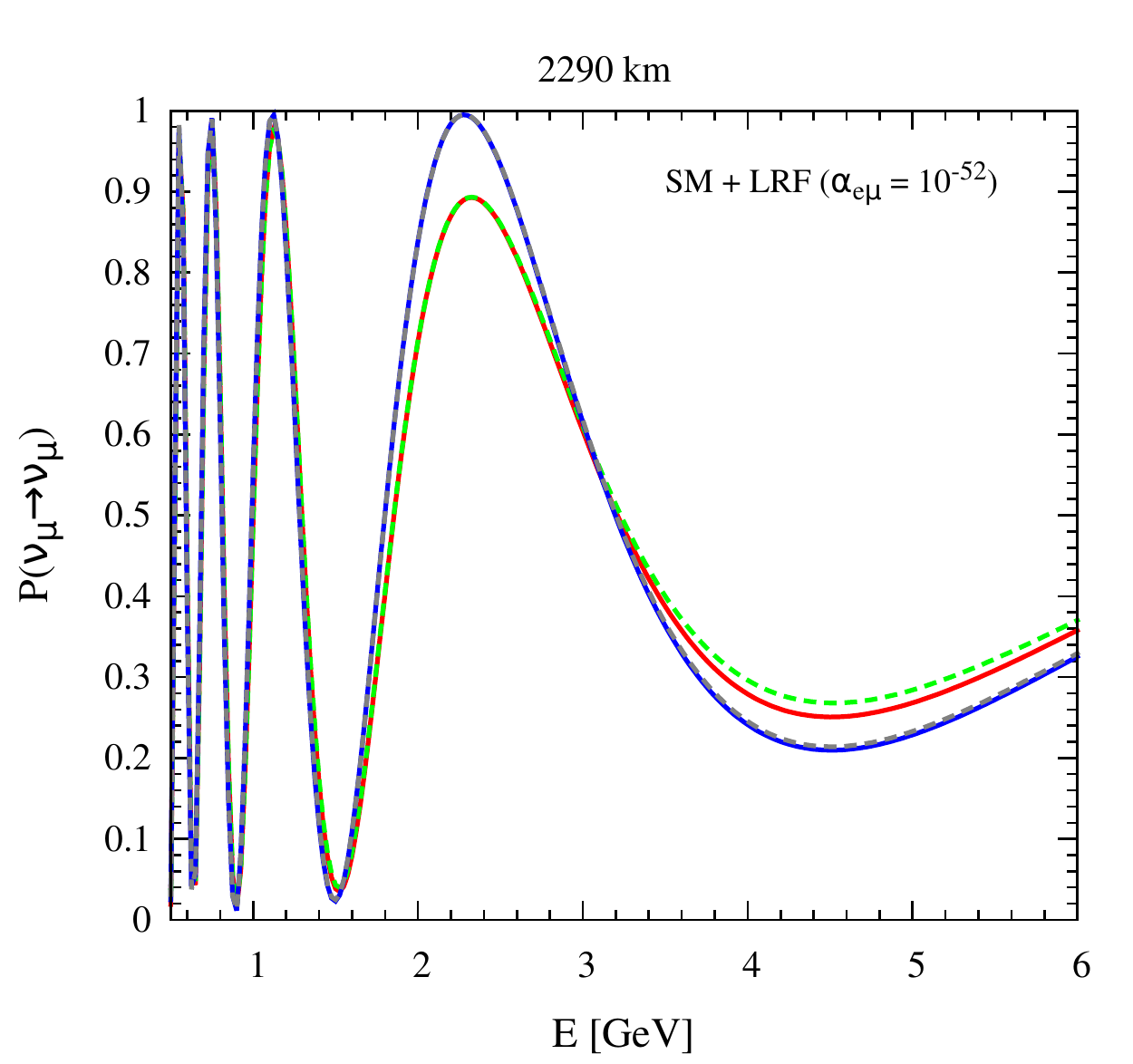}
}
\caption{$\numu \rightarrow \numu$ transition probability as a function 
of neutrino energy $E$ in GeV for 1300 km (2290 km) baseline in 
left (right) panels. The upper panels are for the SM case without 
long-range potential. The lower panels correspond to 
$\alpha_{e\mu}=10^{-52}$. In all the panels, we compare our analytical 
expressions (dashed curves) to the exact numerical results (solid curves) 
for NH and IH. The vacuum values of the oscillation parameters are taken 
from the third column of Table~\ref{tab:benchmark-parameters} and 
we take $\dcp = 0^{\circ}$.}
\label{fig:compare-disappearance-neutrino}
\end{figure}
%-------------------------------------------------------------------------------------------

The neutrino oscillation probabilities in the presence of $V_{CC}$ and $V_{e\mu}$
can be obtained by replacing the vacuum expressions of the elements of the mixing 
matrix $U$ and the mass-square differences $\Delta m_{ij}^2$ with their effective 
`running' values
\begin{equation}
U_{\alpha i}\;\rightarrow\;\tilde{U}_{\alpha i}\;(\theta_{12}\rightarrow\theta_{12}^m,
\theta_{13}\rightarrow\theta_{13}^m, \theta_{23}\rightarrow\theta_{23}^m)\;,\,\,\,
\Delta m_{ij}^2\;\rightarrow\;\Delta m_{ij,m}^2=m_{i,m}^2 - m_{j,m}^2 \;.
\label{eq:vacuum-to-matter}
\end{equation}
Incorporating the modifications due to $V_{CC}$ and $V_{e\mu}$, 
the new transition probability in a CP-conserving scenario can be 
written as
\begin{equation}
P\left(\nu_{\alpha}\rightarrow\nu_{\beta}\right) = \delta_{\alpha\beta} - 
4\sum_{i>j} \tilde{U}_{\alpha i}\tilde{U}_{\beta i}\tilde{U}_{\alpha j}\tilde{U}_{\beta j}
\sin^2\left(\frac{\Delta m_{ij,m}^2L}{4E}\right) \,.
\label{eq:matter-probability}
\end{equation}
Using Eq.~(\ref{eq:matter-probability}), we obtain the following expressions for
the appearance and disappearance channels \cite{Agarwalla:2013tza}
\begin{eqnarray}
P(\numu \rightarrow \nue)
& = & 4\,\tilde{U}_{\mu 2}^2 \tilde{U}_{e 2}^2 \sin^2\frac{\Delta m_{21,m}^2L}{4E}
     +4\,\tilde{U}_{\mu 3}^2 \tilde{U}_{e 3}^2 \sin^2\frac{\Delta m_{31,m}^2L}{4E} \cr
&   & +2\;\tilde{U}_{\mu 3}\tilde{U}_{e 3}\tilde{U}_{\mu 2}\tilde{U}_{e 2}
      \left(4\sin^2\frac{\Delta m_{21,m}^2L}{4E}\sin^2\frac{\Delta m_{31,m}^2L}{4E}\right) \cr
&   & +2\;\tilde{U}_{\mu 3}\tilde{U}_{e 3}\tilde{U}_{\mu 2}\tilde{U}_{e 2}
      \left(\sin\frac{\Delta m_{21,m}^2L}{2E}\sin\frac{\Delta m_{31,m}^2L}{2E} \right)  \;,
\label{eq:prob-numu-to-nue}
\end{eqnarray}
\begin{eqnarray}
P(\numu\rightarrow\numu)
& = & 1 - 4\,\tilde{U}_{\mu 2}^2 \left(1 - \tilde{U}_{\mu 2}^2 \right)
            \sin^2\frac{\Delta m_{21,m}^2L}{4E}
        - 4\,\tilde{U}_{\mu 3}^2 \left(1 - \tilde{U}_{\mu 3}^2 \right)
            \sin^2\frac{\Delta m_{31,m}^2L}{4E} \cr
& &  \phantom{1}
        + 2\, \tilde{U}_{\mu 2}^2 \tilde{U}_{\mu 3}^2
          \left(4\sin^2\frac{\Delta m_{21,m}^2L}{4E}\sin^2\frac{\Delta m_{31,m}^2L}{4E} \right) \cr
& &  \phantom{1}
        + 2\, \tilde{U}_{\mu 2}^2 \tilde{U}_{\mu 3}^2
           \left(\sin\frac{\Delta m_{21,m}^2L}{2E}\sin\frac{\Delta m_{31,m}^2L}{2E} \right)  \;.
\label{eq:prob-numu-to-numu}
\end{eqnarray}

In Fig.~\ref{fig:compare-appearance-neutrino}, we present our approximate 
$\nu_\mu\rightarrow\nu_e$ oscillation probabilities (dashed curves) as a function 
of the neutrino energy against the exact numerical results (solid curves) 
considering $L$ = 1300 km (left panels) and 2290 km (right panels).
We give the plots for both NH and IH in all the panels considering 
line-averaged constant Earth matter densities for both the baselines.
The upper panels are drawn for the SM case ($\alpha_{e\mu}=0$) where
our approximate results match exactly with the numerically obtained
probabilities. In the lower panels, we give the probabilities considering 
$\alpha_{e\mu}=10^{-52}$ and find that our approximate expressions 
work quite well in the presence of long-range potential and can predict
almost accurate $L/E$ patterns of the oscillation probability.
In Fig.~\ref{fig:compare-disappearance-neutrino}, we study the same 
for $\numu \rightarrow \numu$ oscillation channel and find that
our approximate expressions match quite nicely with the numerical results.
Here, we present our analytical results for NH. We can obtain the same
for IH by changing $\Delta m_{31}^2 \rightarrow - \Delta m_{31}^2$.
Following the same procedure and reversing the sign of $V$ in
Eq.~(\ref{eq:new-flavor-Hamiltonian}), we can derive the analytical
expressions for anti-neutrino as well. Note that in this paper, we limit our 
investigation to $L_e-L_{\mu}$ symmetry, though similar procedure
can be adopted for $L_e-L_{\tau}$ symmetry.

%================================================
\subsection{Discussion at the Probability Level -- Neutrino Case}
\label{probability-discusions-neutrino-case}
%================================================

%-----------------------------------------------------------------------------------------------------------------------
\begin{figure}[t]
\centerline{
\includegraphics[width=7.5cm,height=7.0cm]{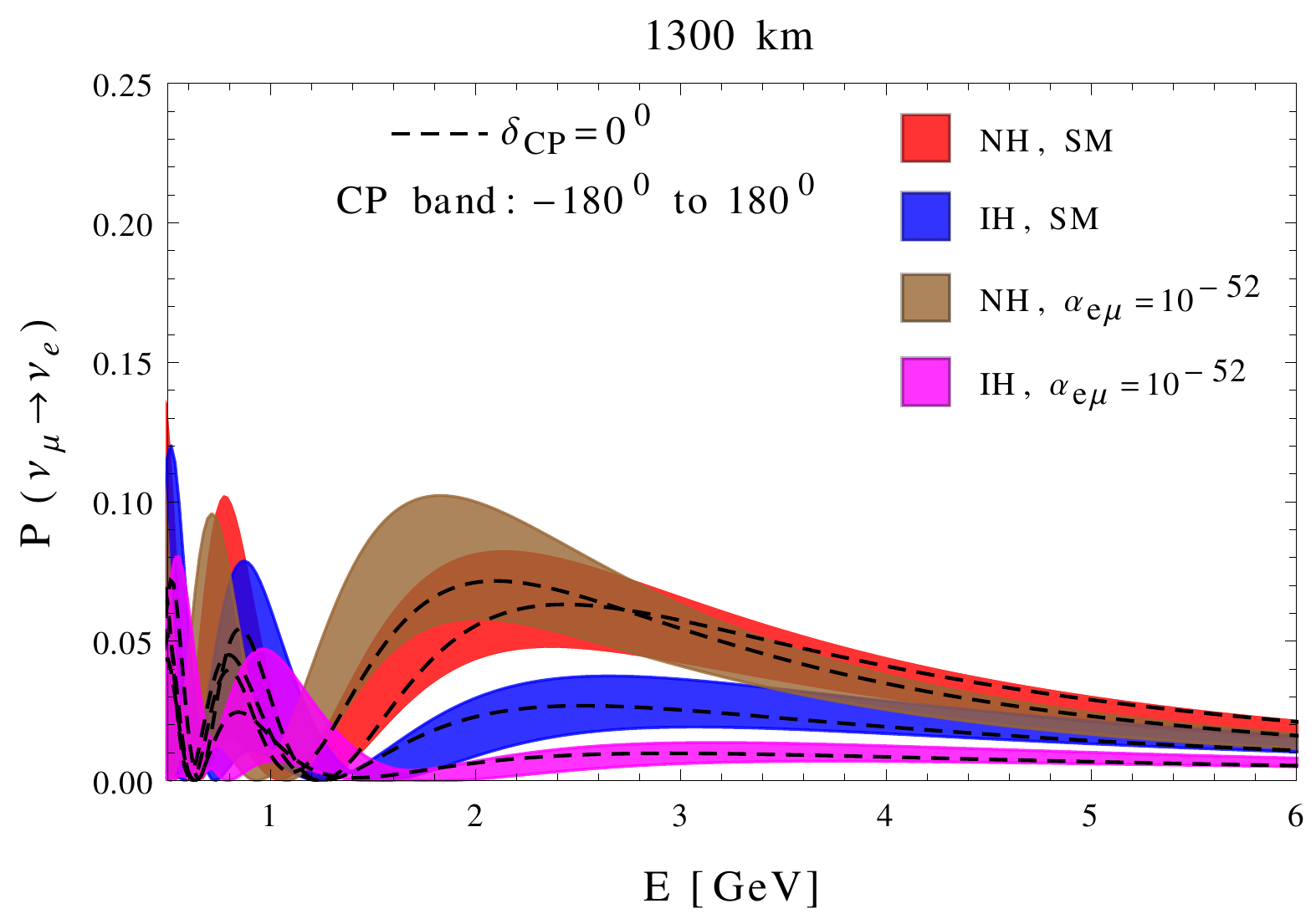}
\includegraphics[width=7.5cm,height=7.0cm]{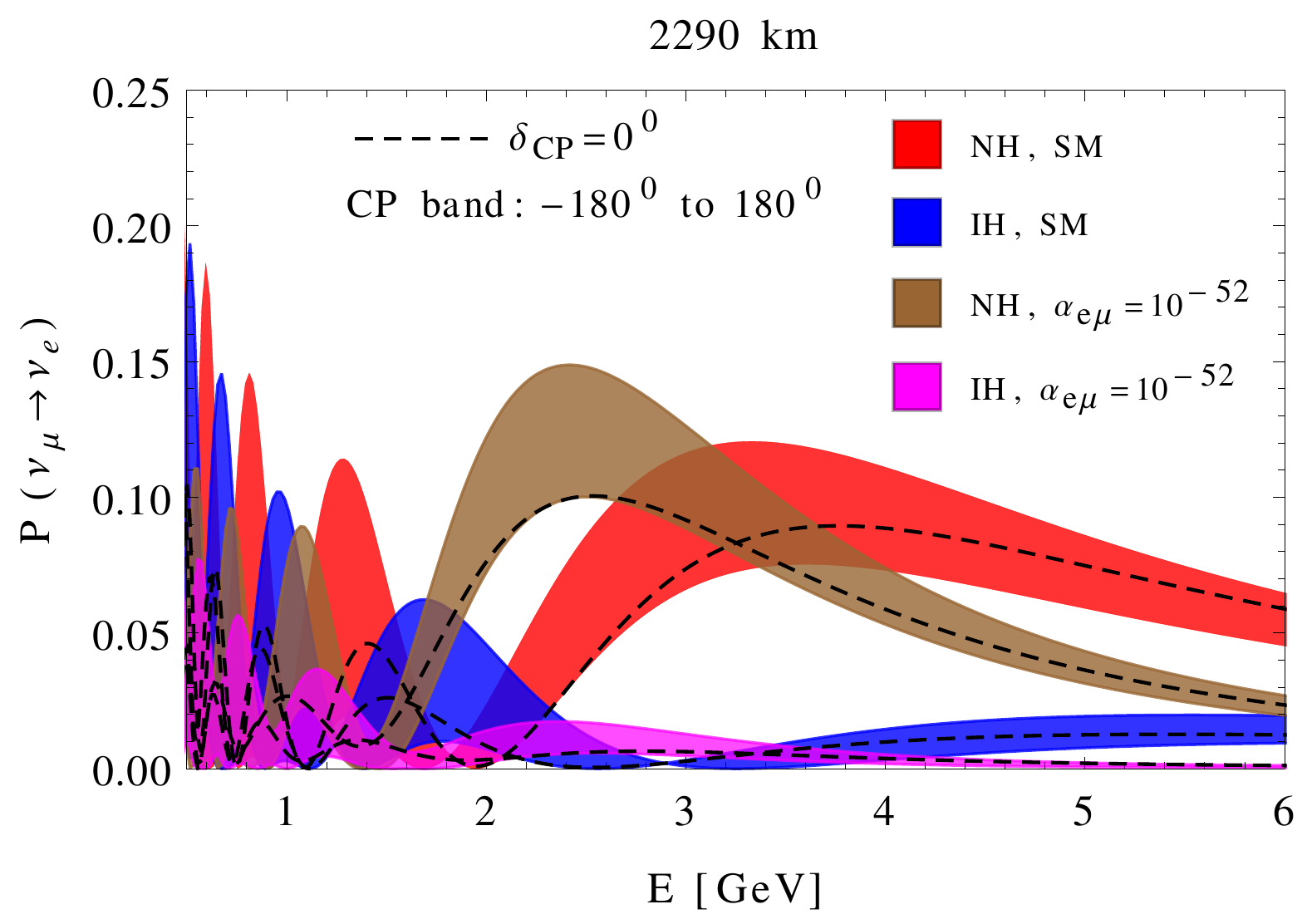}
}
\centerline{
\includegraphics[width=7.5cm,height=7.0cm]{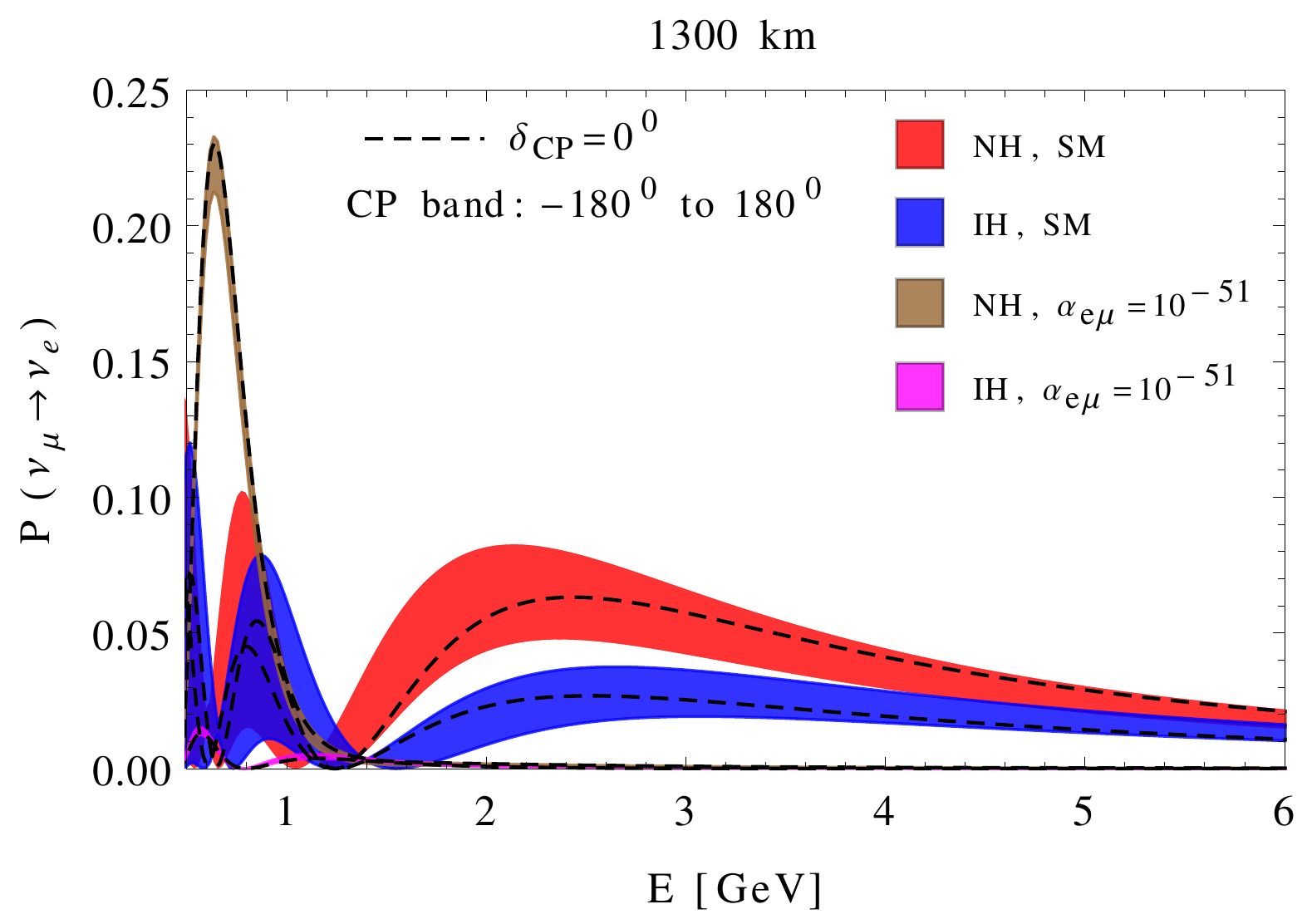}
\includegraphics[width=7.5cm,height=7.0cm]{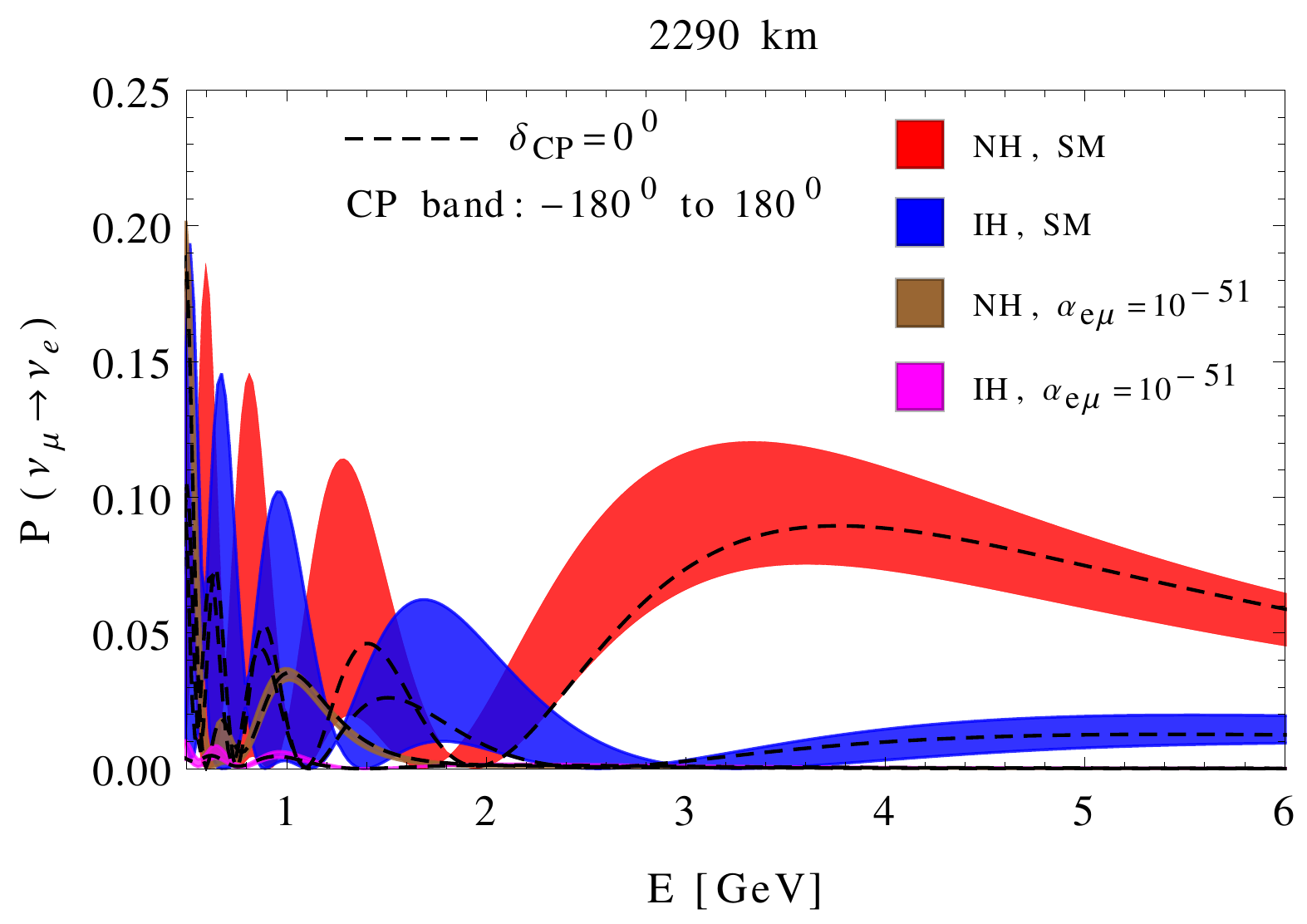}
} 
\caption{The transition probability $P_{\mu e}$ as a function of neutrino energy. The band reflects 
the effect of unknown $\dcp$. Inside each band, the probability for $\dcp = 0^\circ$ case is shown 
by the black dashed line. The left panels (right panels) are for 1300 km (2290 km) baseline.
In each panel, we compare the probabilities for NH and IH with and without long-range potential. 
In the upper (lower) panels, we take $\alpha_{e\mu}=10^{-52}$ ($\alpha_{e\mu}=10^{-51}$)
for the cases with long-range potential.}
\label{fig:neutrino-appearance-probability}
\end{figure}
%-----------------------------------------------------------------------------------------------------------------------

%-----------------------------------------------------------------------------------------------------------------------
\begin{figure}[t]
\centerline{
\includegraphics[width=7.5cm,height=7.0cm]{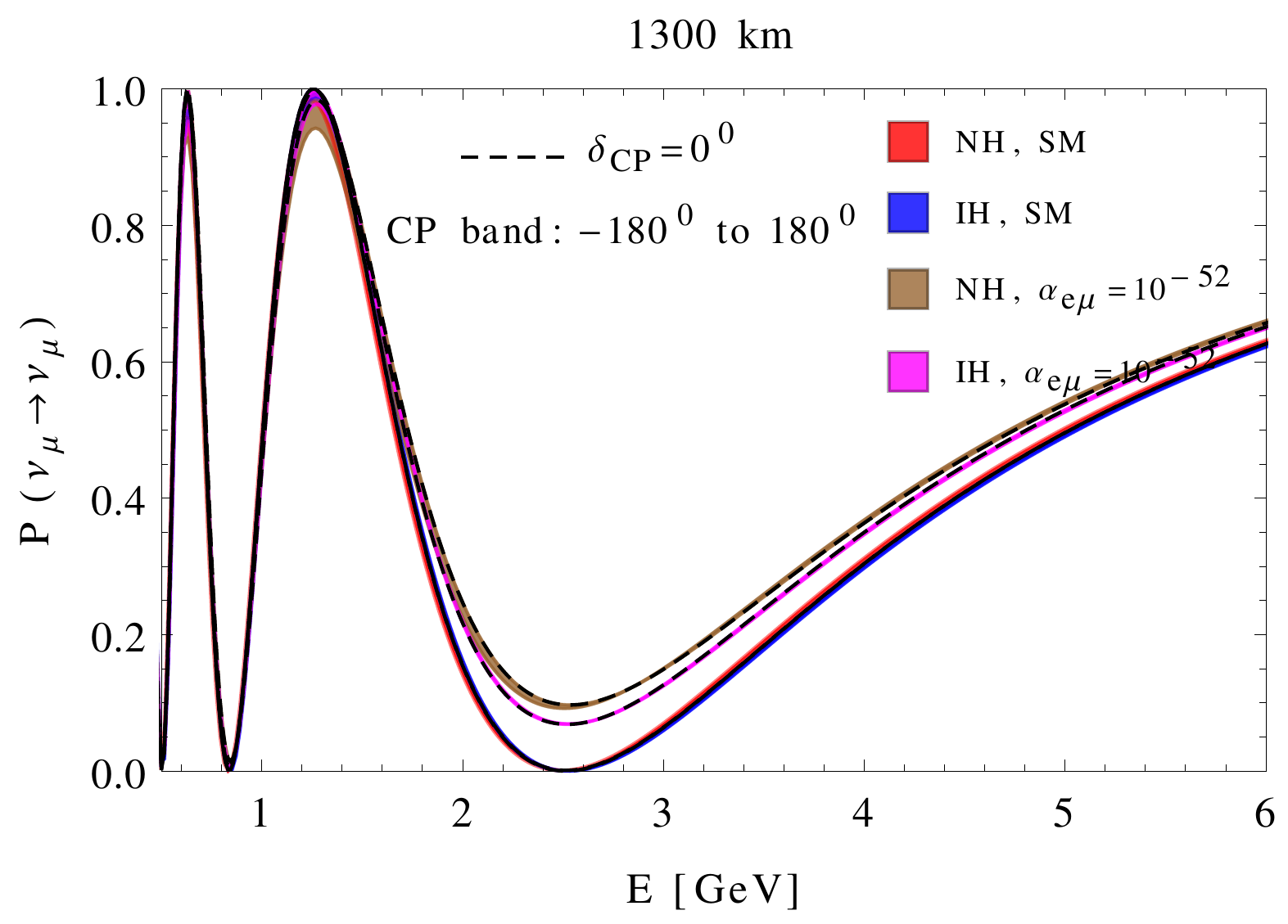}
\includegraphics[width=7.5cm,height=7.0cm]{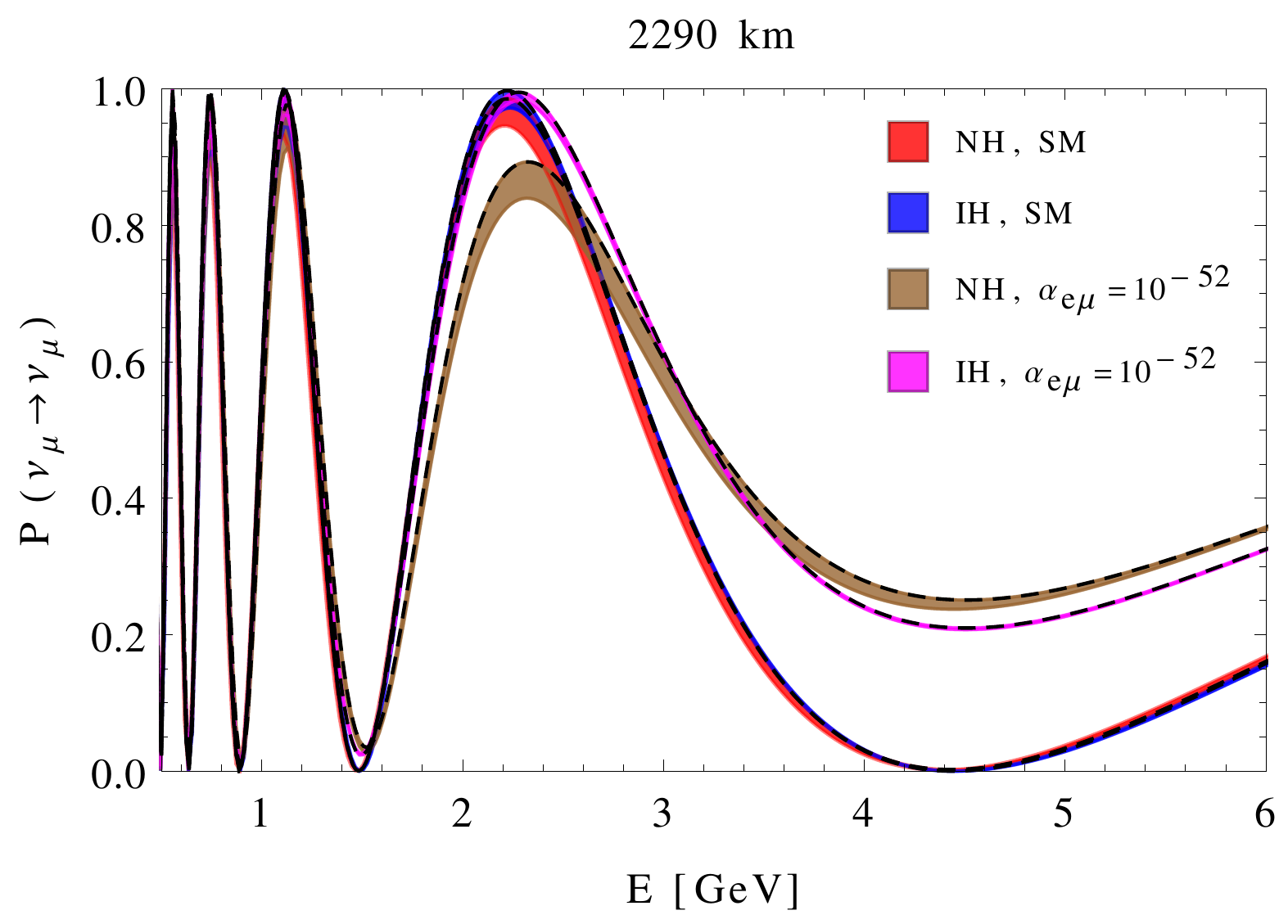}
}
\centerline{
\includegraphics[width=7.5cm,height=7.0cm]{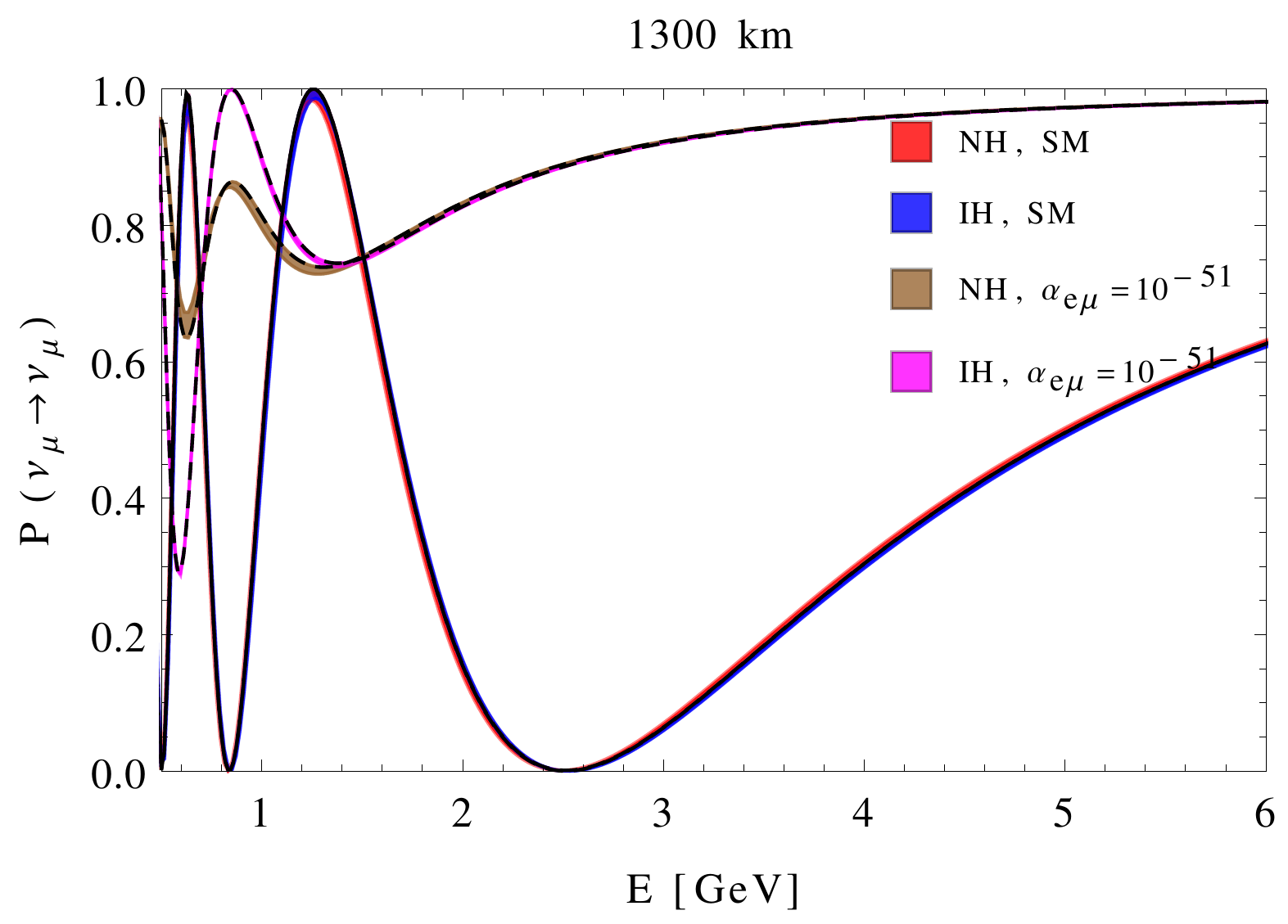}
\includegraphics[width=7.5cm,height=7.0cm]{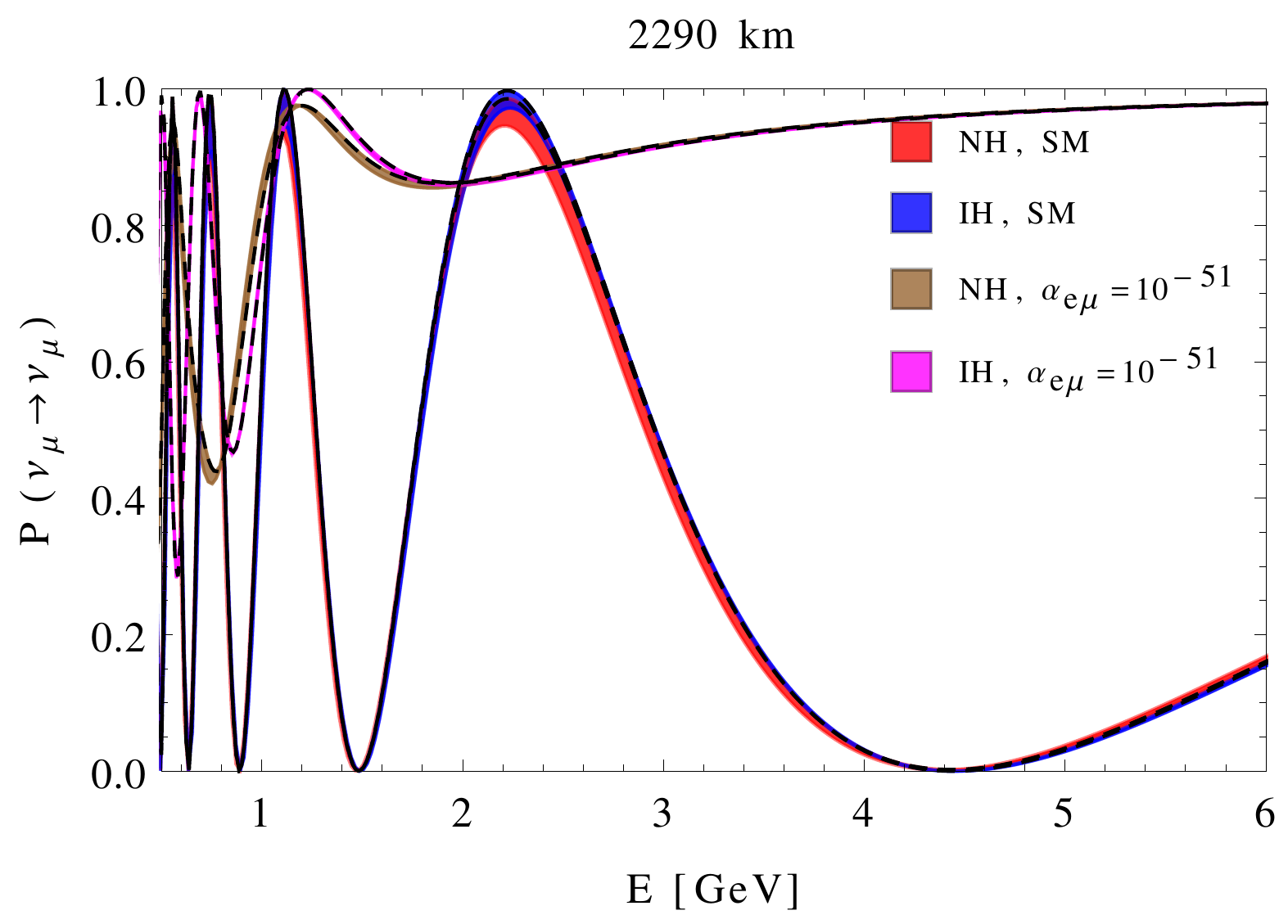}
}
\caption{The transition probability $P_{\mu\mu}$ as a function of neutrino energy. The band reflects 
the effect of unknown $\dcp$. Inside each band, the probability for $\dcp = 0^\circ$ case is shown 
by the black dashed line. The left panels (right panels) are for 1300 km (2290 km) baseline.
In each panel, we compare the probabilities for NH and IH with and without long-range potential. 
In the upper (lower) panels, we take $\alpha_{e\mu}=10^{-52}$ ($\alpha_{e\mu}=10^{-51}$)
for the cases with long-range potential.}
\label{fig:neutrino-disappearance-probability}
\end{figure}
%----------------------------------------------------------------------------------------------------------------------

In this section, we discuss in detail how the long-range potential affects 
the full three-flavor neutrino oscillation probabilities in matter considering 
non-zero values of $\dcp$. In Fig.~\ref{fig:neutrino-appearance-probability}, 
we show the exact numerical transition probability $P_{\mu e}$ as a function 
of the neutrino energy using the line-averaged constant Earth matter 
densities for 1300 km (left panels) and 2290 km (right panels) baselines. 
We vary $\dcp$ within the range $-180^\circ$ to $180^\circ$ and the resultant 
probability is shown as a band, with the thickness of the band reflecting the effect 
of $\dcp$ on $P_{\mu e}$. Inside each band, the probability for $\dcp = 0^\circ$ 
case is shown explicitly by the black dashed line. The left panels (right panels) 
are for 1300 km (2290 km) baseline. In each panel, we compare the probabilities 
for NH and IH with and without long-range potential. In the upper (lower) panels, 
we take $\alpha_{e\mu}=10^{-52}$ ($\alpha_{e\mu}=10^{-51}$) for the cases 
with long-range potential. We study the same for the disappearance 
($\numu \to \numu$) channel in Fig.~\ref{fig:neutrino-disappearance-probability}.
We give the similar plots for the anti-neutrino case in appendix 
\ref{probability-discusions-anti-neutrino-case}.

To explain the behavior of the oscillation probabilities in 
Fig.~\ref{fig:neutrino-appearance-probability} and
Fig.~\ref{fig:neutrino-disappearance-probability} at the qualitative 
level, we can simplify the analytical expressions given in 
Eqs.~(\ref{eq:prob-numu-to-nue}) and (\ref{eq:prob-numu-to-numu})
in the following fashion. The extreme right panel of 
Fig.~\ref{fig:running-mixing-angles} suggests that 
$\sin\theta_{12}^m \rightarrow 1$ and $\cos\theta_{12}^m \rightarrow 0$ 
very quickly as we increase $E$ in the SM case or with non-zero
$\alpha_{e\mu}$. So, we set $\sin\theta_{12}^m \approx 1$ and
$\cos\theta_{12}^m \approx 0$ in 
Eqs.~(\ref{eq:prob-numu-to-nue}) and (\ref{eq:prob-numu-to-numu})
and obtain the following simple expressions:
\begin{equation}
P(\numu \rightarrow \nue) = \sin^2\theta_{23}^m \, \sin^22\theta_{13}^m \, \sin^2\frac{\Delta m_{32,m}^2L}{4E} \,,
\label{eq:omsd-prob-numu-to-nue}
\end{equation}
and
\begin{eqnarray}
P(\numu \rightarrow \numu)
& = & 1 - \sin^22\theta_{23}^m \sin^2\theta_{13}^m \sin^2\frac{\Delta m_{21,m}^2L}{4E} \cr
& & \phantom{1}     
       - \sin^22\theta_{23}^m \cos^2\theta_{13}^m \sin^2\frac{\Delta m_{31,m}^2L}{4E} \cr
& &  \phantom{1}             
       - \sin^4\theta_{23}^m \sin^22\theta_{13}^m \sin^2\frac{\Delta m_{32,m}^2L}{4E} \,.
\label{eq:omsd-prob-numu-to-numu}
\end{eqnarray}
In Fig.~\ref{fig:neutrino-appearance-probability}, we can see that 
for $\alpha_{e\mu} = 10^{-52}$ case (upper panels), the locations of the
first oscillation maxima have been shifted toward lower energies for both
the baselines and also the amplitudes of the first oscillation maxima have
been enhanced when we assume NH. This can be understood from the 
`running' of $\theta_{23}^m$, $\theta_{13}^m$ (extreme left and middle 
panels of Fig.~\ref{fig:running-mixing-angles}), and $\Delta m_{32,m}^2$ 
(left and right panels of Fig.~\ref{fig:running-mass-squared-differences}).
As we go to higher energies, $\theta_{13}^m$ increases and $\theta_{23}^m$
decreases and there is a trade-off between the terms $\sin^2\theta_{23}^m$
and $\sin^22\theta_{13}^m$ in Eq.~(\ref{eq:omsd-prob-numu-to-nue}). 
Also, the value of $\Delta m_{32,m}^2$ ($\Delta m_{31,m}^2$ - $\Delta m_{21,m}^2$)
decreases with energy as $\Delta m_{21,m}^2$ increases by substantial amount 
compared to $\Delta m_{31,m}^2$, which shifts the location of the first oscillation
maxima toward lower energies. For IH, the value of $\theta_{13}^m$ decreases
fast with non-zero $\alpha_{e\mu}$ compared to the SM case, causing a depletion 
in the probabilities over a wide range of energies.
In case of $\alpha_{e\mu} = 10^{-51}$ (lower panels), there is a huge 
suppression in the probabilities at both the baselines over a wide range of energies 
above 1 GeV assuming NH. The main reason behind this large damping in 
the probabilities is that $\theta_{13}^m$ approaches very quickly to $90^{\circ}$
around 1 GeV or so for $\alpha_{e\mu} = 10^{-51}$
(see middle panel of Fig.~\ref{fig:running-mixing-angles}) and 
therefore, $\sin^22\theta_{13}^m \rightarrow 0$, vanishing the 
probability amplitude for $\numu \rightarrow \nue$ oscillation channel. 
Below 1 GeV, $\theta_{13}^m$ runs toward $45^{\circ}$ and therefore,
$\sin^22\theta_{13}^m \rightarrow 1$, causing the enhancement in the 
probabilities. When we take IH, $\theta_{13}^m$ quickly advances 
to zero, causing a huge damping in the probabilities at all the energies. 
These `running' behaviors of $\theta_{23}^m$, $\theta_{13}^m$, and 
the mass-squared differences in the presence of long-range potential 
as discussed above also affect $\numu \rightarrow \numu$ oscillation 
channel (see Fig.~\ref{fig:neutrino-disappearance-probability}) 
which can be explained with the help of 
Eq.~(\ref{eq:omsd-prob-numu-to-numu}).
Next, we discuss how the long-range potential due to $L_e-L_{\mu}$ 
symmetry modifies the expected event spectra and total event rates
of the DUNE and LBNO experiments.

%============================================
\section{Impact of Long-Range Potential at the Event Level}
\label{event-study-with-lrf}
%============================================

We start this section with a brief description of the main experimental features 
of the DUNE and LBNO set-ups that we use in our simulation.

%=========================================
\subsection{Key Features of DUNE and LBNO Set-ups}
\label{set-ups}
%=========================================

The proposed DUNE experiment 
\cite{Diwan:2003bp,Barger:2007yw,Huber:2010dx,Akiri:2011dv,Adams:2013qkq,DUNE} 
in the United States with a baseline of 1300 km from Fermilab to Homestake mine
in South Dakota is planning to build a massive 35 kt liquid argon time projection chamber 
(LArTPC) as the far detector choice. This LArTPC will have excellent kinematic reconstruction 
capability for all the observed particles, rejecting almost all of the large neutral current 
background. We use the detector properties which are given in Table 1 of 
Ref. \cite{Agarwalla:2011hh}. As far as the neutrinos are concerned,
this facility will have a new, high intensity, on-axis neutrino beam, which 
in its initial phase, will operate at a proton beam power of 708 kW, with 
proton energy of 120 GeV, delivering $6 \times 10^{20}$ protons on target 
in 230 days per calendar year. In this work, we have used the latest fluxes being 
considered by the collaboration \cite{mbishai}. We have assumed five years of 
neutrino run and five years of anti-neutrino run to estimate the physics capabilities
of this set-up.

In Europe, the proposed LBNO experiment 
\cite{Agarwalla:2011hh,Stahl:2012exa,::2013kaa,Agarwalla:2014tca}
offers an interesting possibility to address the fundamental unsolved 
issues in neutrino oscillation physics using a baseline of 2290 km 
between CERN and Pyh\"asalmi mine in Finland which enables us
to cover a wide range of $L/E$ choices, mandatory to resolve parameter
degeneracies. The Pyh\"asalmi mine will house a giant 70 kt LArTPC
as a far detector which will observe the neutrinos produced in a 
conventional wide-band beam facility at CERN. The fluxes that 
we use in our simulation have been computed assuming an
exposure of $1.5 \times 10^{20}$ protons on target in 200 days per 
calendar year from the SPS accelerator at 400 GeV with a
beam power of 750 kW \cite{poster}. For LBNO also, we assume 
five years of neutrino run and five years of anti-neutrino run.
We consider the same detector properties as that of DUNE.

%=============================
\subsection{Event Spectrum and Rates}
\label{event-spectrum-rates}
%=============================

In this section, we present the expected event spectra and total event rates
for both the set-ups under consideration in the presence of long-range potential.
We calculate the number of expected electron events\footnote{The number of 
positron events can be estimated using Eq.~(\ref{eq:events}), by considering
appropriate oscillation probability and cross-section. The same is true for 
$\mu^{\pm}$ events.} in the $i$-th energy bin in the detector using the following 
expression
\begin{equation}
N_{i} = \frac{T\, n_n\, \epsilon}{4\pi L^2}~ \int_0^{E_{\rm max}}
dE \int_{E_{A_i}^{\rm min}}^{E_{A_i}^{\rm max}} dE_A \,\phi(E)
\,\sigma_\nue(E) \,R(E,E_A)\, P_{\mu e}(E) \, ,
\label{eq:events}
\end{equation}
where $\phi(E)$ is the neutrino flux, $T$ is the total running time, 
$n_n$ is the number of target
nucleons in the detector, $\epsilon$ is the detector efficiency, and
$R(E,E_A)$ is the Gau\ss ian energy resolution function of the detector.
$\sigma_\nue$ is the neutrino interaction cross-section which has been 
taken from Refs. \cite{Messier:1999kj,Paschos:2001np}, where the
authors estimated the cross-section for water and isoscalar targets.
In order to have LAr cross-sections, we have scaled the inclusive 
charged current cross-sections of water by a factor of 1.06 for neutrino 
and 0.94 for anti-neutrino \cite{zeller,petti-zeller}.
The quantities $E$ and $E_A$ are the true and reconstructed (anti-)neutrino 
energies respectively, and $L$ is the path length.

%---------------------------------------------------------------------------------------------------------------------
\begin{figure}[H]
\centerline{
\includegraphics[width=0.49\textwidth]{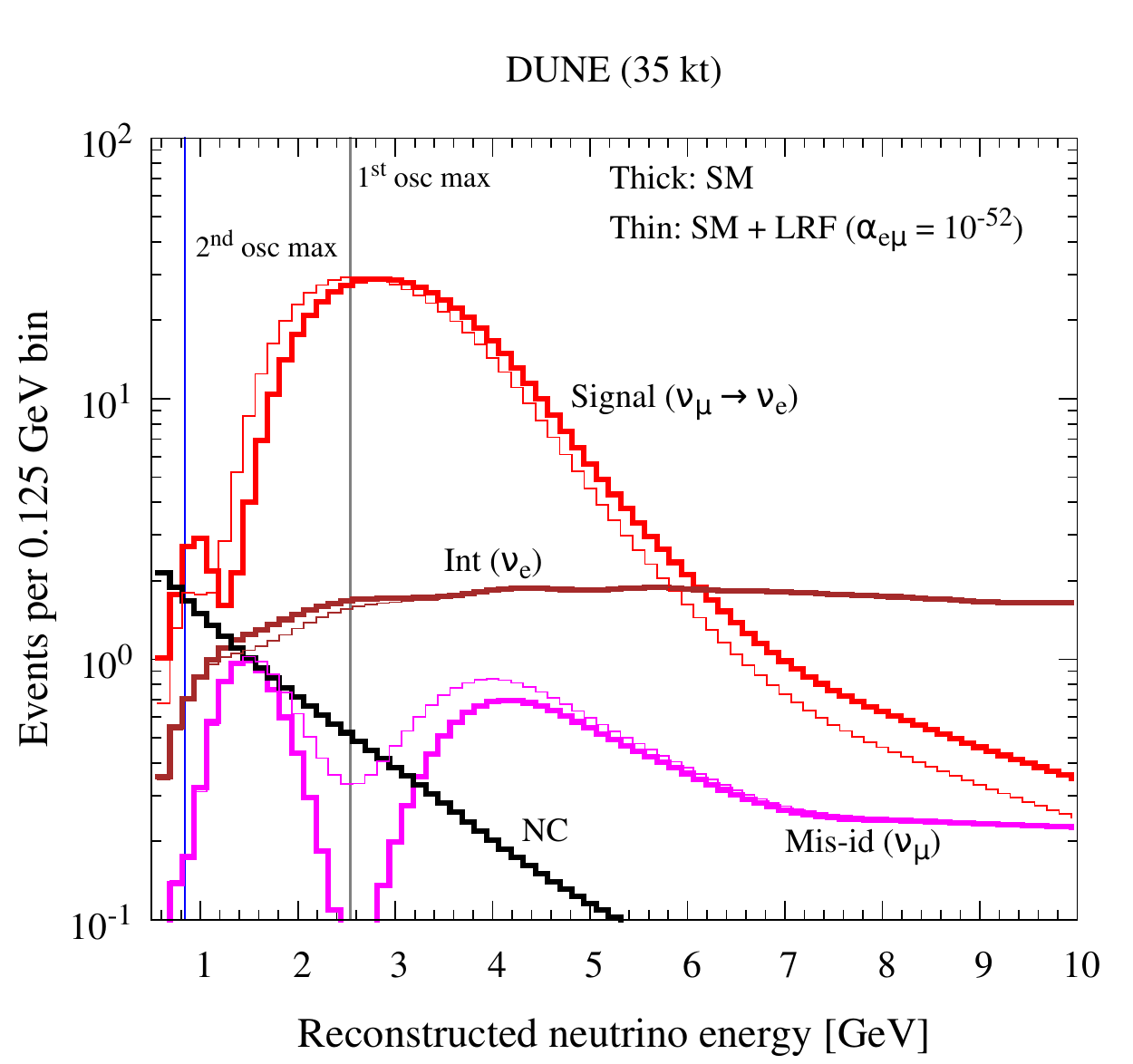}
\includegraphics[width=0.49\textwidth]{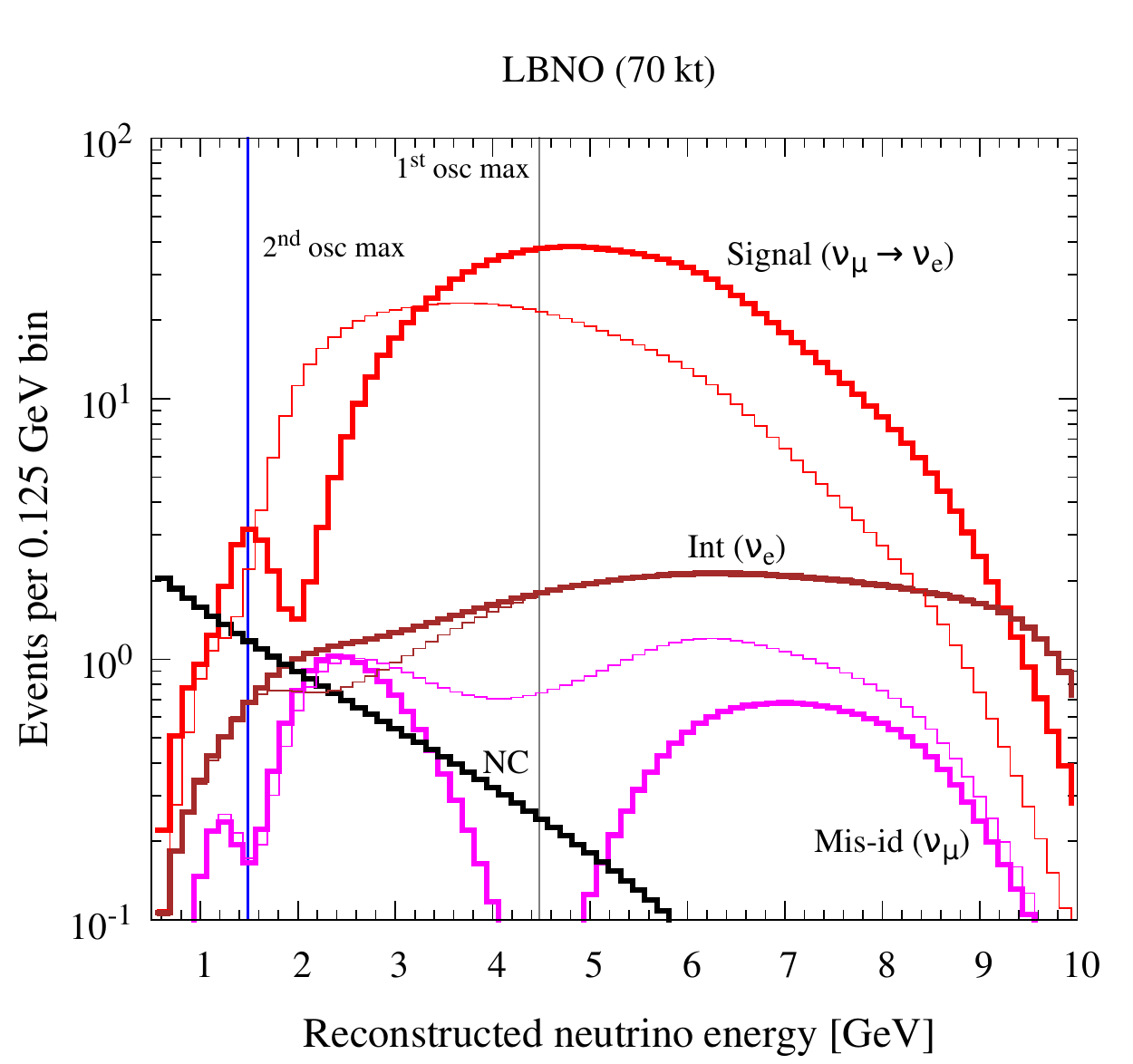}
}
\caption{Expected signal and background event spectra in the $\nue$ appearance 
channel as a function of the reconstructed neutrino energy including the efficiency and 
background rejection capabilities. The left panel is for the DUNE (35 kt) and the right
one is for the LBNO (70 kt). In each panel, the thick lines correspond to the SM case, 
whereas the thin lines are drawn assuming $\alpha_{e\mu} =  10^{-52}$. In both the
panels, the solid grey (blue) vertical lines display the locations of the first (second)
oscillation maxima. We assume $\dcp$ = 0$^{\circ}$ and NH. For other oscillation
parameters, the values are taken from the third column of 
Table~\ref{tab:benchmark-parameters}.}
\label{fig:event-spectrum}
\end{figure}
%--------------------------------------------------------------------------------------------------------------------

In our study, we consider the $\nue$ and $\anue$ appearance channels, where the
backgrounds mainly stem from the the intrinsic $\nue$/$\anue$ contamination of the beam,
the number of muon events which will be misidentified as electron events, and the neutral 
current events. In Fig.~\ref{fig:event-spectrum}, we show the expected signal and background 
event spectra as a function of reconstructed neutrino energy including the efficiency and 
background rejection capabilities. The left panel shows the results for the DUNE set-up 
with 35 kt far detector mass. The right panel displays the same for the LBNO set-up 
with 70 kt far detector. In both the panels, the thick lines correspond to the SM case, 
whereas the thin lines are drawn assuming $\alpha_{e\mu} =  10^{-52}$. 
In both the set-ups, we can clearly see a systematic downward bias in the reconstructed energy 
for the neutral current background events due to the final state neutrino included via the migration 
matrices. The solid grey (blue) vertical lines display the locations of the first (second) oscillation 
maxima. The red solid histogram shows the signal event spectrum. Note that in the presence of long-range
potential, both the signal and background (intrinsic $\nue$ contamination and misidentified muons)
event spectra get modified substantially. For both the baselines, we have considerable number of 
signal events around the second oscillation maximum. But, these event samples are highly contaminated 
with the neutral current and other backgrounds at lower energies, limiting their impact.

%============================================================================
\begin{table}[H]
\begin{center}
{
\newcommand{\mc}[3]{\multicolumn{#1}{#2}{#3}}
\newcommand{\mr}[3]{\multirow{#1}{#2}{#3}}
\begin{adjustbox}{width=1\textwidth}
\begin{tabular}{|c||c|c||c|c|}
\hline
\mr{3}{*}{\bf Channel} & \mc{2}{c||}{\bf DUNE (35 kt)} &\mc{2}{c|}{\bf LBNO (70 kt) } \\
\cline{2-5}
      & Signal & Background &Signal & Background \\
\cline{2-5}
      & CC & Int+Mis-id+NC=Total &CC &Int+Mis-id+NC=Total  \\
\hline \hline
$P_{\mu e}$ (NH, SM) & 590 & 125+29+24=178 &1228 &115+31+29=175 \\
\hline
$P_{\mu e}$ (NH, SM+LRF) & 588 & 123+34+24=181 &786 &112+53+29=194 \\
\hline
$P_{\mu e}$ (IH, SM) & 268 & 129+29+24=182 &220 &126+31+29=186 \\
\hline
$P_{\mu e}$ (IH, SM+LRF) & 108 & 130+33+24=187 &49 &128+50+29=207 \\
\hline \hline
$P_{\bar\mu \bar e}$ (NH, SM) & 116 & 43+10+7=60 &117 &33+11+13=57 \\
\hline
$P_{\bar\mu \bar e}$ (NH, SM+LRF) & 44 & 44+12+7=63 & 22 &34+19+13=66 \\
\hline
$P_{\bar\mu \bar e}$ (IH, SM) & 210 & 42+10+7=59 &484 &30+11+13=54 \\
\hline
$P_{\bar\mu \bar e}$ (IH, SM+LRF) & 220 & 41+12+7=60 & 343 &29+19+13=61 \\
\hline
\end{tabular}
\end{adjustbox}
}
\caption{Comparison of the total signal and background event rates in the 
$\nue$/$\anue$ appearance channel for DUNE (35 kt) and LBNO (70 kt) set-ups.
Here `Int' means intrinsic beam contamination, `Mis-id' means misidentified muon 
events, and `NC' stands for neutral current. For the cases denoted by `SM+LRF',
we take $\alpha_{e\mu} =  10^{-52}$. The results are shown for both NH and IH 
assuming $\dcp$ = 0$^{\circ}$. For both the set-ups, we assume five years of 
neutrino run and five years of anti-neutrino run.}
\label{tab:appearance-signal-background-rates}
\end{center}
\end{table}
%=========================================================================

%=========================================================================
\begin{table}[H]
\begin{center}
{
\newcommand{\mc}[3]{\multicolumn{#1}{#2}{#3}}
\newcommand{\mr}[3]{\multirow{#1}{#2}{#3}}
\begin{tabular}{|c||c|c||c|c|}
\hline
\mr{2}{*}{\bf Channel} & \mc{2}{c||}{\bf DUNE (35 kt)} &\mc{2}{c|}{\bf LBNO (70 kt)} \\
\cline{2-5}
      & Signal & Background &Signal & Background \\
\cline{2-5}
      & CC & NC &CC &NC  \\
\hline \hline
$P_{\mu\mu}$ (NH, SM) & 4889 & 24 & 5222 &29 \\
\hline
$P_{\mu\mu}$ (NH, SM+LRF) & 5806 & 24 & 8949 &29 \\
\hline
$P_{\mu\mu}$ (IH, SM) & 4882 & 24 & 5203 &29 \\
\hline
$P_{\mu\mu}$ (IH, SM+LRF) & 5569 & 24 & 8519 &29 \\
\hline \hline
$P_{\bar\mu\bar\mu}$ (NH, SM) & 1751  & 7 & 1936 &13 \\
\hline
$P_{\bar\mu\bar\mu}$ (NH, SM+LRF) & 2012 & 7 & 3257 &13 \\
\hline
$P_{\bar\mu\bar\mu}$ (IH, SM) & 1752 & 7 &1923  &13 \\
\hline
$P_{\bar\mu\bar\mu}$ (IH, SM+LRF) & 2063 & 7& 3309 &13 \\
\hline
\end{tabular}
}
\caption{Comparison of the total signal and background event rates in the 
$\numu$/$\anumu$ disappearance channel for DUNE (35 kt) and LBNO (70 kt)
set-ups. For the cases denoted by `SM+LRF', we take $\alpha_{e\mu} =  10^{-52}$. 
The results are shown for both NH and IH assuming $\dcp$ = 0$^{\circ}$. 
For both the set-ups, we consider five years of neutrino run and five years 
of anti-neutrino run.}
\label{tab:disappearance-signal-background-rates}
\end{center}
\end{table}
%============================================================================

In Table~\ref{tab:appearance-signal-background-rates}, we show a comparison between the
total signal and background event rates in the $\nue$/$\anue$ appearance channel for 
DUNE (35 kt) and LBNO (70 kt) set-ups. For both the set-ups, we assume five years of 
neutrino run and five years of anti-neutrino run. For the cases denoted by `SM+LRF',
we take $\alpha_{e\mu} =  10^{-52}$. The results are shown for both NH and IH 
assuming $\dcp$ = 0$^{\circ}$. The Earth matter effects play an important role for both the 
baselines which is evident from the fact that in the neutrino channel, the number of expected
events is quite large compared to the IH case and in the anti-neutrino channel, the situation 
is totally opposite where we have larger event rates for IH than for NH.
The relative difference between the number of events for NH and IH is larger for the 
CERN-Pyh{\"a}salmi baseline than the FNAL-Homestake baseline, since the impact of
matter effects is more significant at the 2290 km baseline compared to the 1300 km baseline.  
In the presence of long-range potential with a benchmark choice of $\alpha_{e\mu} =  10^{-52}$, 
qualitatively, the trend remains the same as mentioned above. 
Table~\ref{tab:appearance-signal-background-rates} clearly shows that in all the cases, 
the most dominant contribution to the background comes from the intrinsic $\nue$/$\anue$ 
beam contamination. Note that though the total signal event rate for the DUNE set-up 
in the neutrino mode with NH, does not change much due to long-range potential with 
$\alpha_{e\mu} = 10^{-52}$, but, the shape of the signal event spectrum gets affected by
considerable amount as can be seen from Fig.~\ref{fig:event-spectrum}, which enables us
to place tight constraints on $\alpha_{e\mu}$ as we discuss in the results section.
In our simulation, we also include the information coming from the $\numu$/$\anumu$ 
disappearance channels. For these type of channels, neutral current events are the main
source of background. Table~\ref{tab:disappearance-signal-background-rates} shows
the total signal and background event rates in the $\numu$/$\anumu$ disappearance 
channels for both the set-ups, considering five years of neutrino run and five years 
of anti-neutrino run. For the cases marked by `SM+LRF', we take 
$\alpha_{e\mu} =  10^{-52}$ and we present results for both NH and IH 
assuming $\dcp$ = 0$^{\circ}$. Interestingly, the $\numu$/$\anumu$
disappearance channels are also quite sensitive to the long-range potential
and in all the cases, we see a significant change in the total signal event rates 
with $\alpha_{e\mu} =  10^{-52}$ as compared to the SM case.
Also, the rates are different for NH and IH in the presence of long-range potential.
The $\numu$/$\anumu$ disappearance channels also play an important
role to constrain the atmospheric oscillation parameters in the fit.
 
%===================
\subsection{Bi-events Plot}
\label{bi-events-plot}
%===================

%--------------------------------------------------------------------------------------------------------
\begin{figure}[t]
\centerline{
\includegraphics[width=0.49\textwidth]{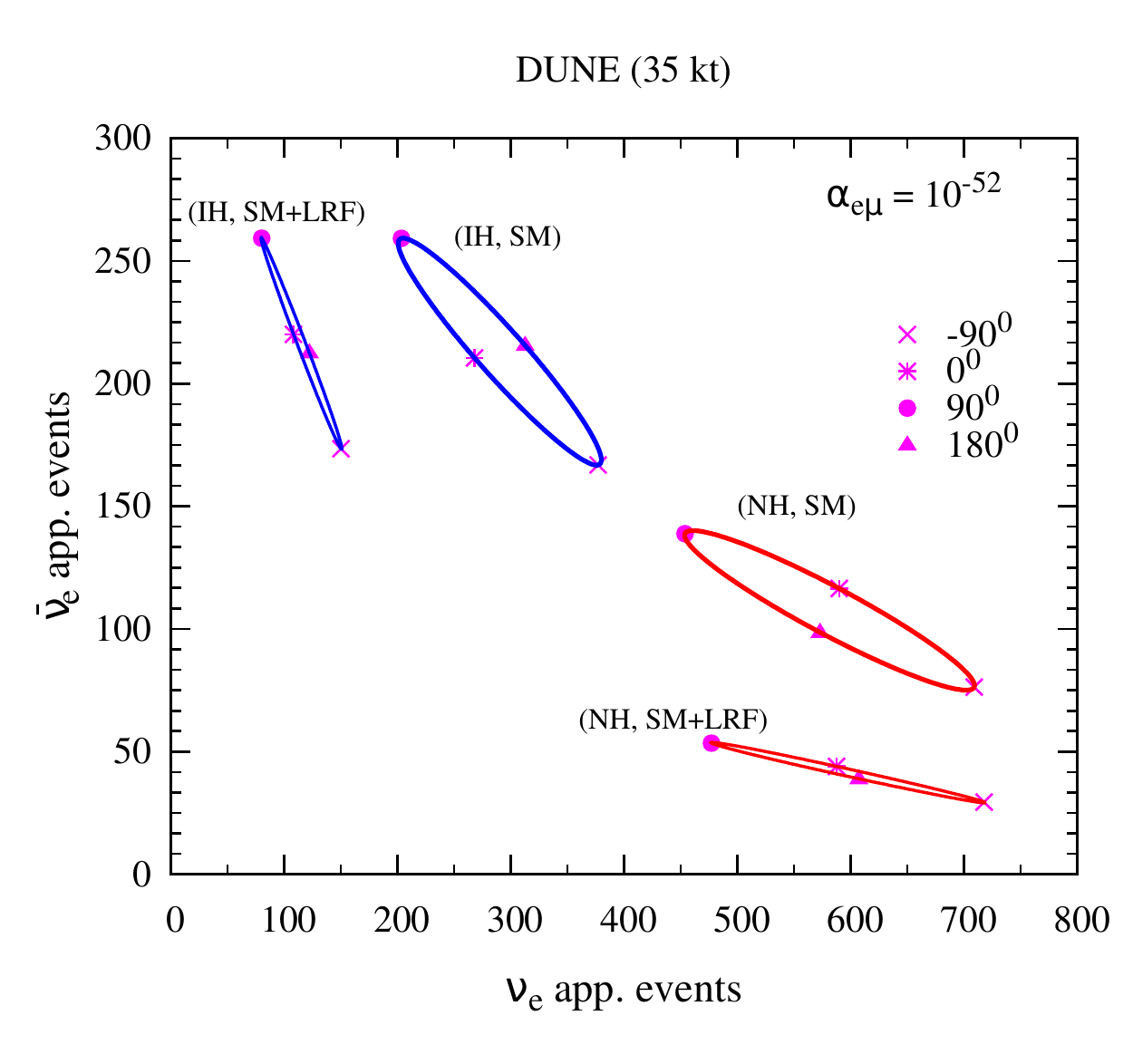}
\includegraphics[width=0.49\textwidth]{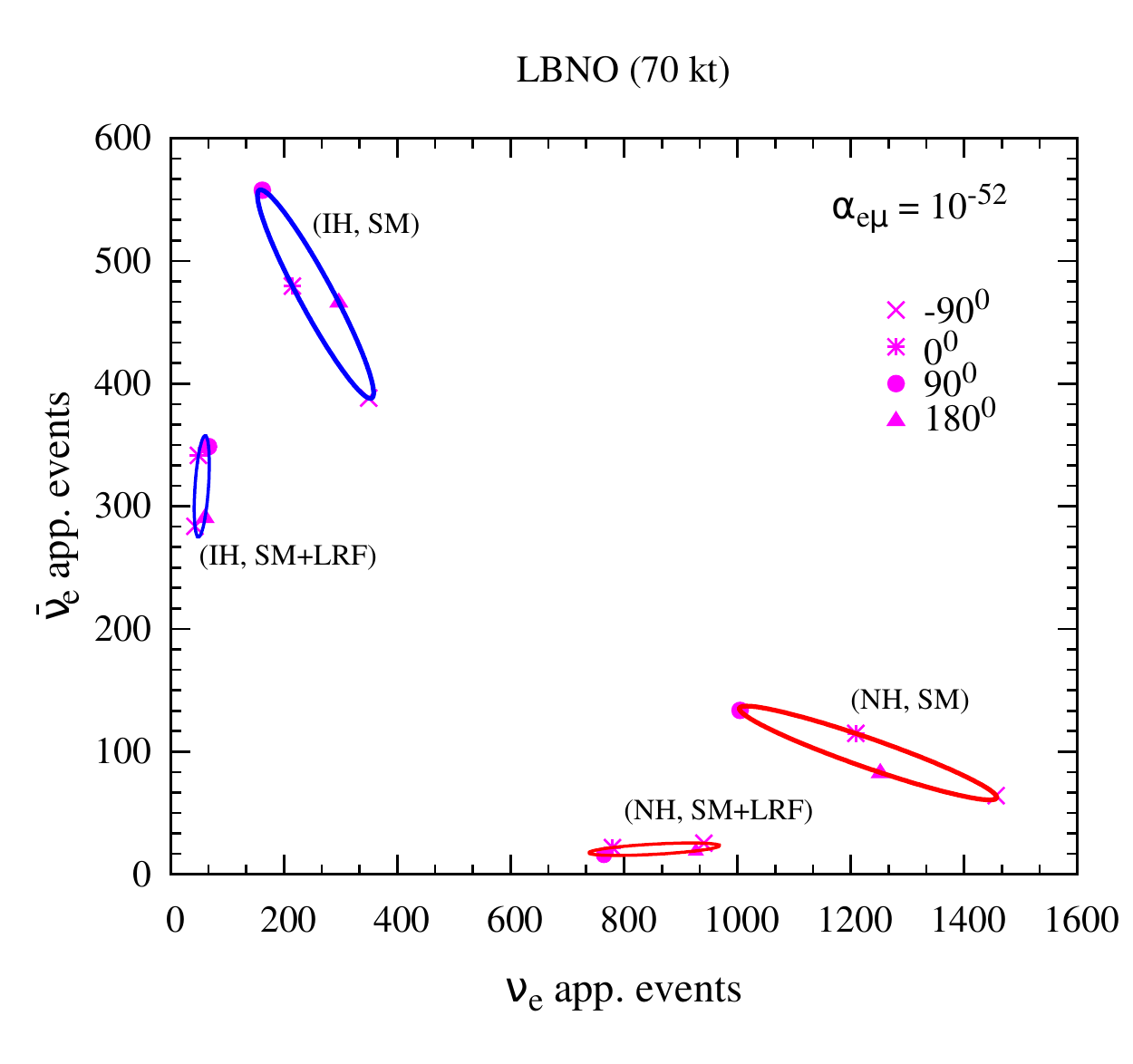}
}
\caption{Bi-events ($\nu_e$ and $\anu_e$ appearance) plots for NH and IH with
and without long-range potential. The ellipses are due to all possible $\dcp$ values. 
For the cases labelled by `SM+LRF', we take $\alpha_{e\mu} =  10^{-52}$.
The left panel is for the DUNE set-up (35 kt) and the right panel is for the LBNO 
set-up (70 kt).} 
\label{fig:bi-events-plot}
\end{figure}
%---------------------------------------------------------------------------------------------------------

In this section, we make an attempt to unravel the impact of long-range potential 
with the help of bi-events plot. In Fig.~\ref{fig:bi-events-plot}, we have plotted
$\nue$ vs. $\bar{\nu}_e$ appearance events, for DUNE (left panel) and 
LBNO (right panel), considering both NH and IH and with and without long-range 
potential. Since $\dcp$ is not known, events are generated for the full range
$[-180^\circ, 180^\circ]$, leading to the ellipses. For the cases labelled by 
`SM+LRF', we take $\alpha_{e\mu} =  10^{-52}$. We generate these plots with
$\sin^2\theta_{23}$ = 0.5 as mentioned in Table~\ref{tab:benchmark-parameters}.
The ellipses in Fig.~\ref{fig:bi-events-plot} suggest that both the set-ups can 
discriminate between NH and IH at high confidence level, irrespective of the choice 
of $\dcp$, and the presence of long-range potential with $\alpha_{e\mu} = 10^{-52}$
does not spoil this picture. We can see from the left panel that for the DUNE set-up,
the anti-neutrino (neutrino) event rates get reduced for NH (IH) with LRF as compared 
to the SM case. But, for CERN-Pyh{\"a}salmi baseline (see right panel) with more 
matter effect, both the neutrino and anti-neutrino event rates get diminished for NH
and IH in the presence of long-range potential as compared to the SM case.
Now, let us make an attempt to understand this behavior. In the presence of long-range 
potential, the locations of the first oscillation maxima shift towards lower energies 
(see the upper panels of Fig.~\ref{fig:neutrino-appearance-probability} and 
Fig.~\ref{fig:anti-neutrino-appearance-probability}), where both the fluxes and the
interaction cross-sections are small. On the contrary, at higher energies, we see 
a suppression in the probability where we have most of the neutrino fluxes and the 
cross-sections are also high at these energies. These opposite bahaviors
are responsible for the large depletions in the event rates. Fig.~\ref{fig:bi-events-plot} 
also portrays that the asymmetries between the neutrino and anti-neutrino appearance 
events are largest for the combinations: (NH, $\dcp = -90^\circ$) and 
(IH, $\dcp = 90^\circ$). One striking feature emerging from both the panels is that 
all the ellipses get shrunk in the presence of long-range potential, reducing the 
differences in the number of events due to the CP-conserving and CP-violating 
phases. It ultimately affects the CP-coverage for the leptonic 
CP-violation searches, where we study the choices of the CP phase, $\dcp$ 
which can be distinguished from both $0^\circ$ and $180^\circ$ at a given 
confidence level. The right panel shows that this effect is more prominent for 
the LBNO set-up, severely limiting its discovery reach for CP-violation which 
we discuss in detail in the results section.

%====================
\section{Simulation Method}
\label{simulation-method}
%====================

In this section, we give a brief description of the numerical technique 
and analysis procedure which we adopt to estimate the physics reach
of the experimental set-ups. We have made suitable changes in the 
GLoBES software \cite{Huber:2004ka,Huber:2007ji} to obtain our results. 
The entire numerical analysis is performed using the full three-flavor oscillation 
probabilities. Unless stated otherwise, we generate our simulated 
data considering the true values of the oscillation parameters given in the 
third column of Table~\ref{tab:benchmark-parameters}.
These choices of the oscillation parameters are well within their 
1$\sigma$ allowed ranges which are obtained in recent global fit 
analysis \cite{Gonzalez-Garcia:2014bfa}. In the fit, we marginalize over 
test $\sa$ and test $\dcp$ in their allowed ranges which are given in the 
fourth column of Table~\ref{tab:benchmark-parameters}, without assuming 
any prior on these parameters. We also marginalize over both the hierarchy
choices in the fit for all the analyses, except for the mass hierarchy 
discovery studies where our goal is to exclude the wrong hierarchy 
in the fit. We keep $\tet$ fixed in the fit as the Daya Bay 
experiment is expected to achieve a relative 1$\sigma$ precision of 
$\sim$ 3\% by the end of 2017 \cite{Zhan:2015aha},
and needless to say that the global oscillation data will severely 
constrain $\tet$ beyond the Daya Bay limit before these future 
experiments will come online. For the atmospheric mass-squared splitting,
we take the true value of $\Delta m_{\mu\mu}^2$ = $\pm$ $2.4 \times 10^{-3}$ 
eV$^2$ where positive (negative) sign is for NH (IH), and we do not marginalize 
over this parameter in the fit since the projected combined data from the currently 
running T2K and NO$\nu$A experiments will be able to improve the precision
in $|\Delta m^2_{\mu\mu}|$ to sub-percent level for maximal $\tmt$ 
\cite{Agarwalla:2013qfa}. On top of the standard three-flavor oscillation parameters,
we have also the LRF parameter $\alpha_{e\mu}$ which enters into the oscillation
probability. As far as the true and test values of $\alpha_{e\mu}$ are concerned, 
we vary our choices in the range $10^{-54}$ to $10^{-51}$, where the 
lower limit\footnote{If the range of the long-range force is equal or larger
than our distance from the galactic center, then the collective long-range 
potential due to all the electrons in the galaxy becomes significant.
In such cases, these experimental set-ups can be sensitive to even 
lower values of $\alpha_{e\mu}$ \cite{PhysRevD.75.093005}.}
corresponds to the cases where the oscillation probabilities {\it almost} 
overlap with the SM cases for the set-ups that we consider in this work. 
We take the upper limit of $\alpha_{e\mu}$ as $10^{-51}$ which covers 
all the existing bounds on this parameter available from the oscillation 
experiments as discussed in section~\ref{current-bounds}. 
Based on the techniques discussed in Refs. \cite{Huber:2002mx,Fogli:2002pt},
we use the following $\chi^2$ function in our statistical analysis:
\begin{equation}
\chi^2= min_{\xi_s, \xi_b}\left[2\sum^{n}_{i=1}
(\tilde{y}_{i}-x_{i} - x_{i} \ln \frac{\tilde{y}_{i}}{x_{i}}) +
\xi_s^2 + \xi_b^2\right ] \,,
\label{eq:chipull}
\end{equation}
where $n$ is the total number of bins and
\begin{equation}
\tilde{y}_{i}(\{\omega, \alpha_{e\mu}\},\{\xi_s, \xi_b\}) = N^{th}_i(\{\omega, \alpha_{e\mu}\})
[1+ \pi^s \xi_s] + N^{b}_i(\{\omega, \alpha_{e\mu}\})[1+ \pi^b \xi_b] \,.
\label{eq:theory-errors}
\end{equation}
Above, $N^{th}_i(\{\omega, \alpha_{e\mu}\})$ is the predicted number 
of signal events in the $i$-th energy bin for a set of oscillation parameters 
$\omega$ and a particular value of $\alpha_{e\mu}$. 
$N_i^b(\{\omega, \alpha_{e\mu}\})$ is the number of background events 
in the $i$-th bin where the charged current backgrounds are dependent on 
$\omega$ and $\alpha_{e\mu}$, and the neutral current backgrounds do not 
depend on the oscillation parameters and $\alpha_{e\mu}$.
The quantities $\pi^s$ and $\pi^b$ in Eq.~(\ref{eq:theory-errors}) are the 
systematic errors on the signal and background respectively. 
For both the set-ups, we consider $\pi^s$ = 5\% and $\pi^b$ = 5\% in the 
form of normalization error for both the appearance and disappearance 
channels. The quantities $\xi_s$ and $\xi_b$ are the ``pulls'' due to the 
systematic error on signal and background respectively. We incorporate
the data in Eq.~(\ref{eq:chipull}) through the variable $x_i=N_i^{ex}+N_i^b$,
where $N_i^{ex}$ is the number of observed charged current signal events 
in the $i$-th energy bin and $N_i^b$ is the background as mentioned earlier. 
To estimate the total $\chi^2$, we add the $\chi^2$ contributions coming from 
all the relevant channels in a given experiment in the following way
\begin{equation}
\chi^2_{\rm total} = \chi^2_{\numu \rightarrow \nue} + \chi^2_{\numu \rightarrow \numu}
+ \chi^2_{\anumu \rightarrow \anue} + \chi^2_{\anumu \rightarrow \anumu} \,,
\label{eq:total-chisq}
\end{equation}
where we assume that all these channels are completely uncorrelated,
all the energy bins in a given channel are fully correlated, and the systematic 
errors on signal and background are fully uncorrelated. Finally, 
$\chi^2_{\textrm{total}}$ is marginalized in the fit over the oscillation parameters, 
both the hierarchy choices, the LRF parameter $\alpha_{e\mu}$ (as needed), 
and the systematic parameters as mentioned above to obtain 
$\Delta\chi^2_{\textrm{min}}$.

%===========
\section{Results}
\label{results}
%===========

In this section, we report our main findings. First, we present the expected constraints 
on $\alpha_{e\mu}$ from the proposed DUNE and LBNO experiments. Next, we quantify
the discovery reach for $\alpha_{e\mu}$ of these future facilities. Then, we address how 
the flavor-dependent LRF, mediated by the extremely light $L_e-L_{\mu}$ gauge boson
can affect the CP-violation searches and the mass hierarchy measurements at these 
upcoming facilities.

%===============================================================
\subsection{Expected Constraints on the Effective Gauge Coupling $\alpha_{e\mu}$}
\label{expected-constraints}
%================================================================

%--------------------------------------------------------------------------------------------
\begin{figure}[t]
\centerline{
\includegraphics[width=0.49\textwidth]{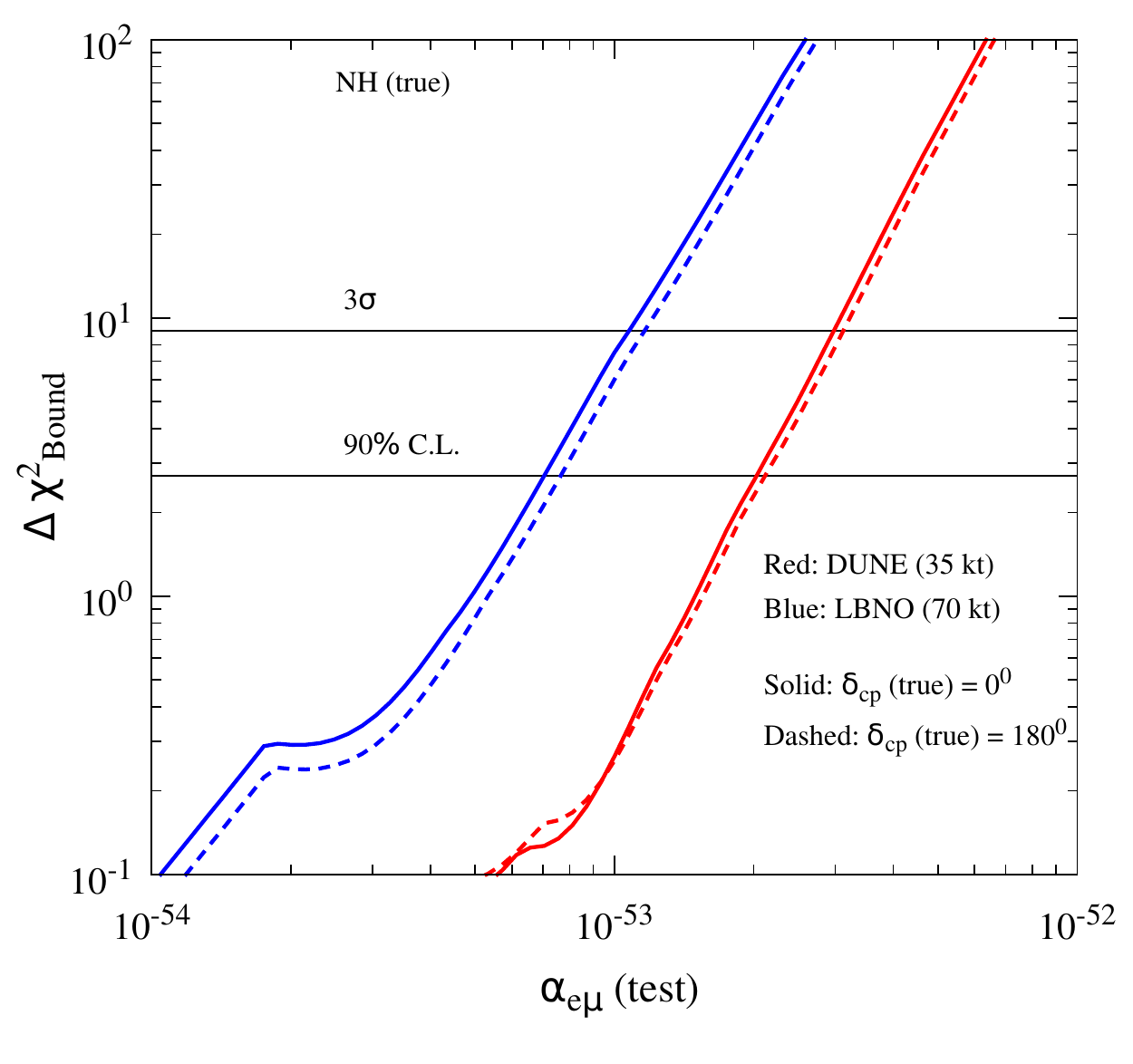}
\includegraphics[width=0.49\textwidth]{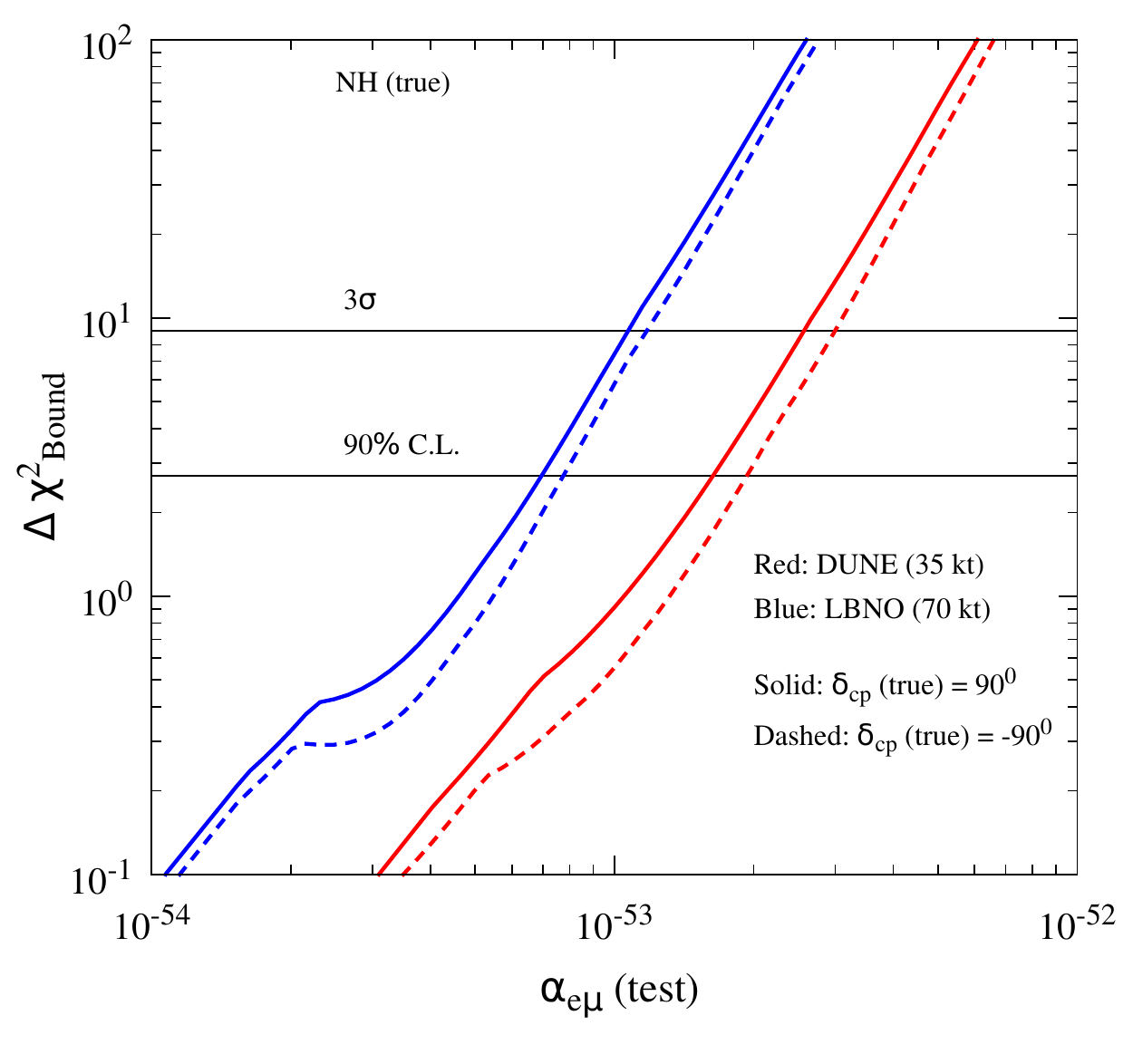}
}
\caption{Expected bounds on $\alpha_{e\mu}$ from the DUNE (35 kt) and LBNO (70 kt)
experiments in the scenarios when the data show no signal of LRF. Results are given for
four different choices of true values of $\dcp$. The left panel is for CP-conserving 
choices: $\dcp$(true) = $0^\circ$ (solid lines), $180^\circ$ (dashed lines).
The right panel is for maximal CP-violating choices: $\dcp$(true) = $90^\circ$ (solid lines), 
$-90^\circ$ (dashed lines). In all the cases, we assume NH as true hierarchy.}
\label{fig:expected-constraints}
\end{figure}
%-------------------------------------------------------------------------------------------

%=========================================================================
\begin{table}[t]
\begin{center}
{
\newcommand{\mc}[3]{\multicolumn{#1}{#2}{#3}}
\newcommand{\mr}[3]{\multirow{#1}{#2}{#3}}
\begin{adjustbox}{width=1\textwidth}
\begin{tabular}{||c|c|c|c||c|c|c|c||}
\hline
Expt. & $\dcp$(true) & 90\% C.L. & 3$\sigma$& Expt. & $\dcp$(true) & 90\% C.L. & 3$\sigma$ \\
\hline
  & \mr{2}{*}{$0^0$} & $2.0\times 10^{-53}$(NH) &  $3.0\times 10^{-53}$(NH) &   & \mr{2}{*}{$0^0$} & $7.0\times 10^{-54}$(NH) &  $1.1\times 10^{-53}$(NH) \\
                                 &    &$2.1\times 10^{-53}$(IH) & $3.1\times 10^{-53}$(NH) & & & $7.8\times 10^{-54}$(IH) & $1.2\times 10^{-53}$(IH) \\ 
\cline{2-4}
\cline{6-8}

  {\bf DUNE}  & \mr{2}{*}{$180^0$} & $2.1\times 10^{-53}$(NH)& $3.1\times 10^{-53}$(NH) & {\bf LBNO} &\mr{2}{*}{$180^0$} &$7.6\times 10^{-54}$(NH) &$1.2\times 10^{-53}$(NH) \\
  {\bf (35 kt)}    &  & $2.0\times 10^{-53}$(IH) & $3.0\times 10^{-53}$(IH) & {\bf (70 kt)} & & $7.0\times 10^{-54}$(IH) & $1.1\times 10^{-53}$(IH) \\
\cline{2-4}
\cline{6-8}    
	    
     & \mr{2}{*}{$90^0$} & $1.7\times 10^{-53}$(NH)& $2.7\times 10^{-53}$(NH) & &\mr{2}{*}{$90^0$} & $7.0\times 10^{-54}$(NH)& $1.1\times 10^{-53}$(NH) \\
	    &  & $1.8\times 10^{-53}$(IH) & $2.8\times 10^{-53}$(IH) & & & $7.0\times 10^{-54}$(IH) & $1.1\times 10^{-53}$(IH) \\	    

\cline{2-4} 
\cline{6-8}
  &\mr{2}{*}{$-90^0$} & $1.9\times 10^{-53}$(NH)& $3.0\times 10^{-53}$(NH) &  & \mr{2}{*}{$-90^0$} & $7.8\times 10^{-54}$(NH) &  $1.2\times 10^{-53}$(NH) \\
           &  &$1.5\times 10^{-53}$(IH) & $2.4\times 10^{-53}$(IH) &  & &$6.0\times 10^{-54}$(IH) & $9.6\times 10^{-54}$(IH) \\
\cline{2-4} 
\cline{6-8}
\hline
\end{tabular}
\end{adjustbox}
}
\caption{The expected bounds on $\alpha_{e\mu}$ from the DUNE (35 kt) and LBNO (70 kt)
experiments if there is no signal of LRF in the data. The results are presented at 90\% and 
3$\sigma$ confidence levels for four different choices of true values of $\dcp$: $0^\circ$, 
$180^\circ$, $90^\circ$, and $-90^\circ$. For each $\dcp$(true) value, we show the results
for both the choices of true hierarchy: NH and IH.}
\label{tab:bounds-on-alpha-emu}
\end{center}
\end{table}
%=========================================================================

\begin{figure}[t]
\centering
\includegraphics[width=0.75\textwidth]{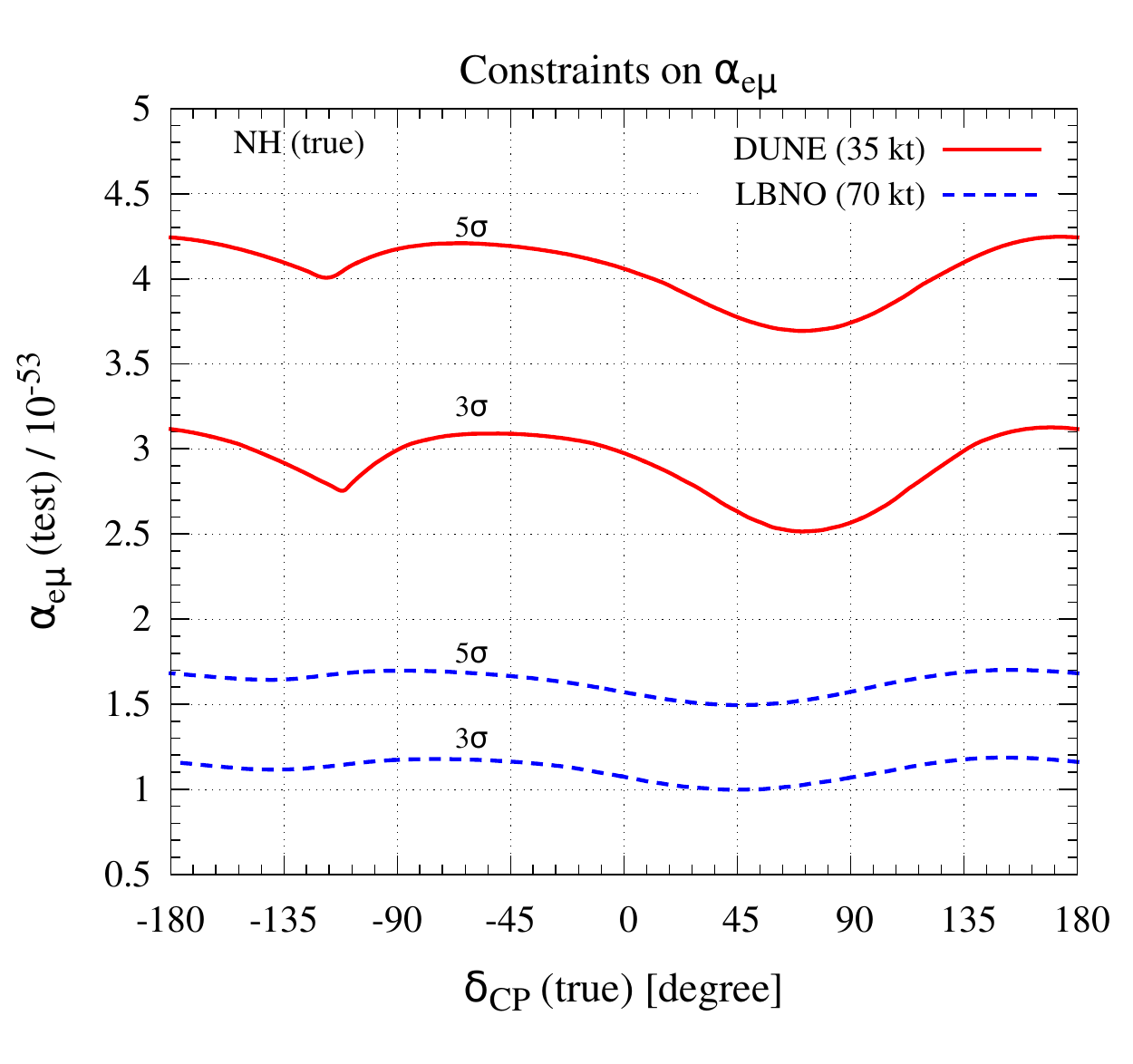}
\caption{Constraints on $\alpha_{e\mu}$ as a function of true value of $\dcp$ 
assuming NH as true hierarchy. Results are shown for DUNE (35 kt) and 
LBNO (70 kt) at 3$\sigma$ and 5$\sigma$ confidence levels.}
\label{fig:alpha-emu-bounds-with-CP-true}
\end{figure}

In this section, we estimate the upper bounds on $\alpha_{e\mu}$ from the proposed 
DUNE and LBNO experiments if there is no signal of LRF in the data.
This performance indicator corresponds to the new upper limit on $\alpha_{e\mu}$
if the experiment does not see a signal of LRF in oscillations. We simulate this situation 
in our analysis by generating the data at $\alpha_{e\mu}$(true) = 0 and fitting it with 
some non-zero value of $\alpha_{e\mu}$ by means of the $\chi^2$ technique as outlined
in section~\ref{simulation-method}. The corresponding $\Delta\chi^2_{\textrm{Bound}}$
obtained after marginalizing over oscillation parameters ($\tmt$, $\dcp$, and mass hierarchy)
and systematic parameters in the fit, is plotted in Fig.~\ref{fig:expected-constraints} as a 
function of $\alpha_{e\mu}$ (test), which gives a measure of the sensitivity reach of 
the DUNE or LBNO set-up to the effective gauge coupling of the LRF. New limits 
are given for four different choices of true values of $\dcp$. The left panel is for 
CP-conserving choices: $\dcp$(true) = $0^\circ$ (solid lines), $180^\circ$ (dashed lines).
The right panel is for maximal CP-violating choices: $\dcp$(true) = $90^\circ$ (solid lines), 
$-90^\circ$ (dashed lines). In all the cases, we assume NH as true hierarchy.
Fig.~\ref{fig:expected-constraints} clearly shows that the LBNO set-up with 
70 kt detector mass can place better limits on $\alpha_{e\mu}$ as compared to the
DUNE set-up with 35 kt detector and the limits are not very sensitive to the choice
of unknown $\dcp$(true) for both the set-ups. Table~\ref{tab:bounds-on-alpha-emu}
lists the precise upper limits on $\alpha_{e\mu}$ which are expected from these 
future facilities if there is no trace of LRF in the data. We present the bounds at
90\% (1.64$\sigma$) and 3$\sigma$ confidence levels\footnote{To calculate this, 
we use the relation $\textrm{n}\sigma = \sqrt{\Delta\chi^2_{\textrm{min}}}$.
In \cite{Blennow:2013oma}, it was shown that the above relation is valid in the 
frequentist method of hypothesis testing.} for four different choices 
of true values of $\dcp$: $0^\circ$, $180^\circ$, $90^\circ$, and $-90^\circ$. 
For each $\dcp$(true) value, we give the results for both NH and IH as true
hierarchy choice. For an example, if $\dcp$(true) = $-90^\circ$ and true hierarchy
is NH, then the 90\% C.L. limit from the DUNE (LBNO) experiment is
$\alpha_{e\mu} < 1.9 \times 10^{-53}~(7.8 \times 10^{-54})$, suggesting that the 
constraint from the LBNO experiment is $\sim$ 2.4 times better than the DUNE
set-up\footnote{We have checked that the larger detector mass (two times) 
in the LBNO set-up compared to the DUNE set-up is partially responsible for 
this improvement in the sensitivity, but also the larger path length with more matter 
effect plays an important role in this direction.}. This future limit from the DUNE 
(LBNO) experiment is $\sim$ 30 (70) times better than the existing 
limit\footnote{This limit is quite old and was derived
in a two-flavor scheme assuming $\tet$ = $0^\circ$ \cite{Joshipura:2003jh}.
One needs to revise this limit using the presently available full data set from 
the SK experiment in light of the non-zero and large $\tet$.} from the SK 
experiment \cite{Joshipura:2003jh} which is also mainly sensitive to the 
atmospheric mass scale like long-baseline experiments.
At 3$\sigma$, we see the same relative improvement in the LBNO experiment 
in constraining $\alpha_{e\mu}$ compared to the DUNE set-up.
Table~\ref{tab:bounds-on-alpha-emu} also suggests that the limits are
not highly dependent on the true choices of mass hierarchy. The variation
in the upper limits on $\alpha_{e\mu}$ is also not significant as we vary
$\dcp$(true) in the range -$180^\circ$ to $180^\circ$ as can be seen from
Fig.~\ref{fig:alpha-emu-bounds-with-CP-true}, where we give the expected 
bounds at 3$\sigma$ and 5$\sigma$ confidence levels from both the 
set-ups assuming NH as true hierarchy. Next, we discuss the discovery
reach for $\alpha_{e\mu}$ if we find a positive signal of LRF in the expected 
event spectra at DUNE and LBNO.

%--------------------------------------------------------------------------------------------
%--------------------------------------------------------------------------------------------

%=====================================
\subsection{Discovery Reach for $\alpha_{e\mu}$}
\label{expected-discovery}
%=====================================

\begin{figure}[t]
\centerline{
\includegraphics[width=0.49\textwidth]{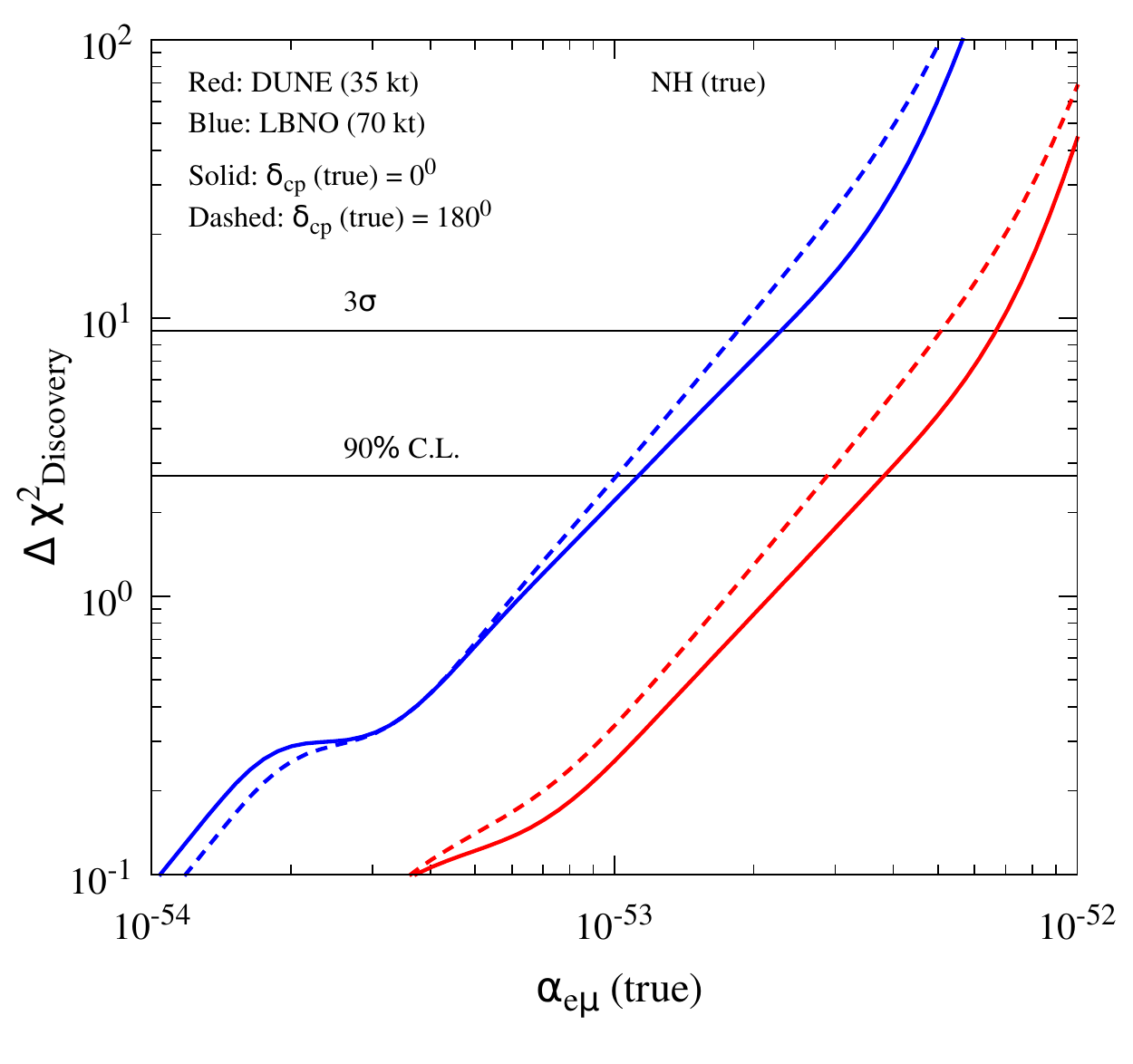}
\includegraphics[width=0.49\textwidth]{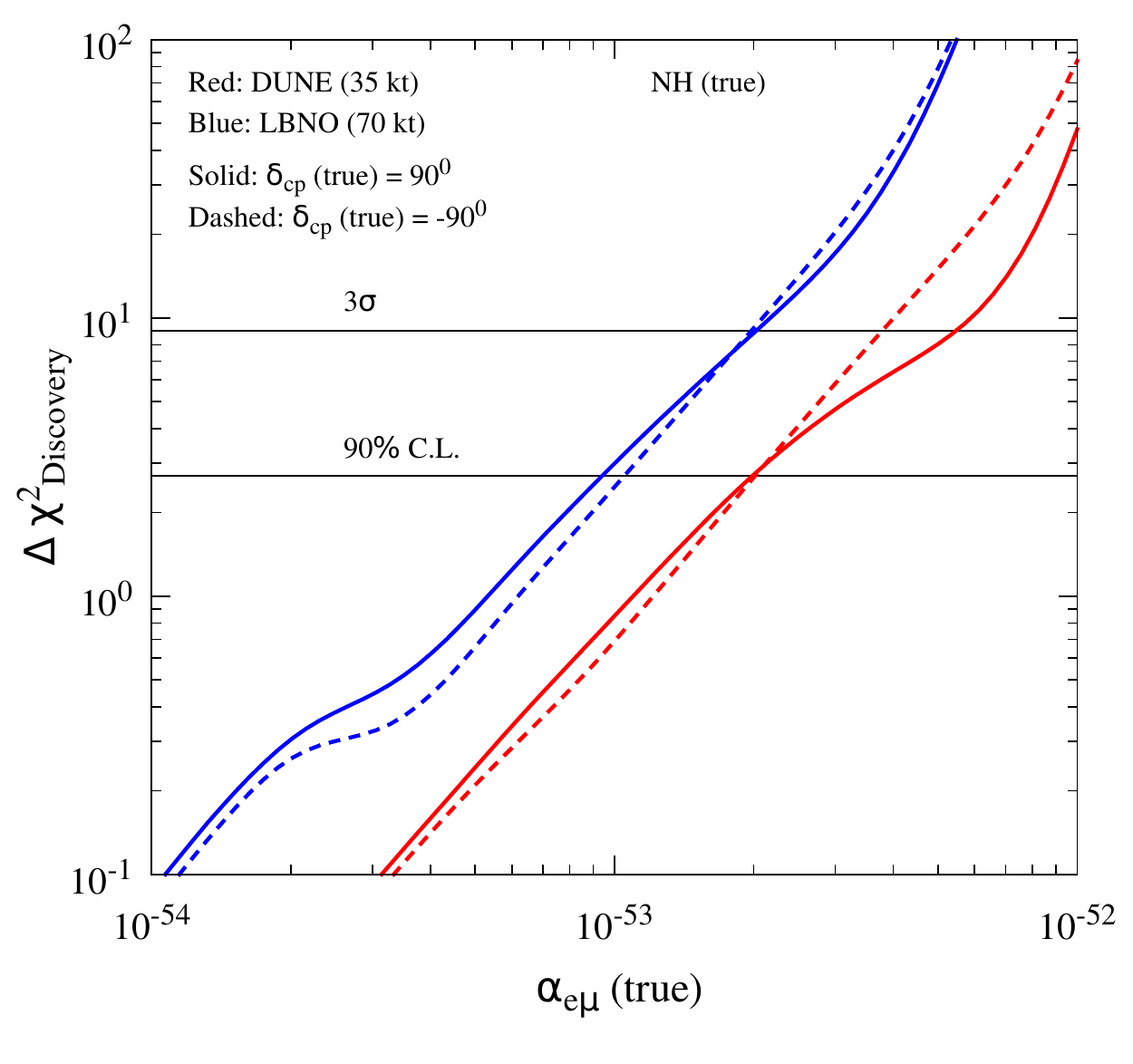}
}
\caption{$\Delta\chi^2_{\textrm{Discovery}}$ showing the discovery reach for $\alpha_{e\mu}$ 
expected from the DUNE (35 kt) and LBNO (70 kt) set-ups in the cases when one finds 
a signal of LRF in the data. Results are given for four different choices of true values 
of $\dcp$. The left panel is for CP-conserving choices: $\dcp$(true) = $0^\circ$ (solid lines), 
$180^\circ$ (dashed lines). The right panel is for maximal CP-violating choices: 
$\dcp$(true) = $90^\circ$ (solid lines), $-90^\circ$ (dashed lines). 
In all the cases, we assume NH as true hierarchy.}
\label{fig:alpha-emu-discovery-reaches}
\end{figure}

How good are our chances of observing a positive signal for LRF and 
hence $\alpha_{e\mu}$ in these proposed facilities? We answer this 
question in terms of the parameter indicator which we call the 
``discovery reach'' of the experiment for $\alpha_{e\mu}$. We define 
this performance indicator as the expected lower limit on true values 
of $\alpha_{e\mu}$ above which the projected data at DUNE or LBNO 
would give us a signal for LRF at a certain confidence level. To find 
these limiting values, we simulate the data for various true values 
of $\alpha_{e\mu}$ and fit it with a predicted event spectrum corresponding 
to $\alpha_{e\mu}$ = 0. We marginalize over $\tmt$, $\dcp$, mass hierarchy,
and systematic parameters in the fit to estimate the resultant 
$\Delta\chi^2_{\textrm{Discovery}}$ which is plotted in 
Fig.~\ref{fig:alpha-emu-discovery-reaches} for DUNE (35 kt) and LBNO (70 kt)
set-ups, considering four different choices of true values of $\dcp$. 
In the left panel, we take the CP-conserving choices: $\dcp$(true) = $0^\circ$ 
(solid lines), $180^\circ$ (dashed lines). In the right panel, we consider the 
maximal CP-violating choices: $\dcp$(true) = $90^\circ$ (solid lines), $-90^\circ$ 
(dashed lines). In all the cases, we take NH as true hierarchy. The nature of the
curves in Fig.~\ref{fig:alpha-emu-discovery-reaches} are quite similar to the curves
which are shown in Fig.~\ref{fig:expected-constraints}, and LBNO with 70 kt
detector has better discovery reach for $\alpha_{e\mu}$ as compared to DUNE
with 35 kt, like in the case of constraints on $\alpha_{e\mu}$.
In Table~\ref{tab:discoveries-on-alpha-emu}, we give the precise lower limits on
true values of $\alpha_{e\mu}$ which can be separated from $\alpha_{e\mu}$ = 0
in the fit at 90\% and 3$\sigma$ confidence levels. The results are given for
both the set-ups and for four different choices of true values of $\dcp$: $0^\circ$, 
$180^\circ$, $90^\circ$, and $-90^\circ$. For each $\dcp$(true) value, we show 
the results for both the choices of true hierarchy: NH and IH. If we compare the
entries in Table~\ref{tab:discoveries-on-alpha-emu} and 
Table~\ref{tab:bounds-on-alpha-emu}, then we can see that the values of the 
discovery reach for $\alpha_{e\mu}$ are slightly different than the constraints 
on $\alpha_{e\mu}$ at a given confidence level and for the same choices of 
true oscillation parameters. Also, we can see that the values of discovery reach
are marginally dependent on the choices of true $\dcp$ and mass hierarchy 
as we have seen for the constraints in the previous section.

%--------------------------------------------------------------------------------------------
%-------------------------------------------------------------------------------------------

%=========================================================================
\begin{table}[t]
\begin{center}
{
\newcommand{\mc}[3]{\multicolumn{#1}{#2}{#3}}
\newcommand{\mr}[3]{\multirow{#1}{#2}{#3}}
\begin{adjustbox}{width=1\textwidth}
\begin{tabular}{||c|c|c|c||c|c|c|c||}
\hline
Expt. & $\rm{\delta_{CP}(true)}$ & 90\% C.L. & 3$\sigma$& Expt. & $\rm{\delta_{CP}(true)}$ & 90\% C.L. & 3$\sigma$ \\
\hline
  & \mr{2}{*}{$0^0$} & $3.8\times 10^{-53}$(NH) &  $6.5\times 10^{-53}$(NH) &   & \mr{2}{*}{$0^0$} & $1.2\times 10^{-53}$(NH) &  $2.2\times 10^{-53}$(NH) \\
                                 &    &$3.7\times 10^{-53}$(IH) & $7.2\times 10^{-53}$(IH) & & & $1.4\times 10^{-53}$(IH) & $2.6\times 10^{-53}$(IH) \\ 
\cline{2-4}
\cline{6-8}
{\bf DUNE} & \mr{2}{*}{$180^0$} & $2.9\times 10^{-53}$(NH) & $5.0\times 10^{-53}$(NH) & {\bf LBNO} &\mr{2}{*}{$180^0$} &$9.5\times 10^{-54}$(NH) &$1.9\times 10^{-53}$(NH)  \\
	{\bf (35 kt)}    &  & $4.0\times 10^{-53}$(IH) & $7.8\times 10^{-53}$(IH) & {\bf (70 kt)} & & $1.4\times 10^{-53}$(IH) & $2.6\times 10^{-53}$(IH) \\	 
\cline{2-4} 
\cline{6-8}
     & \mr{2}{*}{$90^0$} & $2.0\times 10^{-53}$(NH)& $5.4\times 10^{-53}$(NH) & &\mr{2}{*}{$90^0$} &$9.0\times 10^{-54}$(NH) & $2.0\times 10^{-53}$(NH)  \\
	    &  & $2.3\times 10^{-53}$(IH) &$4.4\times 10^{-53}$(IH)  & & & $1.0\times 10^{-53}$(IH) & $2.0\times 10^{-53}$(IH)\\	    
\cline{2-4} 
\cline{6-8}
 &\mr{2}{*}{$-90^0$} & $2.0\times 10^{-53}$(NH) & $3.7\times 10^{-53}$(NH) &  & \mr{2}{*}{$-90^0$} & $1.0\times 10^{-53}$(NH) &  $2.0\times 10^{-53}$(NH) \\
           &  & $1.8\times 10^{-53}$(IH) & $4.8\times 10^{-53}$(IH) & & &$7.6\times 10^{-54}$(IH) &$1.5\times 10^{-53}$(IH) \\   
\cline{2-4} 
\cline{6-8}

\hline
\end{tabular}
\end{adjustbox}
}
\caption{The discovery reach for $\alpha_{e\mu}$ as expected from the DUNE (35 kt) 
and LBNO (70 kt) experiments if the data show a signal of LRF. The results are 
presented at 90\% and 3$\sigma$ confidence levels for four different choices of 
true values of $\dcp$: $0^\circ$, $180^\circ$, $90^\circ$, and $-90^\circ$. 
For each $\dcp$(true) value, we show the results for both the choices of 
true hierarchy: NH and IH.}
\label{tab:discoveries-on-alpha-emu}
\end{center}
\end{table}
%=========================================================================

%======================================================
\subsection{How Robust are CP-violation Searches in Presence of LRF?}
\label{CPV-Impact}
%======================================================

%--------------------------------------------------------------------------------------------
\begin{figure}[t]
\centerline{
\includegraphics[width=0.49\textwidth]{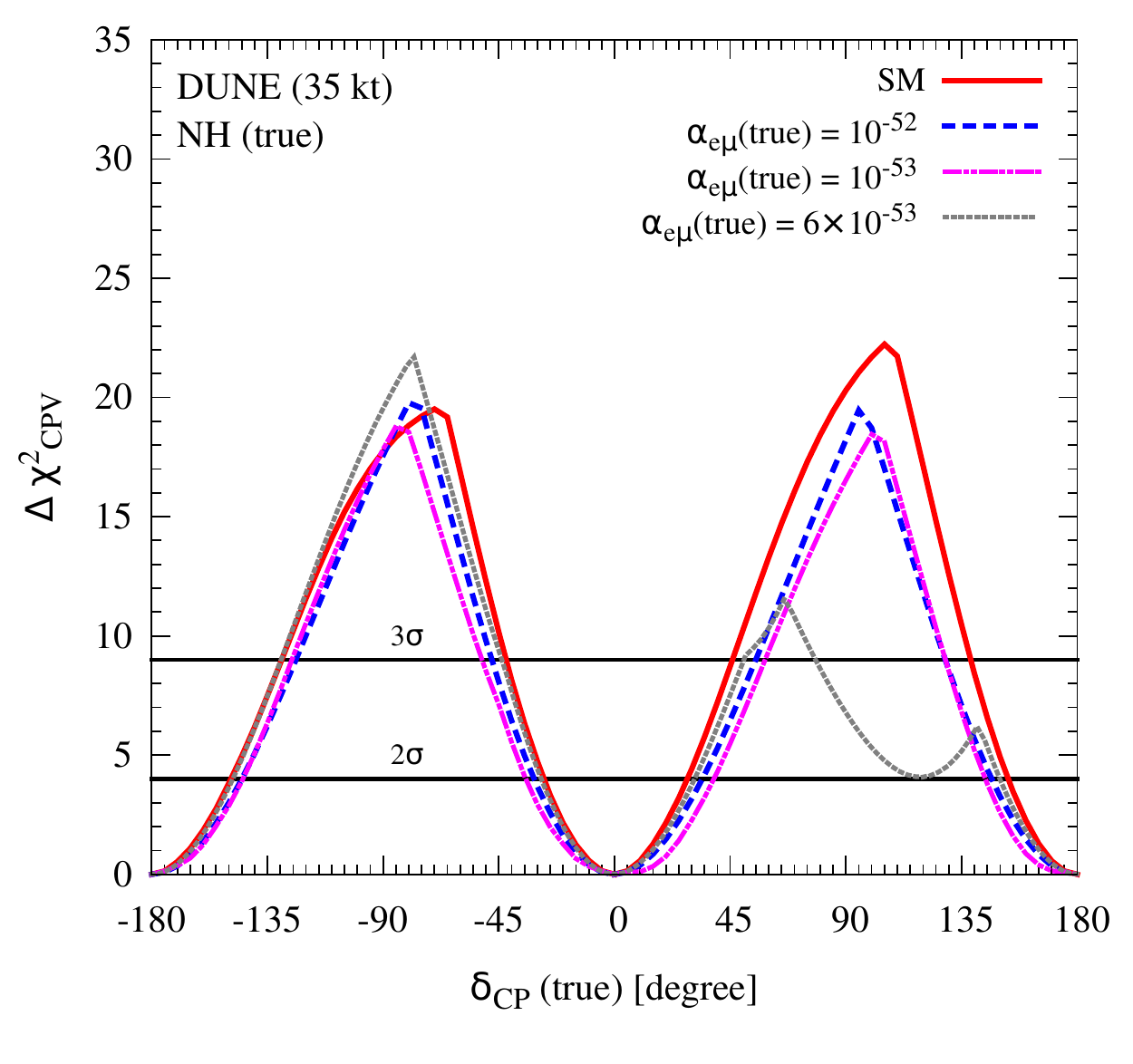}
\includegraphics[width=0.49\textwidth]{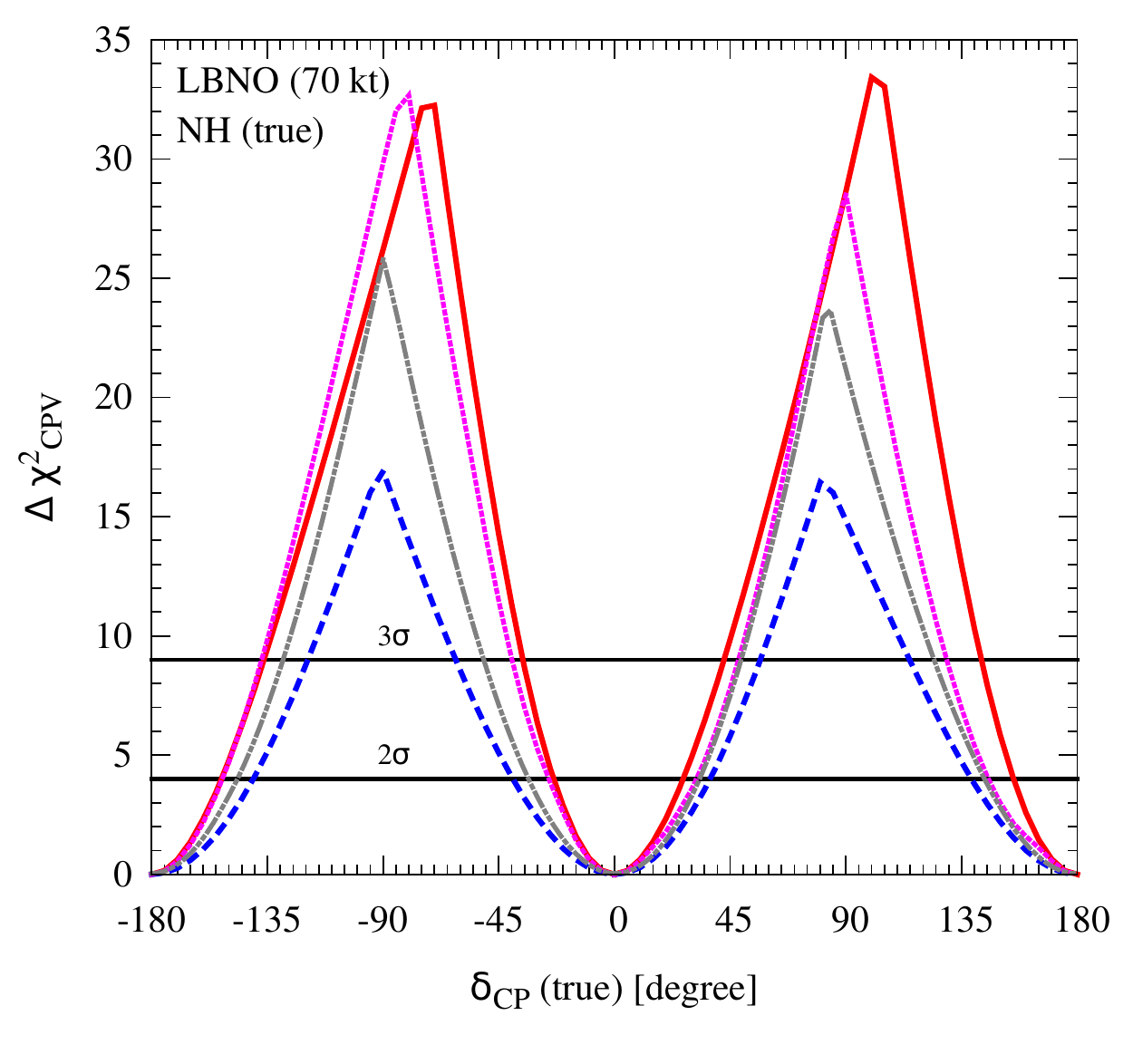}
}
\caption{CP-violation discovery reach as a function of true value of $\dcp$
assuming NH as true hierarchy. In the left panel, we show the results for DUNE (35 kt)
and the right panel is for LBNO (70 kt). For the `SM' case, $\alpha_{e\mu}$ = 0 in the 
data and also in the fit. For each $\dcp$(true), we also give the results generating the 
data with three different true values of $\alpha_{e\mu}$ which are mentioned in the 
figure legends. In all these three cases, in the fit, we marginalize over test values of 
$\alpha_{e\mu}$ in its allowed range. The rest of the simulation details are exactly 
similar to the `SM' case (see text for details).}
\label{fig:deltachisq-for-CPV-discovery}
\end{figure}
%-------------------------------------------------------------------------------------------

%=========================================================================
\begin{table}[t]
\begin{center}
{
\newcommand{\mc}[3]{\multicolumn{#1}{#2}{#3}}
\newcommand{\mr}[3]{\multirow{#1}{#2}{#3}}
\begin{tabular}{|c|c||c|c||c|c|}
\hline
\mc{2}{|c||}{\mr{2}{*}{True Hierarchy}} & \mc{2}{c||}{\bf DUNE (35 kt)} & \mc{2}{c|}{\bf LBNO (70 kt)} \\
\cline{3-6}
\mc{2}{|c||}{} & SM & $\alpha_{e\mu}$(true) = $10^{-52}$ & SM & $\alpha_{e\mu}$(true) = $10^{-52}$ \\
\hline
\mr{2}{*} {2$\sigma$ C.L.} & NH (true) & 0.67 & 0.62 & 0.71 & 0.56 \\
                                          & IH (true) & 0.68 & 0.59 & 0.73 & 0.44 \\
\hline
\mr{2}{*} {3$\sigma$ C.L.} & NH (true) & 0.48 & 0.41 & 0.55 & 0.30 \\
                                          & IH (true) & 0.53 & 0.37 & 0.60 & 0.12 \\
\hline
\end{tabular}
}
\caption{Fraction of $\dcp$(true) for which a discovery is possible for CP-violation from 
DUNE (35 kt) and LBNO (70 kt) set-ups at 2$\sigma$ and 3$\sigma$ confidence levels.
We show the coverage in $\dcp$(true) for both the choices of true hierarchy: NH and IH.
For the `SM' cases, we consider $\alpha_{e\mu}$ = 0 in the data and also in the fit.
We also give the results generating the data with $\alpha_{e\mu}$(true) = $10^{-52}$
and in the fit, we marginalize over test values of $\alpha_{e\mu}$ in its allowed range. 
The rest of the simulation details are exactly similar to the `SM' case 
(see text for details).}
\label{tab:CPV-with-SM-and-SM+LRF}
\end{center}
\end{table}
%=========================================================================

This section is devoted to study how the long-range potential due to $L_e-L_{\mu}$ 
symmetry affects the CP-violation search which is the prime goal of these future
facilities. Can we reject both the CP-conserving values of $0^\circ$, $180^\circ$
at a given confidence level? The performance indicator ``discovery reach of
leptonic CP-violation'' addresses this question and obviously, this measurement 
becomes extremely difficult for the $\dcp$ values which are close to $0^\circ$ and
$180^\circ$. In Fig.~\ref{fig:deltachisq-for-CPV-discovery}, we present the 
CP-violation discovery reach of DUNE (left panel) and LBNO (right panel) as a 
function of true value of $\dcp$ assuming NH as true hierarchy. 
In this plot, we generate our predicted event spectrum (data) considering the true 
value of $\dcp$ as shown in the x-axis, along with the other true values of the oscillation 
parameters given in the third column of Table~\ref{tab:benchmark-parameters}.
Then, we estimate the various theoretical event spectra assuming the test $\dcp$ 
to be the CP-conserving values $0^\circ$ and $180^\circ$, and by varying 
simultaneously $\tmt$ in its 3$\sigma$ allowed range and both the choices of 
mass hierarchy. We calculate the $\Delta\chi^2$ between each set of predicted 
and theoretical event spectra using the procedure described in 
section~\ref{simulation-method}. The smallest of all such $\Delta\chi^2$ values: 
$\Delta\chi^2_{\textrm{CPV}}$ is plotted in Fig.~\ref{fig:deltachisq-for-CPV-discovery}
as a function of $\dcp$(true) in the range -$180^\circ$ to $180^\circ$.
In both the panels, the solid red lines depict the `SM' case where $\alpha_{e\mu}$ = 0 
in the data and also in the fit. For each $\dcp$(true), we also give the results 
generating the data with three different true values of $\alpha_{e\mu}$ which are 
mentioned in the figure legends. In all these three cases, in the fit, we also 
marginalize over test values of $\alpha_{e\mu}$ in the range $10^{-54}$ to $10^{-51}$
along with the other three-flavor oscillation parameters as discussed before.
Fig.~\ref{fig:deltachisq-for-CPV-discovery} clearly shows that the CP-violation 
discovery reach can be altered by substantial amount as compared to the `SM'
case depending on the true choice of $\alpha_{e\mu}$. In case of 
$\alpha_{e\mu}$(true) = $6 \times 10^{-53}$, we see a large suppression in the
CP-violation discovery reach of DUNE (left panel) in the range 
$45^\circ \leq \dcp {\rm (true)} \leq 135^\circ$. We have checked that this mainly 
happens due to the marginalization over $\tmt$ in the fit where we vary $\sa$ 
over a wide range (0.38 to 0.64) without imposing any prior on it. In case 
$\alpha_{e\mu}$(true) = $10^{-52}$, the LBNO set-up (right panel) suffers 
a large depletion in the CP-violation discovery reach which can be easily 
explained with the help of bi-events plot (Fig.~\ref{fig:bi-events-plot}) 
shown in section~\ref{bi-events-plot}. In the right panel of Fig.~\ref{fig:bi-events-plot}, 
we have seen a large reduction in the $\nu$ and $\anu$ event rates 
for CERN-Pyh{\"a}salmi baseline with $\alpha_{e\mu}$ = $10^{-52}$ 
and this is true for both NH and IH. The differences in the number of events 
for the CP-conserving and CP-violating phases get reduced as the ellipses 
in Fig.~\ref{fig:bi-events-plot} get shrunk in the presence of LRF, severely 
deteriorating the CP-violation discovery reach of LBNO as can be seen from
the right panel of Fig.~\ref{fig:deltachisq-for-CPV-discovery}.
Table~\ref{tab:CPV-with-SM-and-SM+LRF} also validates this result, where we 
compare the precise fraction of $\dcp$(true) for which a discovery is possible 
for CP-violation from LBNO (70 kt) and DUNE (35 kt) at 2$\sigma$ and 
3$\sigma$ confidence levels. For LBNO set-up with true NH and
$\alpha_{e\mu}$(true) = $10^{-52}$, the coverage in $\dcp$(true) at 3$\sigma$
C.L. reduces to 30\% from 55\% as we have in the `SM' case. In case of true IH,
the impact of long-range potential is even more dramatic for these future facilities. 
At 3$\sigma$ with $\alpha_{e\mu}$(true) = $10^{-52}$, their CP-violation reach 
is quite minimal: only 12\% for LBNO and 37\% for DUNE while in the `SM' 
framework, the coverage is 60\% for LBNO and 53\% for DUNE. Since, the 
sign of the long-range potential $V_{e\mu}$ is opposite for neutrino and 
anti-neutrino, it affects the neutrino and anti-neutrino oscillation probabilities 
in different fashion. This feature introduces fake CP-asymmetry like the 
SM matter effect and severely limits the CP-violation search in these 
long-baseline facilities which can be clearly seen from 
Table~\ref{tab:CPV-with-SM-and-SM+LRF}.

%--------------------------------------------------------------------------------------------
\begin{figure}[t]
\centerline{
\includegraphics[width=0.49\textwidth]{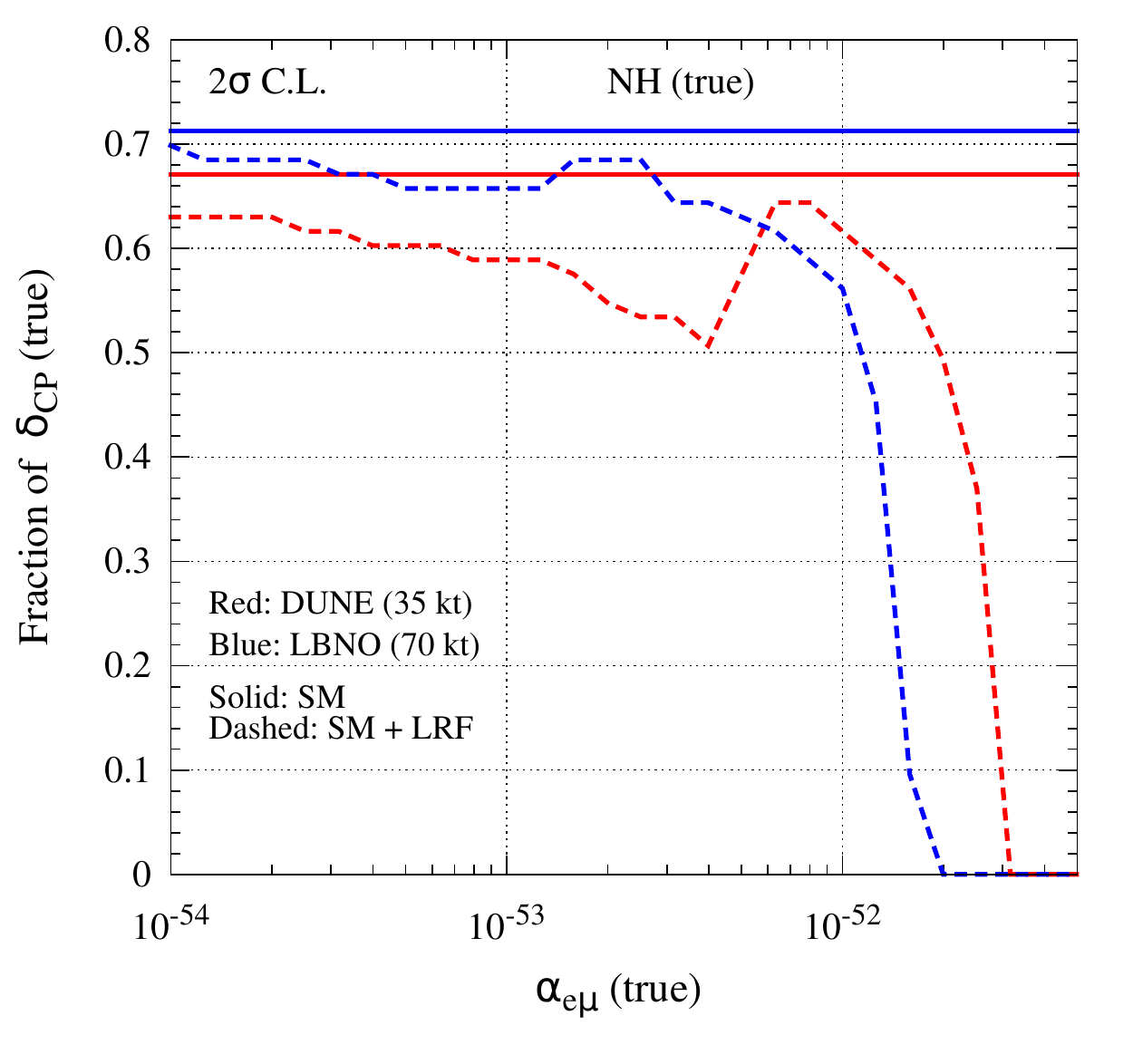}
\includegraphics[width=0.49\textwidth]{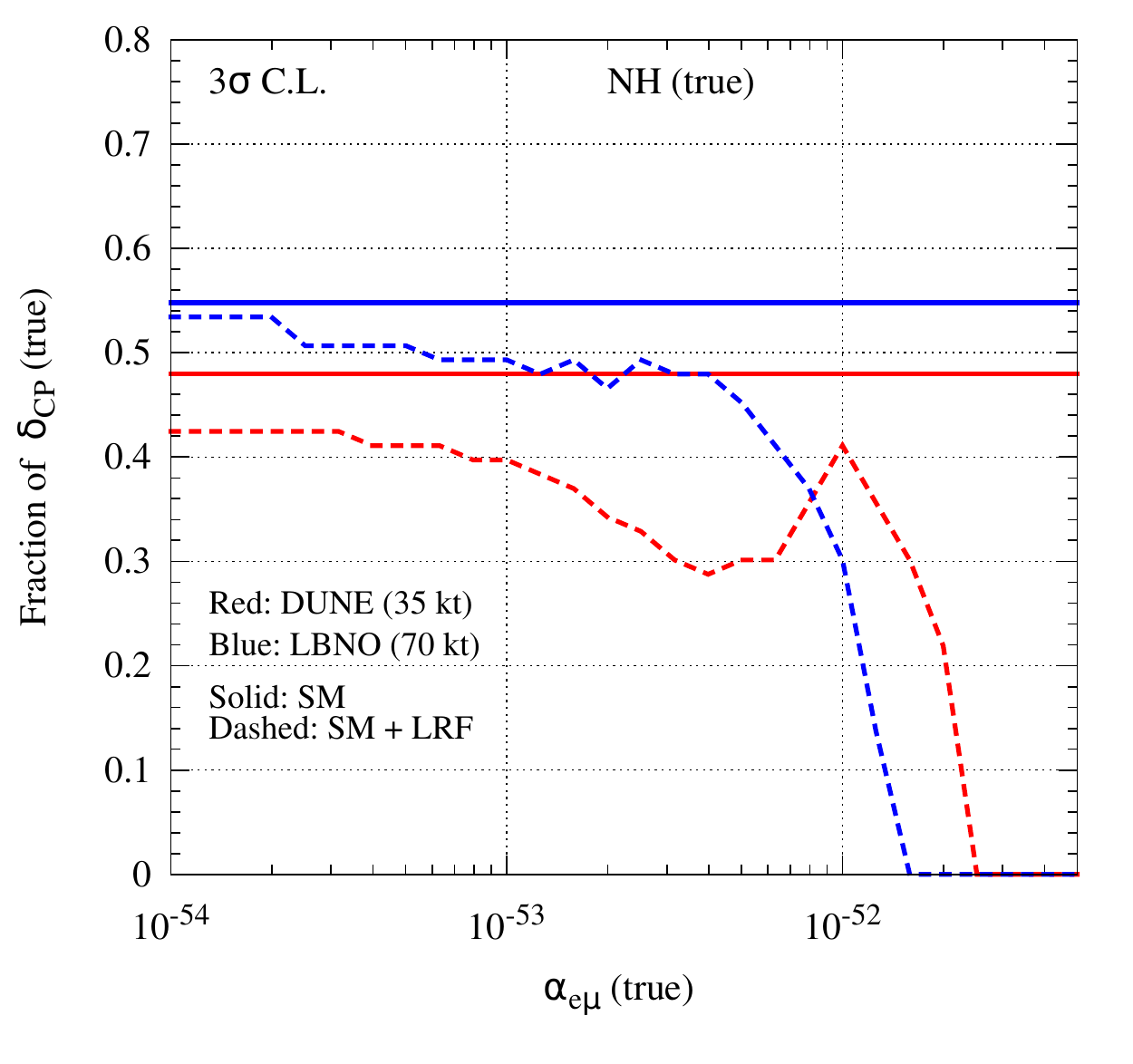}
}
\caption{Fraction of $\dcp$(true) for which a discovery is possible for CP-violation
is plotted as a function of true value of $\alpha_{e\mu}$ assuming NH as true 
hierarchy. In each panel, we compare the performances 
of DUNE (35 kt) and LBNO (70 kt) which are shown by red and blue lines 
respectively. We give the results at 2$\sigma$ (left panel) and 3$\sigma$ 
(right panel) confidence levels. In both the panels, the solid lines portray 
the `SM' scenario where $\alpha_{e\mu}$ = 0 in the data and also in the fit.
For the `SM+LRF' case (dashed lines), the data is generated with the true 
value of $\alpha_{e\mu}$ as shown in the x-axis and in the fit, we marginalize
over test values of $\alpha_{e\mu}$ in its allowed range. The rest of the simulation 
details are exactly similar to the `SM' case (see text for details).}
\label{fig:CPV-coverage}
\end{figure}
%------------------------------------------------------------------------------------------------

Finally, to see the complete picture, the fraction of $\dcp$(true) for which a 
discovery is possible for CP-violation is shown in Fig.~\ref{fig:CPV-coverage}
as a function of true value of $\alpha_{e\mu}$ assuming NH as true 
hierarchy. In each panel, we compare the performances of DUNE (35 kt) and 
LBNO (70 kt) which are shown by red and blue lines respectively. 
We give the results at 2$\sigma$ (left panel) and 3$\sigma$ (right panel) confidence 
levels. In both the panels, the solid horizontal lines depict the `SM' case where 
$\alpha_{e\mu}$ = 0 in the data and also in the fit. For the `SM+LRF' case (dashed lines), 
the data is generated with the true value of $\alpha_{e\mu}$ as shown in the x-axis and 
in the fit, we marginalize over test values of $\alpha_{e\mu}$ in its allowed range. 
The rest of the simulation details are exactly similar to the `SM' case as discussed 
before. For the values close to $\alpha_{e\mu}$(true) = $10^{-54}$, the event spectra 
in the data is almost similar to the `SM' case, but since we allow $\alpha_{e\mu}$ 
to vary in the fit in the range $10^{-54}$ to $10^{-51}$ along with the other three-flavor 
oscillation parameters as discussed before, we see a small suppression in the fraction 
of $\dcp$(true). In both the panels, around $\alpha_{e\mu}$(true) = $6 \times 10^{-53}$, 
the CP-violation discovery reach of DUNE deteriorates substantially, which we also 
observe in Fig.~\ref{fig:deltachisq-for-CPV-discovery}, and the marginalization 
over $\tmt$ is mainly responsible for this as we have already discussed.
Once $\alpha_{e\mu}$(true) approaches toward $10^{-52}$, the coverages 
in $\dcp$(true) for which CP-violation can be observed, shrink very rapidly for 
both the set-ups, and ultimately around $\alpha_{e\mu}$(true) = 
$2 \times 10^{-52}$, the coverages almost become zero. We can understand 
this feature from our discussions in section~\ref{analytical-expressions}, where 
we have seen that in the presence of $V_{e\mu}$, as we increase $E$, 
$\theta_{13}^m$ quickly approaches toward $45^{\circ}$ 
(see middle panel of Fig.~\ref{fig:running-mixing-angles}), and the resonance
occurs at much lower energies as compared to the SM case. Finally, 
$\theta_{13}^m$ reaches to $90^{\circ}$ as we further increase $E$,
and the $\numu \to \nue$ oscillation probabilities vanish for most
of the energies where we have significant amount of neutrino flux. 
It causes a huge suppression in the event rates and as a result, 
the sensitivity goes to zero. Next, we turn our attention to the 
mass hierarchy discovery potential of DUNE and LBNO.

%==============================================
\subsection{Impact of LRF on Mass Hierarchy Measurements}
\label{MH-Impact}
%==============================================

%---------------------------------------------------------------------------------------------------------------------------------------
\begin{figure}[t]
\centerline{
\includegraphics[width=0.33\textwidth]{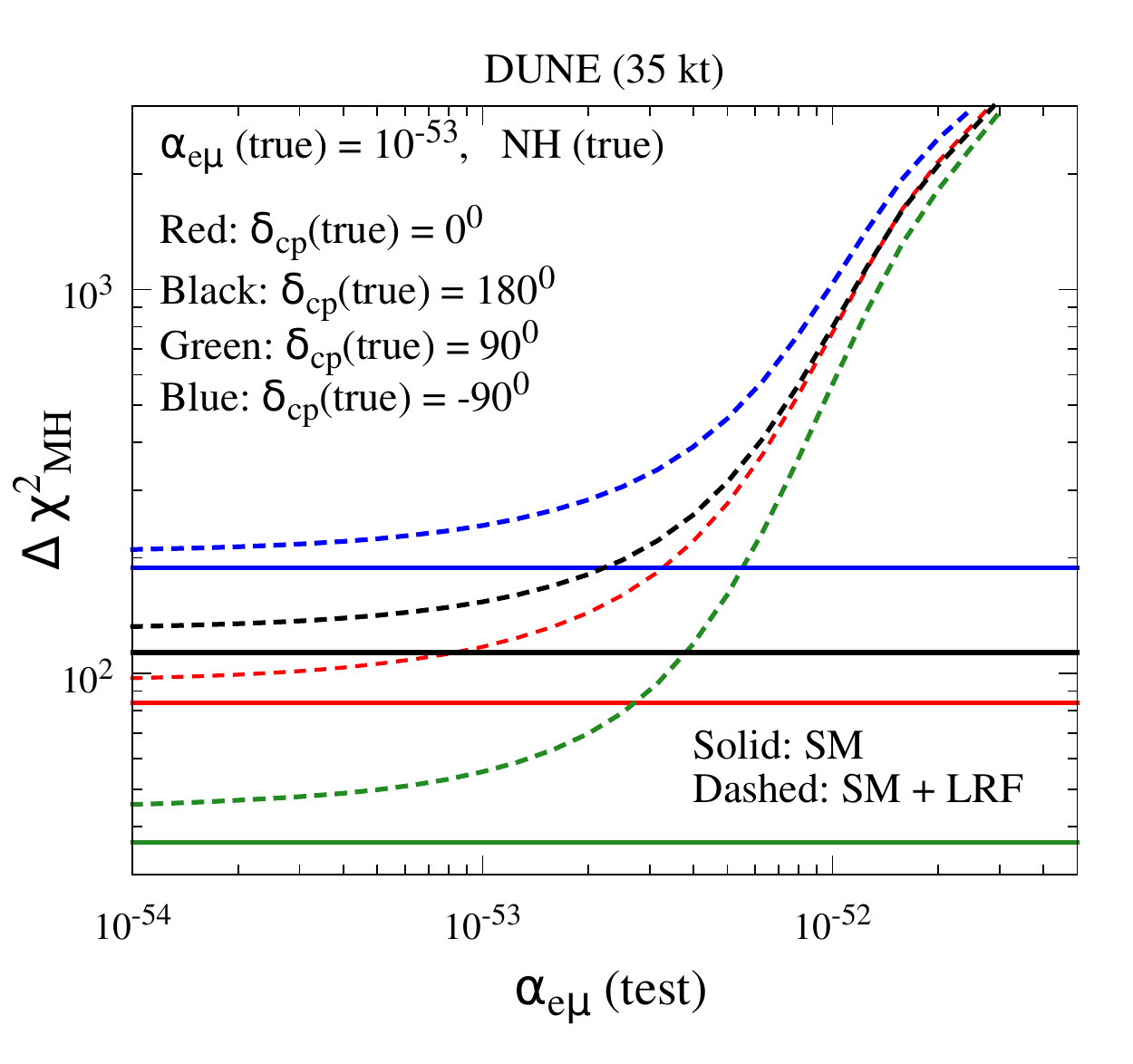}
\includegraphics[width=0.33\textwidth]{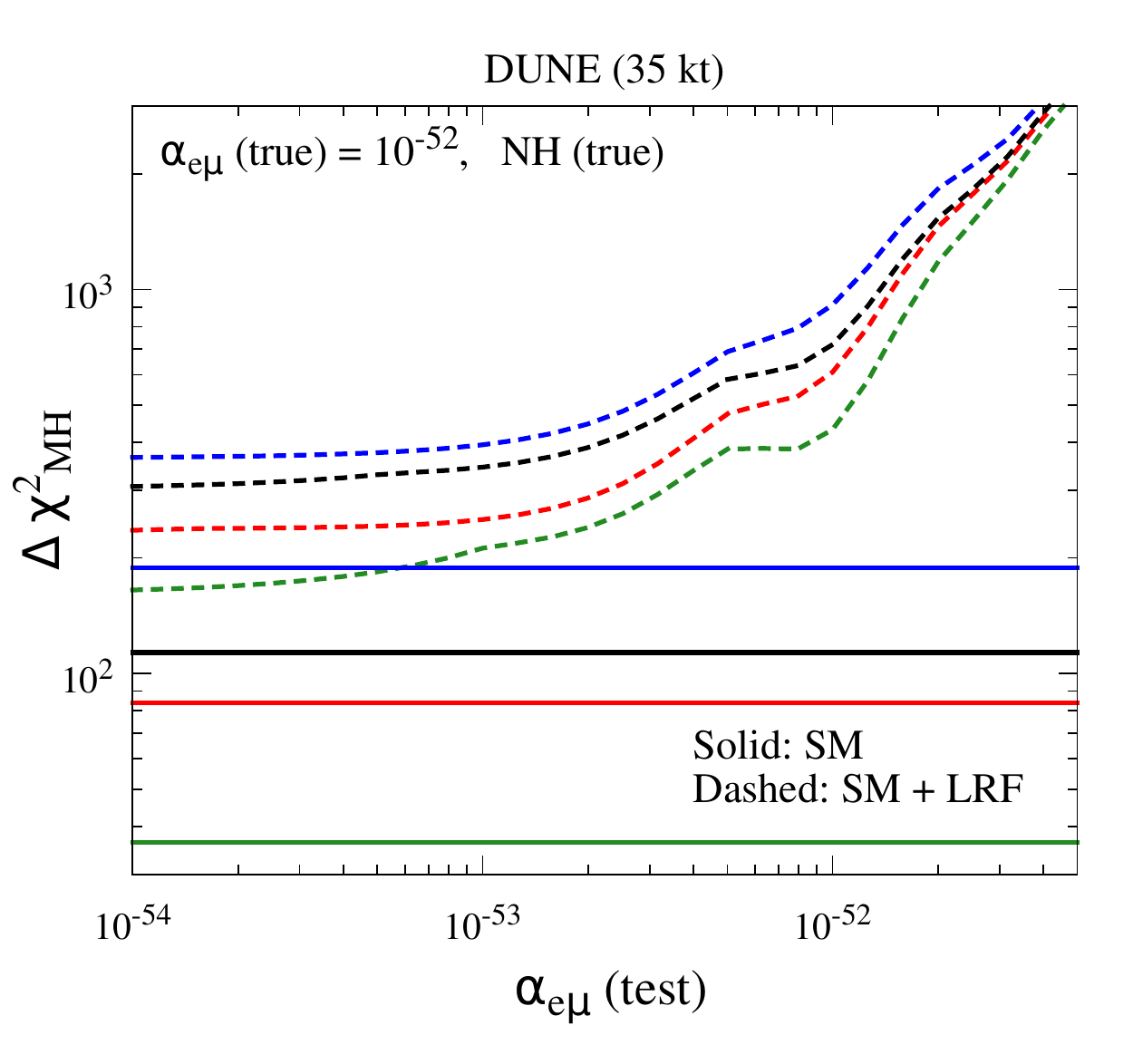}
\includegraphics[width=0.33\textwidth]{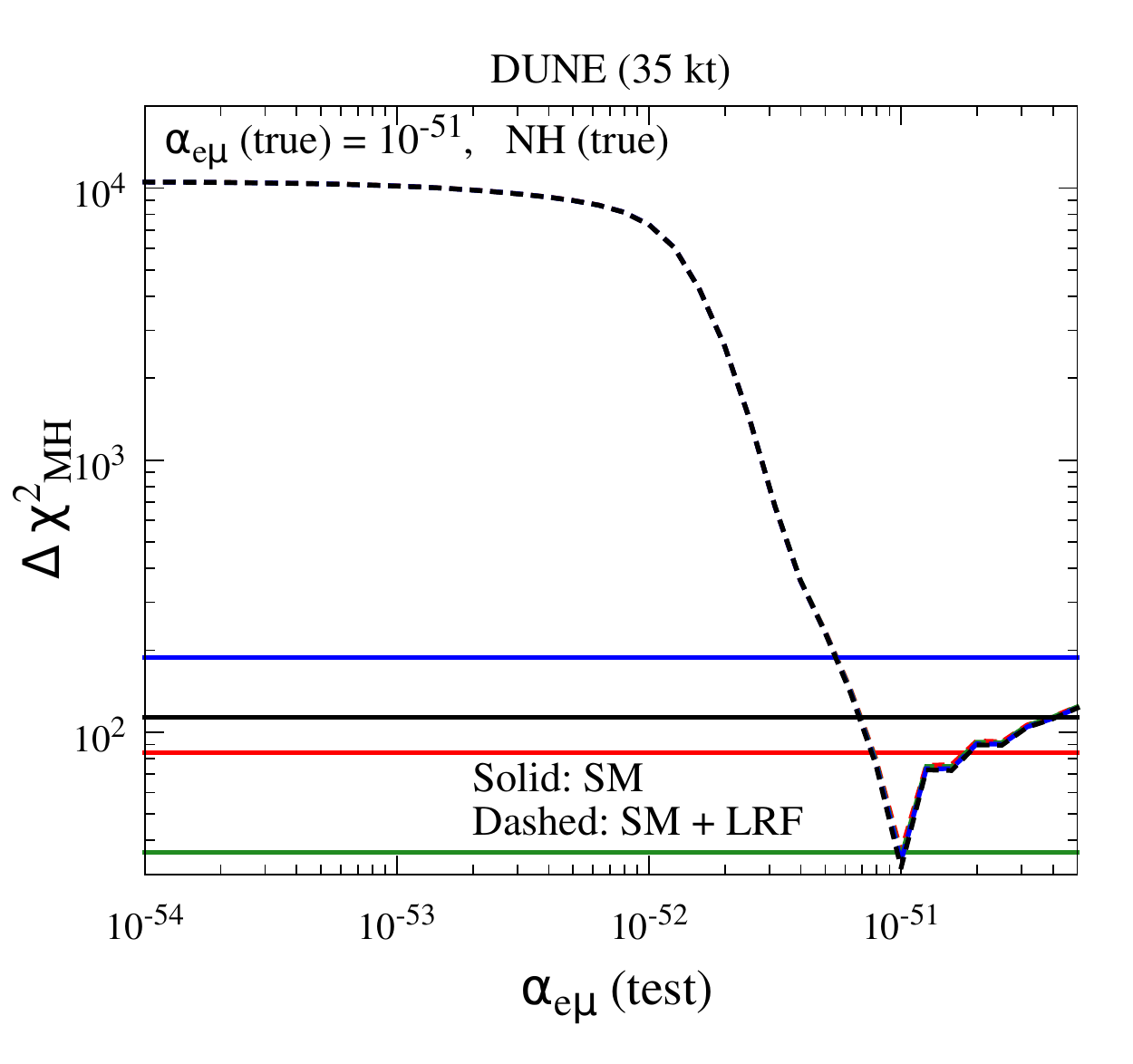}
}
\centerline{
\includegraphics[width=0.33\textwidth]{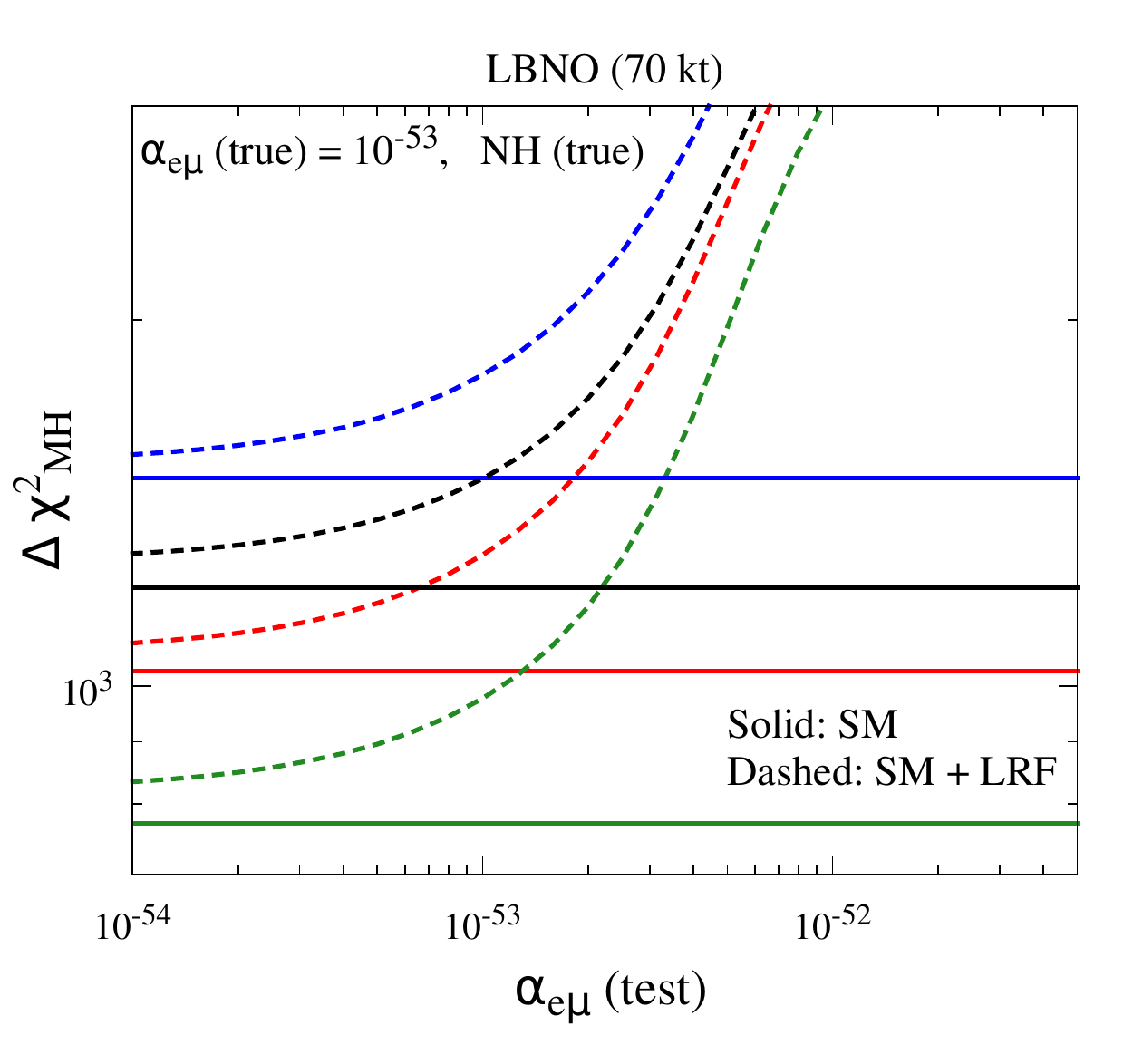}
\includegraphics[width=0.33\textwidth]{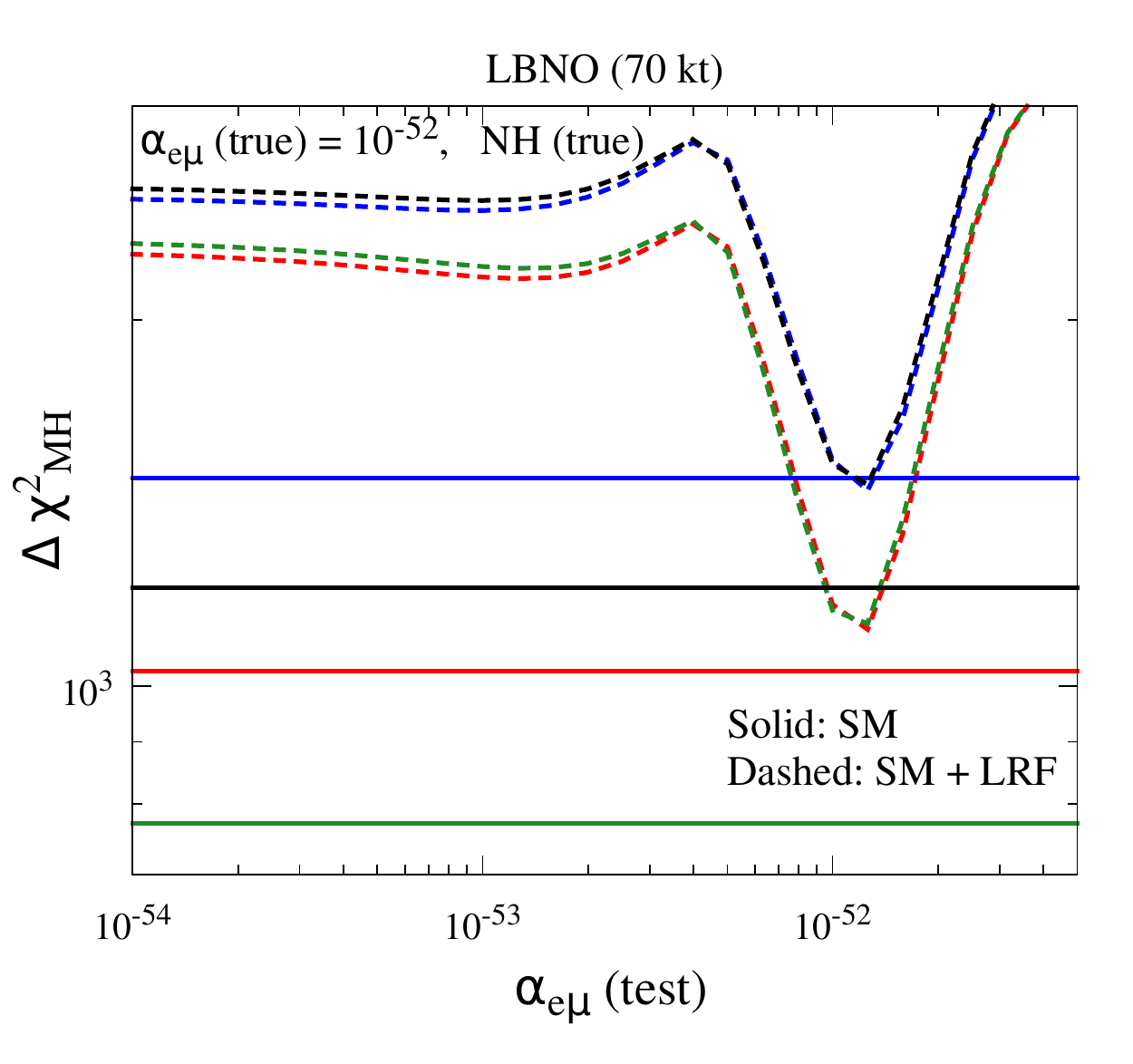}
\includegraphics[width=0.33\textwidth]{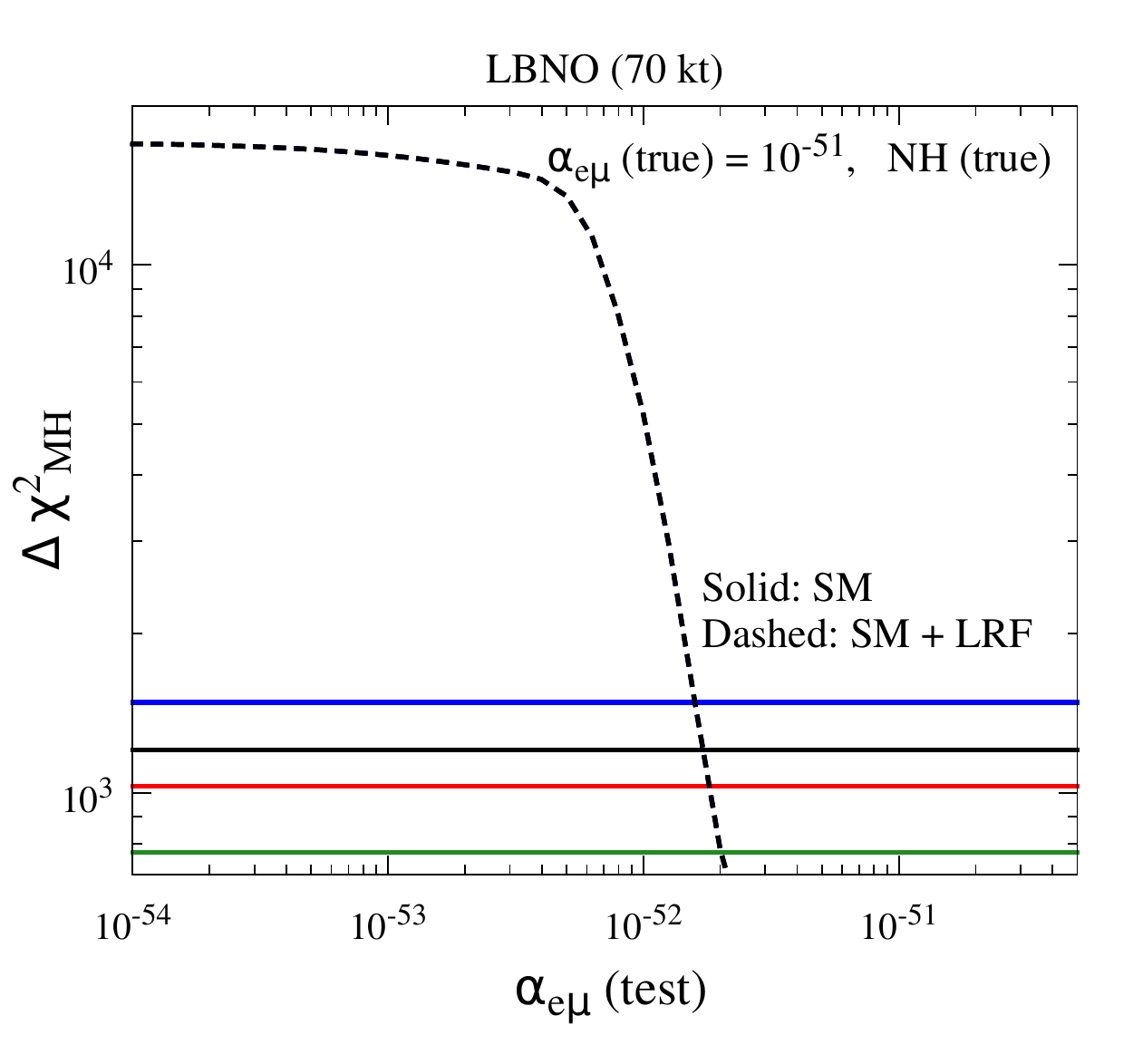}
} 
\caption{Discovery reach for mass hierarchy as a function of test $\alpha_{e\mu}$ assuming NH 
in the data and IH in the fit. The upper (lower) panels are for DUNE (LBNO). We give the results
for four different true values of $\dcp$ in each panel. The solid horizontal lines in each panel show 
the `SM' case where $\alpha_{e\mu}$ = 0 in the data and also in the fit. For the `SM+LRF' case 
(dashed lines), the data is generated with the true value of $\alpha_{e\mu}$ = $10^{-53}$ 
(left panels), $10^{-52}$ (middle panels), and $10^{-51}$ (right panels). Then, in the fit, we vary 
the test values of $\alpha_{e\mu}$ while marginalizing over $\tmt$ and $\dcp$. The rest of the 
simulation details are exactly similar to the `SM' case (see text for details). Also, note that the 
ranges in the x-axis and y-axis are different in some of the panels.}
\label{fig:MH-NH}
\end{figure}
%----------------------------------------------------------------------------------------------------------------------------------------

%----------------------------------------------------------------------------------------------------------------------------------------------
\begin{figure}[t]
\centerline{
\includegraphics[width=0.33\textwidth]{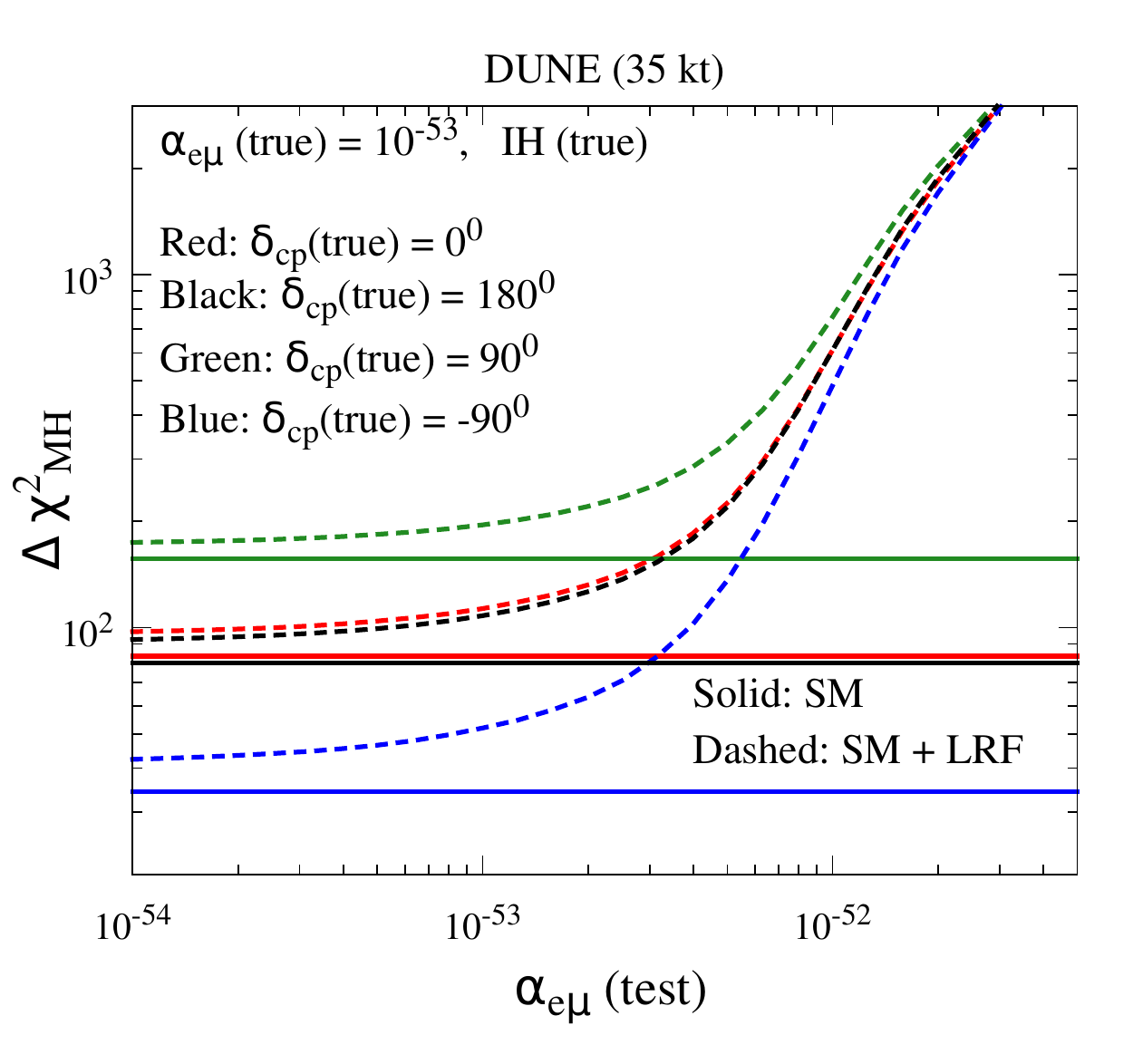}
\includegraphics[width=0.33\textwidth]{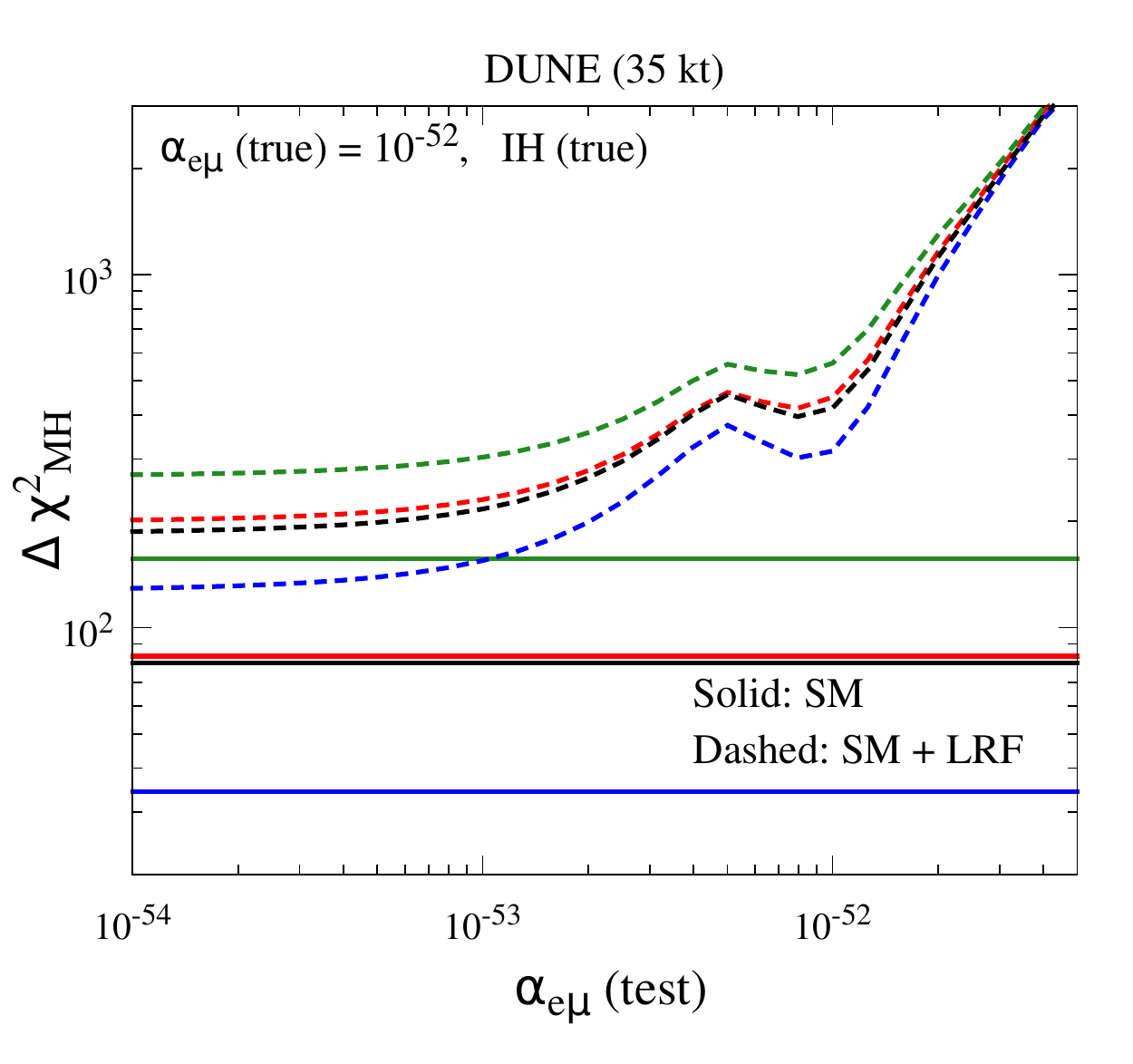}
\includegraphics[width=0.33\textwidth]{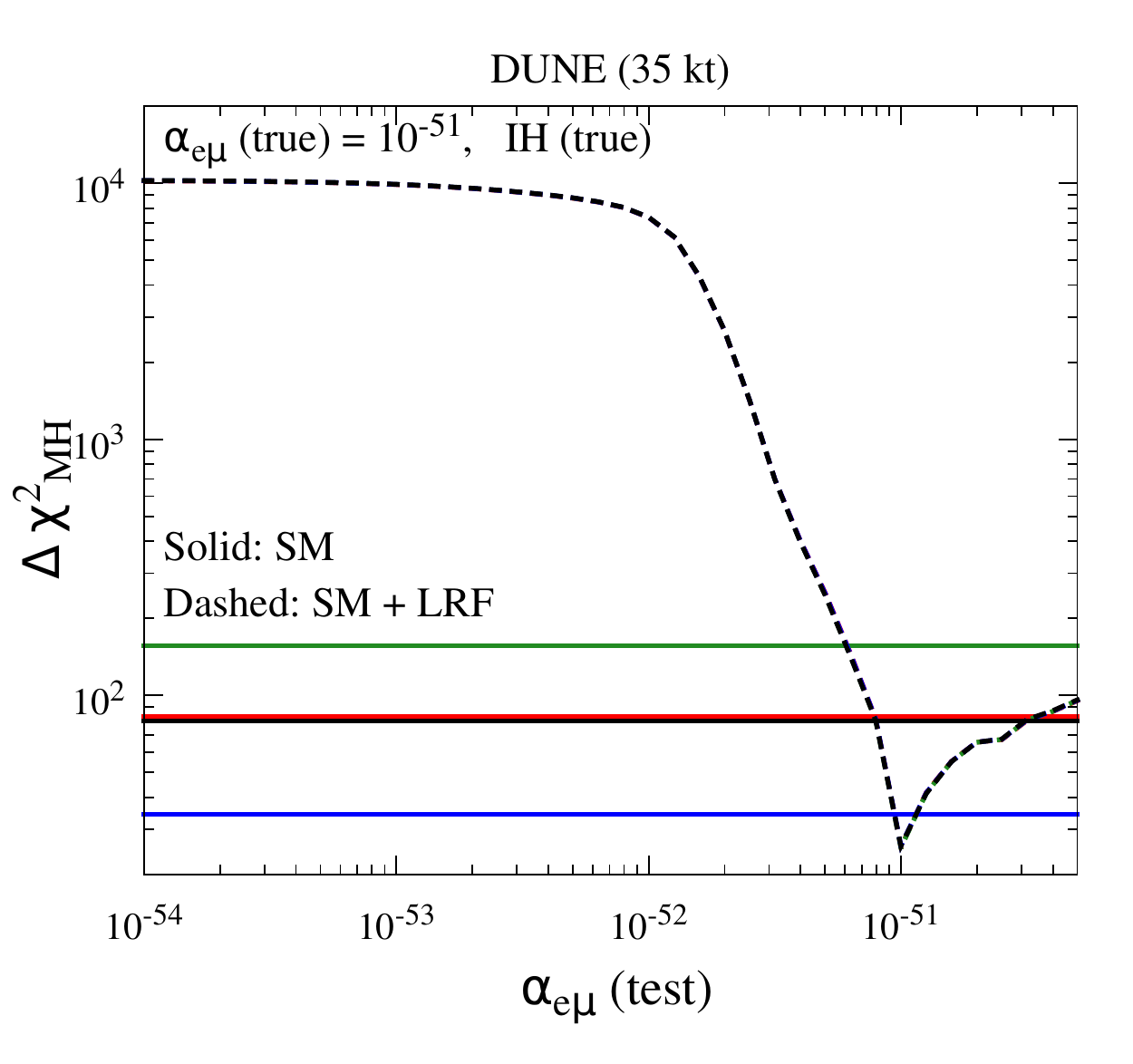}
}
\centerline{
\includegraphics[width=0.33\textwidth]{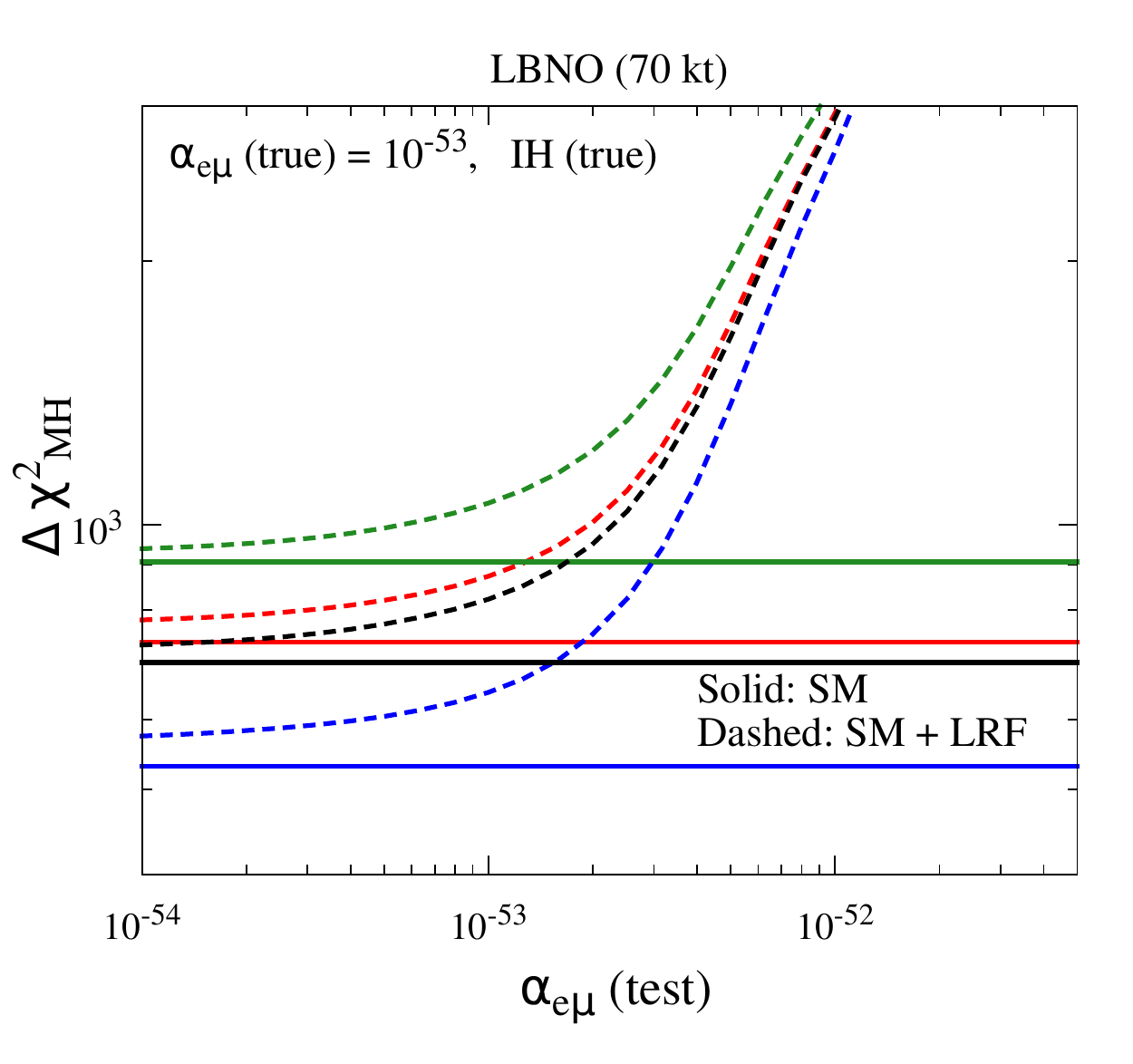}
\includegraphics[width=0.33\textwidth]{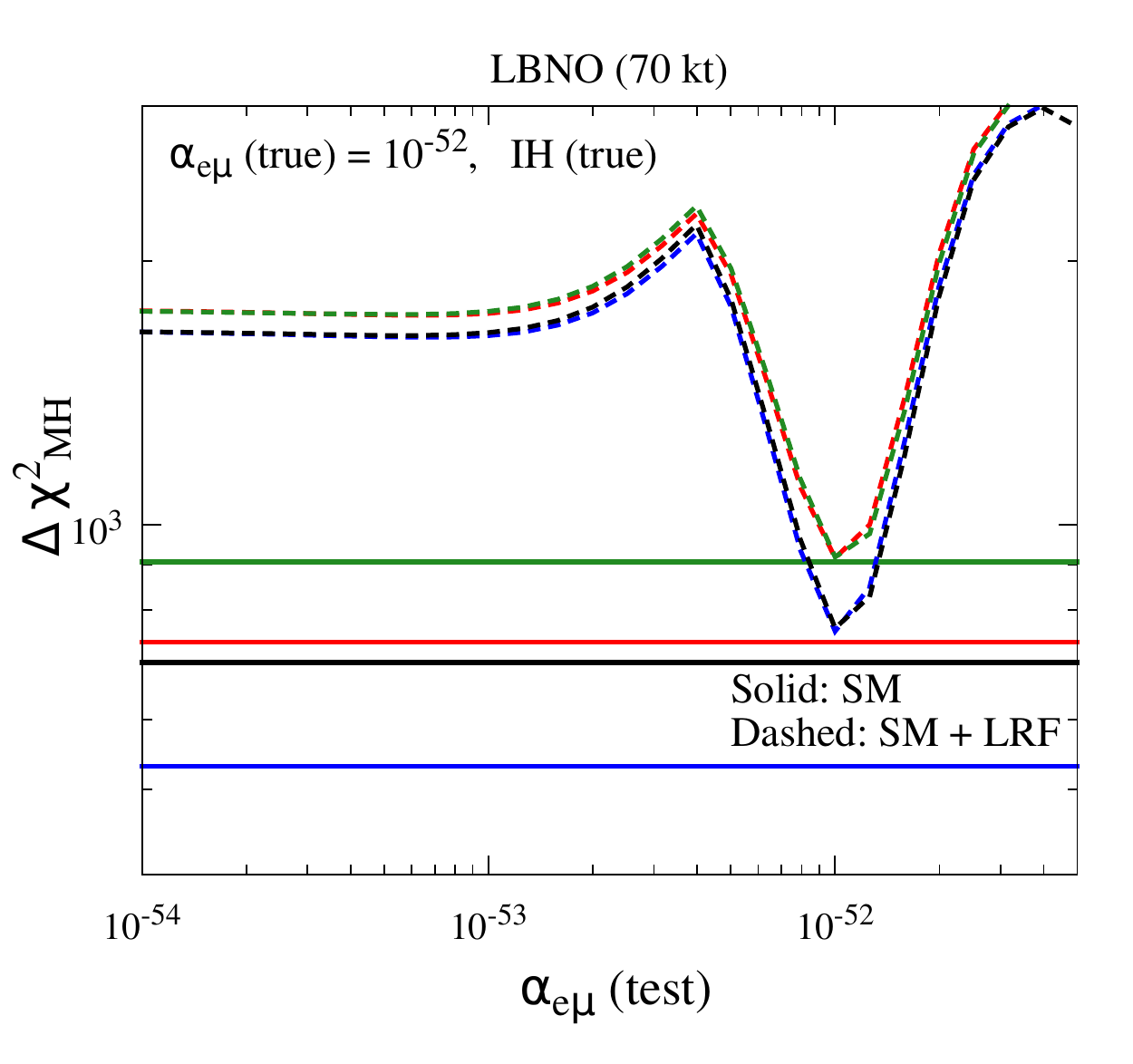}
\includegraphics[width=0.33\textwidth]{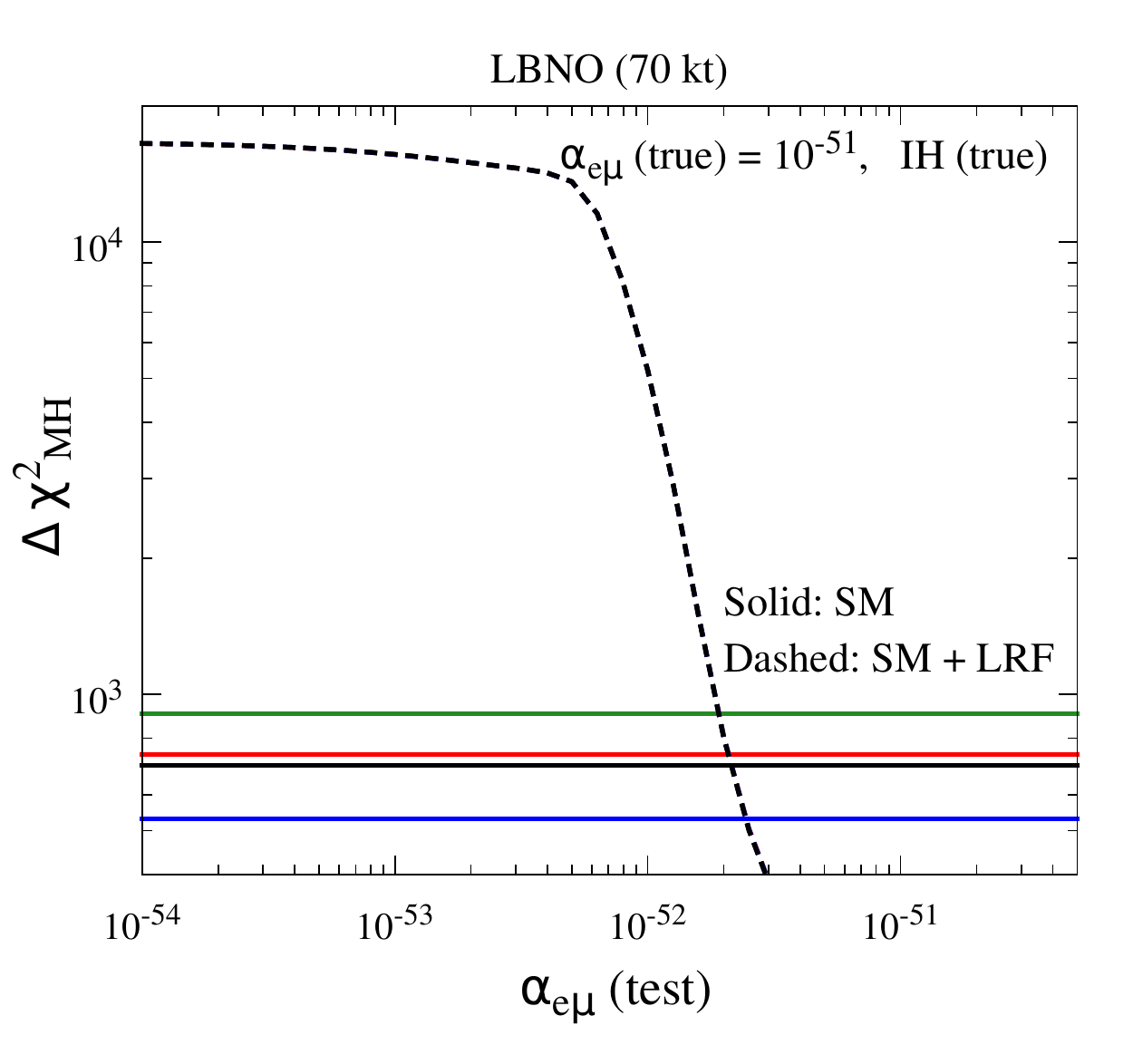}
} 
\caption{Discovery reach for mass hierarchy as a function of test $\alpha_{e\mu}$ assuming IH 
in the data and NH in the fit. The upper (lower) panels are for DUNE (LBNO). We give the results
for four different true values of $\dcp$ in each panel. The solid horizontal lines in each panel show 
the `SM' case where $\alpha_{e\mu}$ = 0 in the data and also in the fit. For the `SM+LRF' case 
(dashed lines), the data is generated with the true value of $\alpha_{e\mu}$ = $10^{-53}$
(left panels), $10^{-52}$ (middle panels), and $10^{-51}$ (right panels). Then, in the fit, we vary 
the test values of $\alpha_{e\mu}$ while marginalizing over $\tmt$ and $\dcp$. The rest of the 
simulation details are exactly similar to the `SM' case (see text for details). Also, note that the 
ranges in the x-axis and y-axis are different in some of the panels.}
\label{fig:MH-IH}
\end{figure}
%--------------------------------------------------------------------------------------------------------------------------------------------------

The large Earth matter effects at both the DUNE and LBNO baselines enhance the 
separation between the oscillation spectra of NH and IH, and hence, we have large 
differences in the event rates for NH and IH, leading to unprecedented sensitivity toward 
neutrino mass hierarchy. Now, it would be quite interesting to see how robust are these
measurements in the presence of LRF? A `discovery' of the mass hierarchy is a discrete
measurement and is defined as the ability to exclude any degenerate solution for the 
wrong (fit) hierarchy at a given confidence level. For hierarchy sensitivity, we first
assume NH to be the true hierarchy and we choose a true value of $\dcp$ 
and $\alpha_{e\mu}$. We compute the NH event spectrum for these assumptions 
and the other true values of the oscillation parameters (see the third column of 
Table~\ref{tab:benchmark-parameters}) and label it to be data. Then, we estimate 
the various theoretical event spectra assuming IH and a test value of 
$\alpha_{e\mu}$ as shown in the x-axis of Fig.~\ref{fig:MH-NH}, and by varying 
simultaneously test $\tmt$ in its 3$\sigma$ allowed range and test $\dcp$ in the full
allowed range (-$180^\circ$ to $180^\circ$). Next, we compute the $\Delta\chi^2$ 
between each set of predicted and theoretical event spectra using the numerical 
technique described in section~\ref{simulation-method}. The smallest of all such 
$\Delta\chi^2$ values: $\Delta\chi^2_{\textrm{MH}}$ is shown in 
Fig.~\ref{fig:MH-NH} as a function of $\alpha_{e\mu}$(test) for given choices
of $\alpha_{e\mu}$(true) and $\dcp$(true). As mentioned above, we always assume 
NH in the data and IH in the fit while generating the curves in Fig.~\ref{fig:MH-NH}.
The upper panels portray the performance of DUNE (35 kt), while the lower panels 
are for LBNO (70 kt). In each panel, the results are given for four different choices 
of $\dcp$(true) and the solid horizontal lines depict the `SM' case where 
$\alpha_{e\mu}$ is zero in the data and also in the fit. For the `SM+LRF' case 
(dashed lines), the data is generated with the true value of $\alpha_{e\mu}$ 
as mentioned in the top part of each panel, and in the fit, we vary the test values 
of $\alpha_{e\mu}$ while marginalizing over $\tmt$ and $\dcp$. The rest of the 
simulation details are exactly similar to the `SM' case as mentioned above.
Though in Fig.~\ref{fig:MH-NH}, we have shown the results for three benchmark 
values of $\alpha_{e\mu}$(true) = $10^{-53}$ (left panels), $10^{-52}$ 
(middle panels), and $10^{-51}$ (right panels), but, we have checked that
for DUNE, the mass hierarchy sensitivity always stays above the standard 
expectations irrespective of $\dcp$(true) provided the true value of 
$\alpha_{e\mu} < 5 \times 10^{-52}$. There is a large suppression 
in the appearance event rates when we generate the data with a true value 
of $\alpha_{e\mu}$ around $5 \times 10^{-52}$, and if we further increase the 
value of $\alpha_{e\mu}$(true), the statistical strength of the data reduces very 
rapidly, and the sensitivity goes below the standard expectation. The upper right 
panel in Fig.~\ref{fig:MH-NH} clearly shows this behavior.
In case of LBNO, the mass hierarchy discovery reach never goes below 
the `SM' value irrespective of $\dcp$(true) if the true choice of $\alpha_{e\mu}$ 
is smaller than $10^{-52}$. Once we consider the true value of 
$\alpha_{e\mu} \geq 10^{-52}$, the appearance event rates get reduced
in data by considerable amount, causing a significant drop in the sensitivity
which can be clearly seen from the lower middle and right panels in 
Fig.~\ref{fig:MH-NH}. In Fig.~\ref{fig:MH-IH}, we generate the data
with IH and fit it with NH. We see almost similar behavior in all the panels
of Fig.~\ref{fig:MH-IH} as we have noticed in Fig.~\ref{fig:MH-NH}.

%==========================
\section{Summary and Conclusions}
\label{summary-conclusions}
%==========================

Flavor-dependent long-range leptonic forces mediated by the extremely 
light and neutral bosons associated with gauged $L_e-L_{\mu}$ or
$L_e-L_{\tau}$ symmetries, constitute a minimal extension of the SM 
preserving its renormalizability and can lead to interesting 
phenomenological consequences. For an example, the electrons
inside the Sun can generate a flavor-dependent long-range potential 
$V_{e\mu/e\tau}$ at the Earth surface which can give rise to non-trivial 
three neutrino mixing affects in terrestrial experiments, and could influence 
the neutrino propagation through matter. The sign of this potential is opposite
for anti-neutrinos, and affects the neutrino and anti-neutrino oscillation
probabilities in different fashion. This feature invokes fake CP-asymmetry
like the SM matter effect and can severely affect the leptonic CP-violation
searches in long-baseline experiments. In this paper for the first time, 
we have investigated in detail the possible impacts of these long-range 
flavor-diagonal neutral current interactions in the oscillations of neutrinos 
and anti-neutrinos in the context of future high-precision superbeam 
facilities, DUNE and LBNO. The key point here is that for long-baseline 
neutrinos, $\Delta m^2/2E \sim 2.5 \times 10^{-13}$ eV 
(assuming $\Delta m^2 \sim 2.5 \times 10^{-3}$ eV$^2$ and 
$E \sim$ 5 GeV) which is comparable to $V_{e\mu}$ even for 
$\alpha_{e\mu} \sim 10^{-52}$, and can influence the long-baseline 
experiments significantly. For the Fermilab-Homestake (1300 km) 
and CERN-Pyh\"asalmi (2290 km) baselines, the Earth matter potentials 
are also around $10^{-13}$ eV (see Table~\ref{tab:compare}), suggesting 
that $V_{CC}$ can also interfere with $V_{e\mu}$ and 
$\Delta m_{31}^2/2E$, having substantial impact on the oscillation 
probability. We have explored these interesting possibilities in detail 
in this work. In this paper, we have presented all the results considering 
the $L_e-L_{\mu}$ symmetry. Similar analysis can be performed
for the $L_e-L_{\tau}$ symmetry which we will present elsewhere.

We have derived approximate analytical expressions for the effective
neutrino oscillation parameters to study how they `run' as functions
of the neutrino energy in the presence of both long-range and Earth
matter potentials. We have also obtained a compact and simple
expression for the resonance energy, where $\tet$ becomes $45^{\circ}$ 
in the presence of both $V_{CC}$ and $V_{e\mu}$. We have observed 
that in the presence of $V_{e\mu}$, as we increase the neutrino energy, 
$\theta_{13}$ in matter quickly approaches toward $45^{\circ}$, and 
the resonance occurs at much lower energies as compared to the 
SM case. Finally, $\theta_{13}$ in matter reaches to $90^{\circ}$ 
as we further increase the energy, causing a large suppression in 
the appearance probability for most of the energies where we have 
significant amount of neutrino flux for both the set-ups. As a result, 
the event rates get reduced which can be clearly seen from the 
bi-events plot in Fig.~\ref{fig:bi-events-plot}.

As the long-range potential due to gauged $L_e-L_{\mu}$ symmetry
can change the standard oscillation picture of these future facilities 
significantly, we can expect to place strong constraints on $\alpha_{e\mu}$
if these experiments do not observe a signal of LRF in oscillations. 
For an example, if $\dcp$(true) is $-90^\circ$ and true hierarchy is NH, 
then the expected bound from the DUNE (35 kt) set-up at 90\% C.L. is 
$\alpha_{e\mu} < 1.9 \times 10^{-53}$. The same from the LBNO (70 kt) 
experiment is $\alpha_{e\mu} < 7.8 \times 10^{-54}$, suggesting that 
the constraint from LBNO is 2.4 times better than DUNE. 
This future limit from the DUNE (LBNO) experiment is almost 30 (70) 
times better than the existing bound from the SK experiment 
\cite{Joshipura:2003jh}. We have noticed that these future limits 
on $\alpha_{e\mu}$ from DUNE and LBNO are not very sensitive to 
the true choice of $\dcp$ and mass hierarchy. We have also estimated
the discovery reach for $\alpha_{e\mu}$ if we find a positive signal of 
LRF in the expected event spectra at DUNE and LBNO. We have found 
that the spectral information on the signal and background events is quite
crucial to constrain/discover this new long-range force.

We have also studied in detail the CP-violation discovery reach of DUNE 
(35 kt) and LBNO (70 kt) in the presence of LRF. We have seen that the 
CP-violation measurements can be deteriorated by considerable amount 
as compared to the standard expectation depending on the true value 
of $\alpha_{e\mu}$. At 3$\sigma$ with $\alpha_{e\mu}$(true) = $10^{-52}$ 
and true NH, the coverage in $\dcp$(true) for which a discovery is possible 
for CP-violation is 41\% (30\%) for DUNE (LBNO) while in the standard case, 
the coverage is 48\% for DUNE and 55\% for LBNO. In case of true IH,
the impact of long-range potential is even more striking for these future 
facilities. As an example, if $\alpha_{e\mu}$(true) = $10^{-52}$, their chances 
of establishing CP-violation are quite minimal: only for 37\% (12\%) values of 
$\dcp$(true), DUNE (LBNO) can reject both the CP-conserving values 
$0^\circ$ and $180^\circ$ in the fit at 3$\sigma$, while in the `SM' 
framework, DUNE (LBNO) can do so for 53\% (60\%) values of true
$\dcp$. As the true value of $\alpha_{e\mu}$ approaches toward 
$10^{-52}$, the coverages in $\dcp$(true) for which CP-violation 
can be observed, diminish very quickly for both the set-ups, and 
ultimately around $\alpha_{e\mu}$(true) = $2 \times 10^{-52}$, 
the coverages almost become zero.

Finally, we have asked the question, how robust are mass hierarchy 
measurements in these future facilities in the presence of LRF?
In the standard case, due to the large Earth matter effects at the 
Fermilab-Homestake and CERN-Pyh\"asalmi baselines, both 
DUNE and LBNO can resolve the issue of mass hierarchy at very 
high confidence level. Now, if LRF exists in Nature, then for DUNE, 
the mass hierarchy sensitivity remains above the standard expectations 
provided the true value of $\alpha_{e\mu} < 5 \times 10^{-52}$. 
In case of LBNO, the mass hierarchy discovery reach does not
go below the `SM' value as long as the true value of $\alpha_{e\mu}$
is smaller than $10^{-52}$.

%==========================
\subsubsection*{Acknowledgments}
%==========================

S.K.A. would like to thank Anjan S. Joshipura and Subhendra Mohanty 
for useful discussions on long-range forces.
S.K.A. was supported by the DST/INSPIRE Research Grant [IFA-PH-12],
Department of Science \& Technology, India. 
S.K.A. is grateful to the Mainz Institute for Theoretical Physics (MITP) for its hospitality 
and its partial support during the completion of this work.
S.S.C. would like to thank the organizers of the XXI DAE-BRNS High Energy Physics Symposium 
2014 for giving an opportunity to present the preliminary results of this work.

%=======================
\begin{appendix}

%================================================
\section{Discussion at the Probability Level -- Anti-Neutrino Case}
\label{probability-discusions-anti-neutrino-case}
%================================================

%-----------------------------------------------------
\begin{figure}[H]
\centerline{
\includegraphics[width=7.5cm,height=7.0cm]{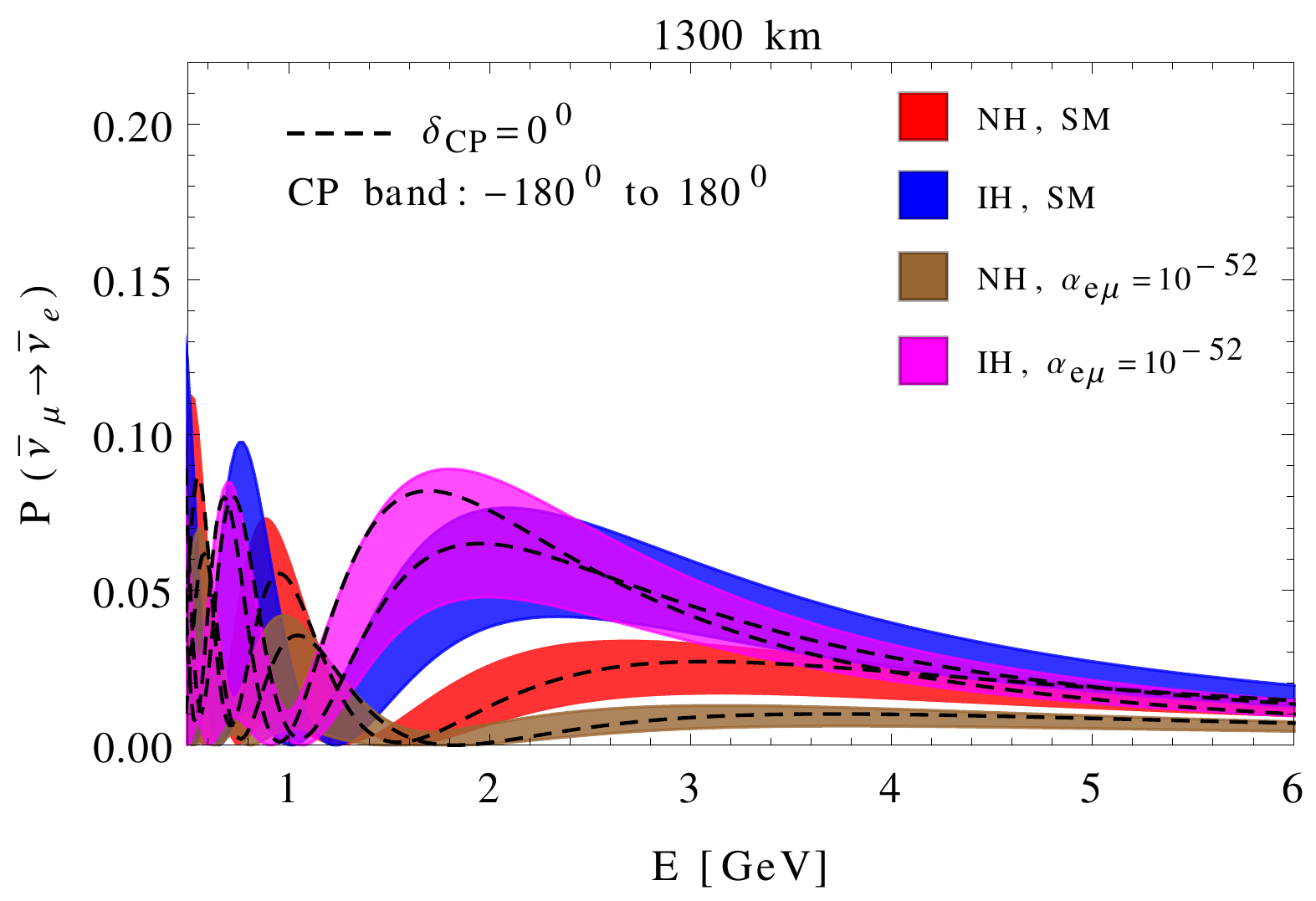}
\includegraphics[width=7.5cm,height=7.0cm]{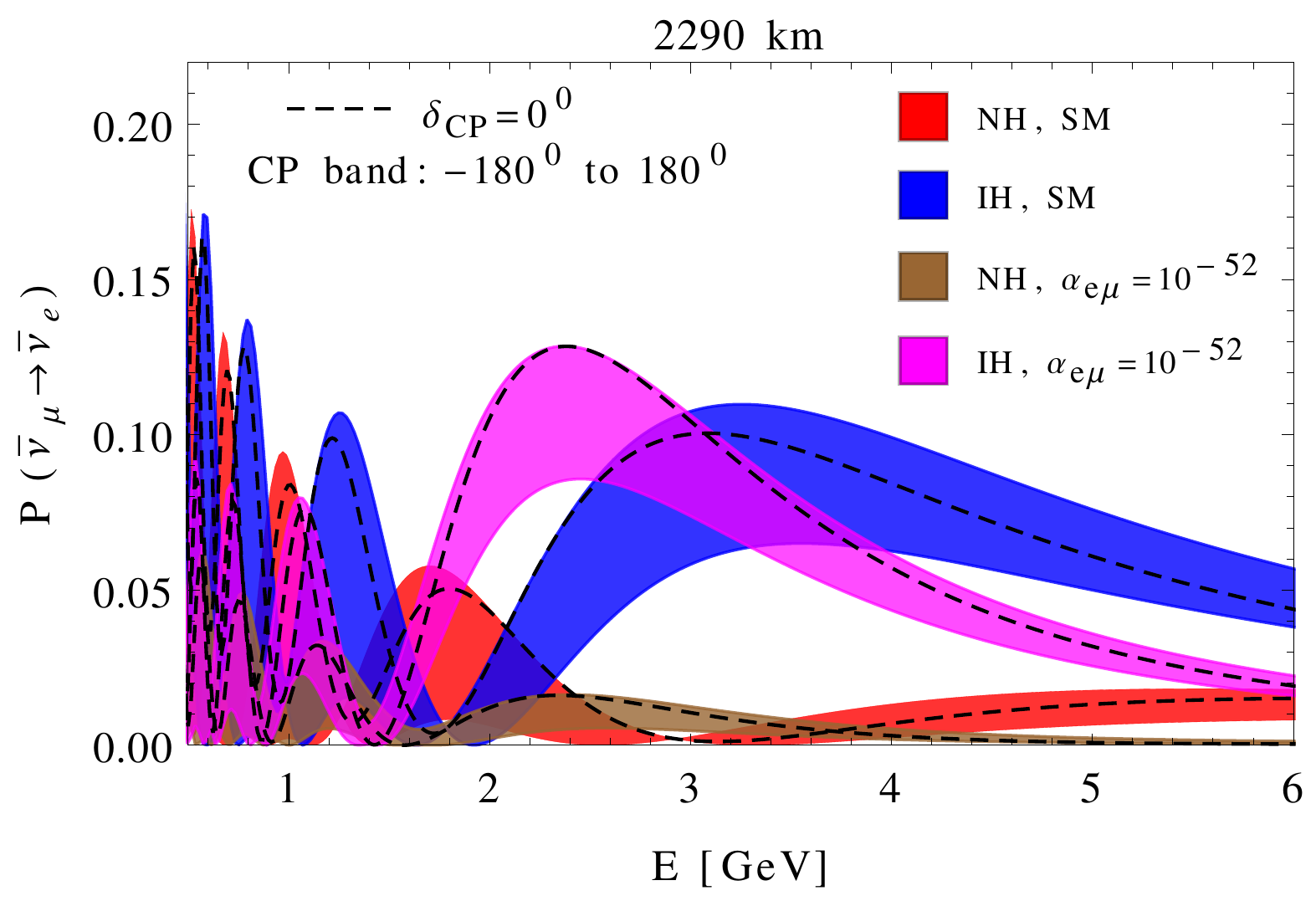}
}
\centerline{
\includegraphics[width=7.5cm,height=7.0cm]{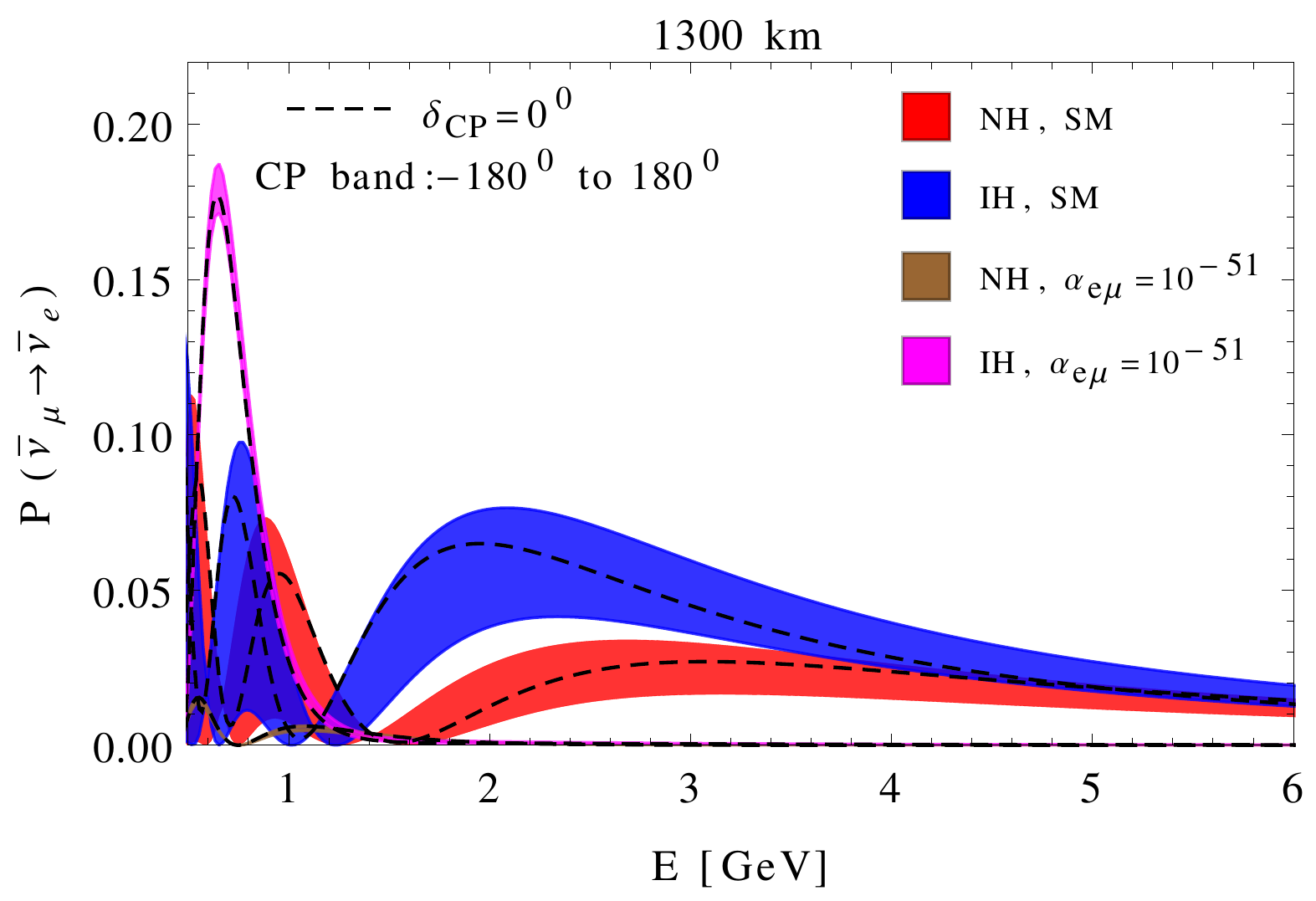}
\includegraphics[width=7.5cm,height=7.0cm]{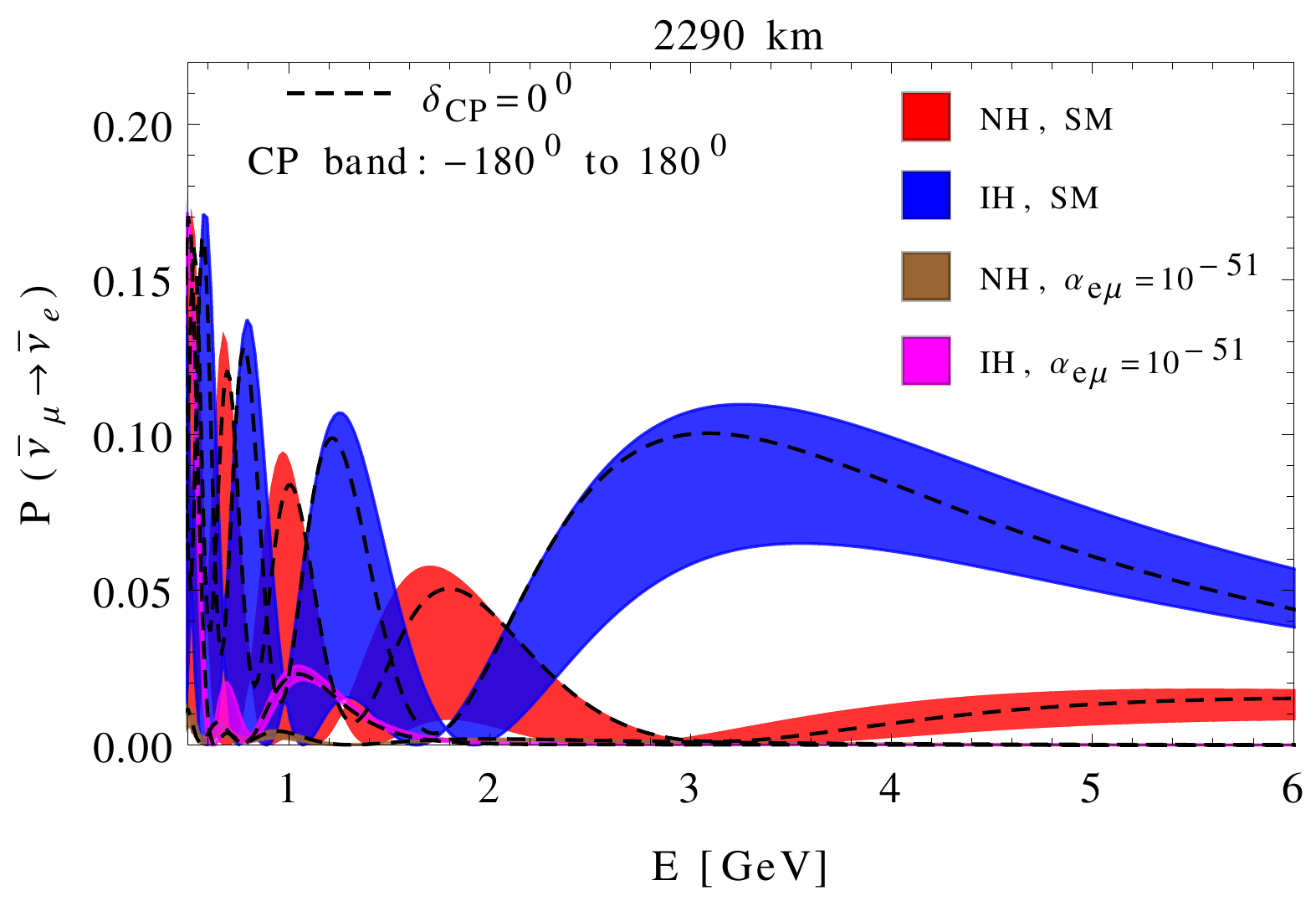}
}
\caption{The transition probability $P_{\bar\mu \bar e}$ as a function 
of anti-neutrino energy. The band reflects the effect of unknown $\dcp$. 
Inside each band, the probability for $\dcp = 0^\circ$ case is shown 
by the black dashed line. The left panels (right panels) are for 1300 km 
(2290 km) baseline. In each panel, we compare the probabilities for 
NH and IH with and without long-range potential. In the upper (lower) panels, 
we take $\alpha_{e\mu}=10^{-52}$ ($\alpha_{e\mu}=10^{-51}$) for the cases 
with long-range potential.}
\label{fig:anti-neutrino-appearance-probability}
\end{figure}
%--------------------------------------------------------

In Fig.~\ref{fig:anti-neutrino-appearance-probability}, we plot the exact numerical 
transition probability $\anumu \to \anue$ as a function of anti-neutrino energy.
The band shows the impact of unknown $\dcp$. Inside each band, the probability 
for $\dcp = 0^\circ$ case is shown by the black dashed line. The left panels 
(right panels) are drawn for 1300 km (2290 km) baseline. In each panel, we compare 
the probabilities for NH and IH with and without long-range potential. 
In the upper (lower) panels, we consider $\alpha_{e\mu}=10^{-52}$ 
($\alpha_{e\mu}=10^{-51}$) for the cases with long-range potential.

%---------------------------------------------------------
\begin{figure}[H]
\centerline{
\includegraphics[width=7.5cm,height=7.0cm]{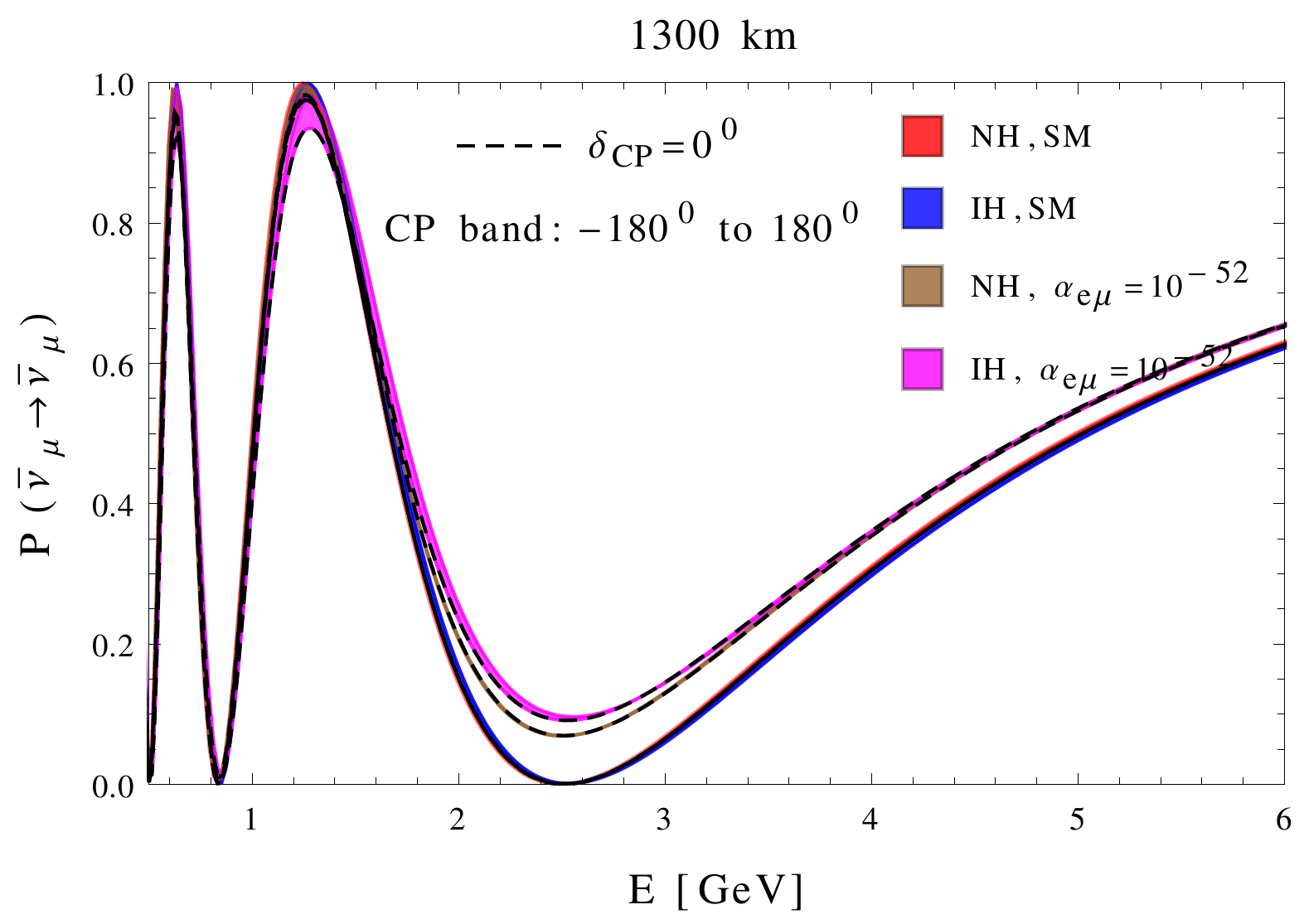}
\includegraphics[width=7.5cm,height=7.0cm]{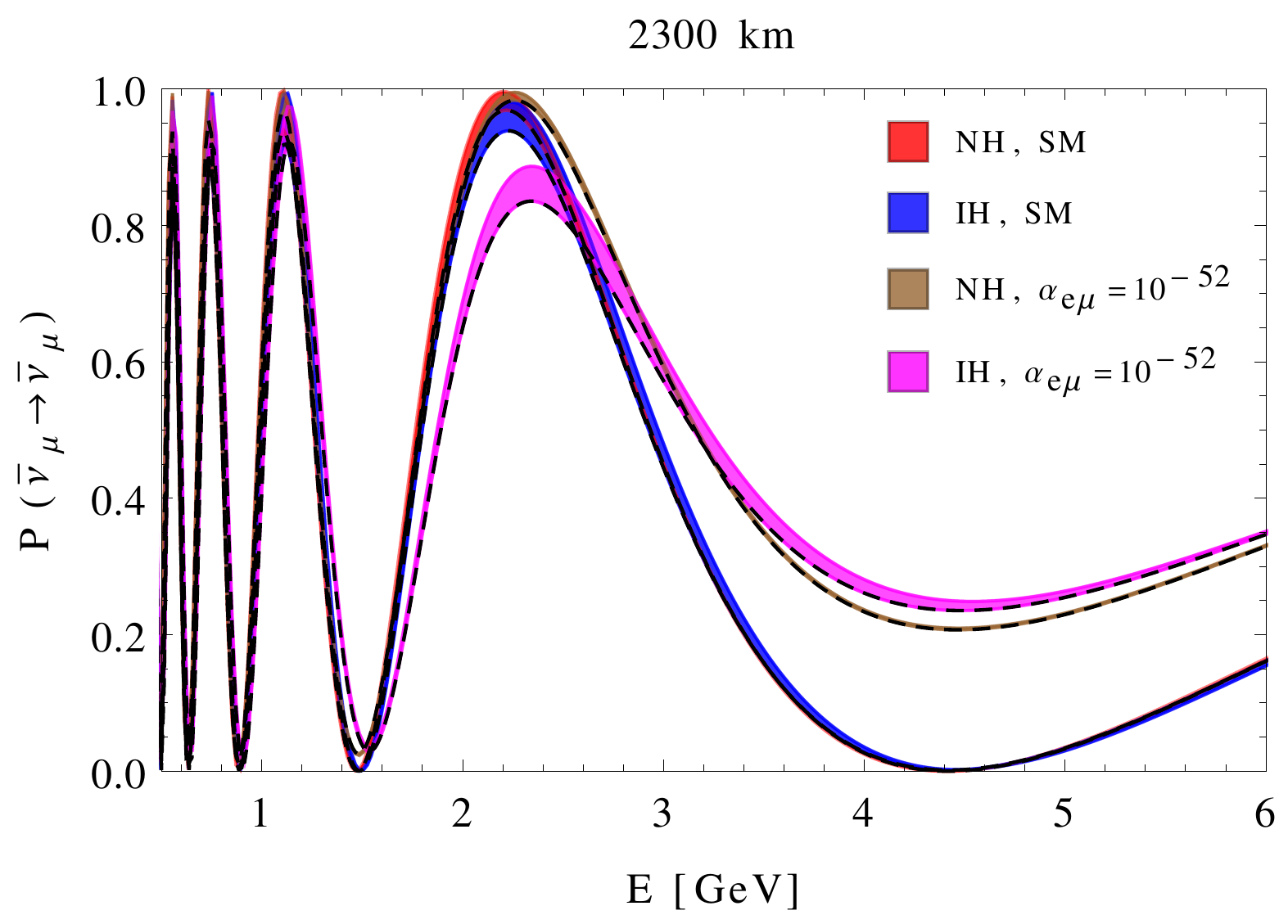}
}
\centerline{
\includegraphics[width=7.5cm,height=7.0cm]{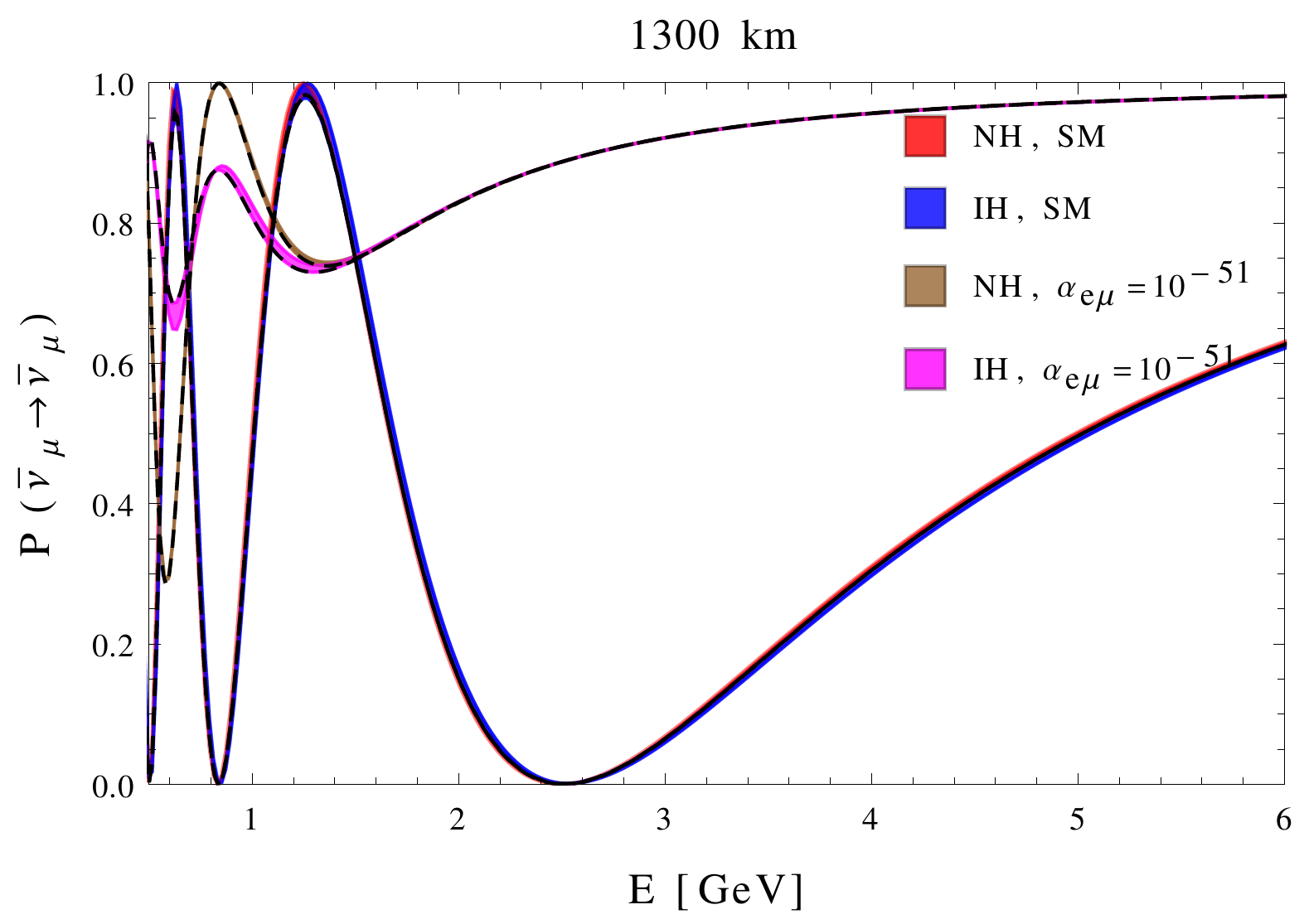}
\includegraphics[width=7.5cm,height=7.0cm]{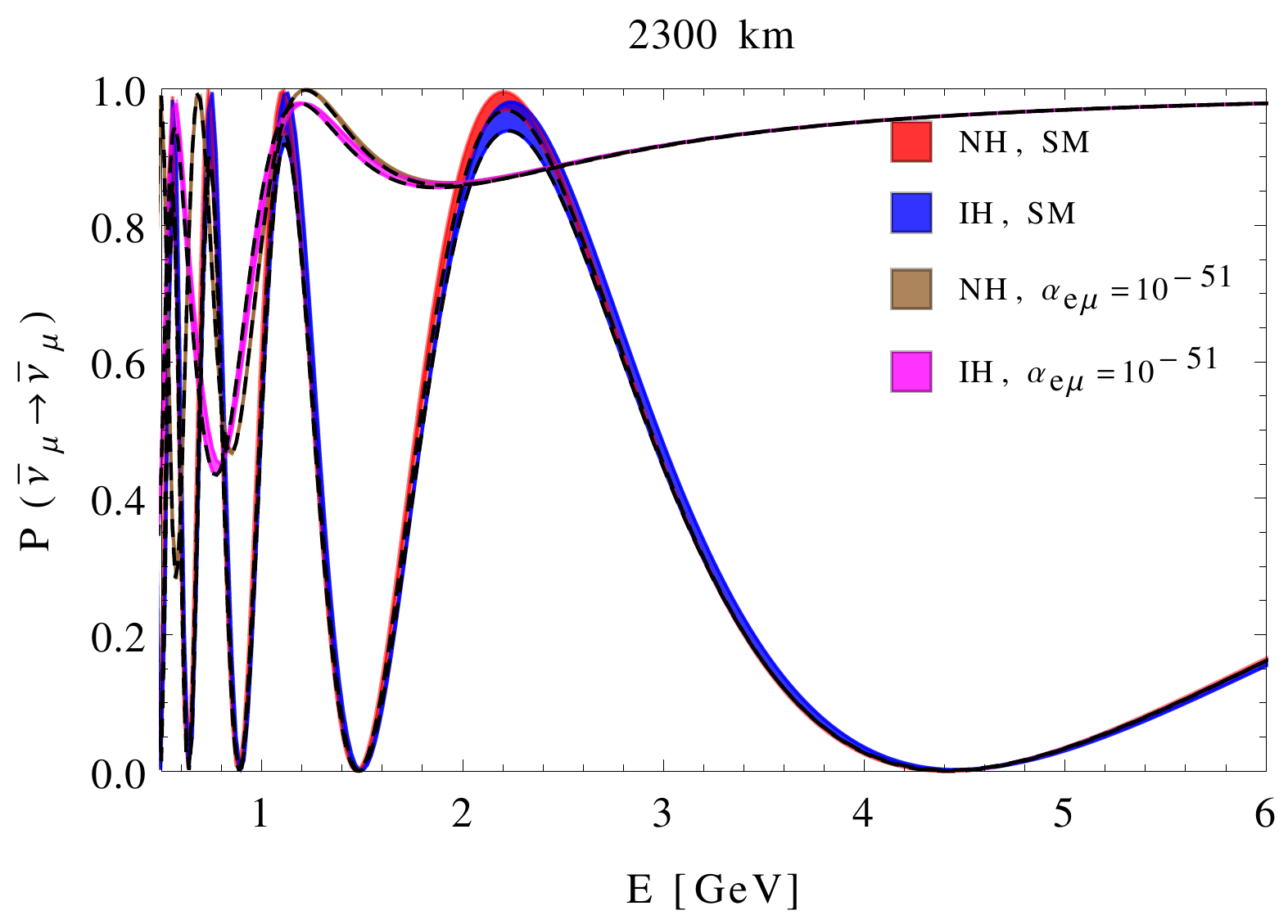}
}
\caption{The transition probability $P_{\bar\mu\bar\mu}$ as a function 
of anti-neutrino energy. The band reflects the effect of unknown $\dcp$. 
Inside each band, the probability for $\dcp = 0^\circ$ case is shown 
by the black dashed line. The left panels (right panels) are for 1300 km 
(2290 km) baseline. In each panel, we compare the probabilities for NH 
and IH with and without long-range potential. In the upper (lower) panels, 
we take $\alpha_{e\mu}=10^{-52}$ ($\alpha_{e\mu}=10^{-51}$) 
for the cases with long-range potential.}
\label{fig:anti-neutrino-disappearance-probability}
\end{figure}
%--------------------------------------------------------

Fig.~\ref{fig:anti-neutrino-disappearance-probability} shows the 
exact numerical $\anumu \to \anumu$ disappearance probability
as a function of anti-neutrino energy. The thin band portrays the 
mild impact of unknown $\dcp$. Inside each band, the probability 
for $\dcp = 0^\circ$ case is given by the black dashed line. The 
left panels (right panels) are drawn for 1300 km (2290 km) 
baseline. In each panel, we compare the probabilities for NH and IH 
with and without long-range potential. In the upper (lower) panels, 
we consider $\alpha_{e\mu}=10^{-52}$ ($\alpha_{e\mu}=10^{-51}$) 
for the cases with long-range potential.

\end{appendix}
%=================================

\bibliographystyle{JHEP}
\bibliography{nsi-lrf-references}

\end{document}